\documentclass[journal]{IEEEtran}
\usepackage{amsmath,amsfonts}
\usepackage{amssymb}
\usepackage{bm}
\usepackage{relsize}
\usepackage{algorithmic}
\usepackage{algorithm}
\usepackage{array}
\usepackage{subfigure}
\usepackage{subfloat}
\usepackage{textcomp}
\usepackage{stfloats}
\usepackage{url}
\usepackage{verbatim}
\usepackage{graphicx}
\usepackage{cite}
\usepackage{color}
\usepackage{makecell}
\usepackage{multirow}
\usepackage{enumitem}
\usepackage[normalem]{ulem}
\newtheorem{theorem}{\underline{Theorem}}
\newtheorem{proposition}{\underline{Proposition}}

\newtheorem{lemma}{\underline{Lemma}}
\begin{document}
\renewcommand{\figurename}{Fig.}

\title{Optimal Beamforming for Integrated Sensing and Communication Exploiting Prior Distribution Information: How Many Beams are Needed?}

\author{Jiayi Yao,~\IEEEmembership{Graduate Student Member,~IEEE}, and Shuowen Zhang,~\IEEEmembership{Senior Member,~IEEE }
	\thanks{This paper was presented in part at the IEEE Global Communications Conference (Globecom) Workshops, Taipei, Dec. 2025 \cite{yao2025beamforming}.}
	\thanks{The authors are with the Department of Electrical and Electronic Engineering, The Hong Kong Polytechnic University, Hong Kong SAR, China (e-mail: jiayi.yao@connect.polyu.hk; shuowen.zhang@polyu.edu.hk).}}

\vspace{-5mm}
\maketitle

\vspace{-5mm}

\begin{abstract}
	This paper studies a multi-target multi-user integrated sensing and communication (ISAC) system where a multi-antenna base station (BS) communicates with multiple single-antenna users in the downlink and senses the \emph{unknown} and \emph{random} angle information of multiple targets based on their reflected echo signals at the BS receiver as well as their \emph{prior distribution information}. We focus on a general transmit beamforming structure with both communication beams and \emph{dedicated sensing beams}, whose design is highly non-trivial as more sensing beams provide more flexibility in multi-target sensing, but introduce extra interference to multi-user communication. This paper aims to unveil bounds on the minimum number of dedicated sensing beams needed to achieve the optimal trade-off between sensing and communication, and devise numerical algorithms that find optimal beamforming solutions with small numbers of sensing beams which yield low algorithmic and implementational complexity. To this end, we first characterize the sensing performance via deriving the \emph{periodic posterior Cram\'er-Rao bound (PCRB)} as a lower bound of the mean-cyclic error (MCE), which is more accurate than the conventional mean-squared error for sensing periodic parameters. Then, we optimize the beamforming to minimize the \emph{maximum periodic PCRB} among all targets to ensure fairness, subject to individual communication rate constraints at multiple users. Despite the non-convexity of this problem, we propose a general construction method for the \emph{optimal solution} by leveraging the semi-definite relaxation (SDR) technique, and derive a \emph{general bound} on the number of dedicated sensing beams needed. Moreover, we unveil the specific structures of the optimal solution in various practical cases, where \emph{tighter bounds} on the number of sensing beams needed are derived (e.g., we analytically prove that \emph{no} or \emph{at most one} sensing beam is needed under stringent rate constraints or with homogeneous targets) together with lower-complexity algorithms that find the optimal solution with small numbers of sensing beams. Next, we extend the above studies to the beamforming optimization problem for minimizing the \emph{sum periodic PCRB} among multiple targets subject to individual user rate constraints. Numerical results validate our derived theoretical bounds and the effectiveness of our proposed beamforming designs.
\end{abstract}


\vspace{-0mm}
\section{Introduction}
The sixth-generation (6G) wireless networks are anticipated to achieve a quantum leap in communication data rate and incorporate sensing as a new function \cite{saad2019vision}. To this end, integrated sensing and communication (ISAC) has attracted increasing research interests in both academia and industry, as it enables the realization of sensing and communication functionalities in a single system to reduce hardware cost and improve spectrum utilization efficiency \cite{liu2022integrated}. It is anticipated that ISAC will support various applications such as smart transportation, smart city, and smart home \cite{zhang2021enabling}. To realize the full potential of ISAC, many new design issues need to be fully investigated, ranging from transceiver architecture and frame structure designs, waveform design, to joint coding design \cite{liu2022survey}. Among them, transmit beamforming design which aims to strike an optimal balance between sensing and communication by the shared use of spatial and power resources is an imperative problem that is critical to the performance of ISAC.
\vspace{-8mm}
\subsection{Motivation for Posterior Cram\'er-Rao Bound (PCRB) based Beamforming Optimization}
\vspace{-1mm}
Existing literature on transmit beamforming design for ISAC mostly focused on the genie-aided case where the parameters to be sensed are assumed to be \emph{known}, for which the $\textit{Cram\'er-Rao bound (CRB)}$ has been adopted as the sensing performance metric \cite{van2004detection,forsythe2005waveform,bekkerman2006target,li2007range,liu2021cramer,liu2020joint,ma2022covert,hua2023secure,hua2023optimal,hua2023mimo}. Specifically, CRB is a lower bound of the sensing mean-squared error (MSE) with any unbiased estimator, which is a function of the \emph{exact} values of the parameters to be sensed. Along this line, \cite{li2007range} studied the multiple-input multiple-output (MIMO) radar waveform optimization with multiple sensing targets, and \cite{liu2021cramer} studied the transmit beamforming design to minimize the CRB for the MSE in sensing both point target and extended target under user communication quality constraints, where the optimal beamforming solutions for both point and extended target cases were obtained in a single-user system. With the optimal beamforming design that minimizes the CRB, a genie-aided lower bound of the MSE with any unbiased estimator can be obtained, which is useful for sensing performance evaluation.

\begin{figure}[t]
	\centering
	\includegraphics[width=8.5cm]{Fig_1_v2.png}
	\vspace{-4mm}
	\caption{Framework for prior-based beamforming optimization and sensing.}
	\vspace{-2mm}
	\label{PCRB_CRB_block_diagram}
\end{figure}

In practice, the parameters to be sensed are \emph{unknown} before beamforming and sensing are performed, while their probability density functions (PDFs) can be known \emph{a priori} based on target appearance pattern and historic data \cite{xu2023mimo,xu2023mimo1,yao2024optimal,wang2024hybrid,xu2024integrated,hou2023optimal,zheng2025beyond,attiah2026uplink,attiah2026many,liu2025ris}. In this case, $\textit{posterior Cram\'er-Rao bound (PCRB)}$ \cite{xu2023mimo,xu2023mimo1,yao2024optimal,wang2024hybrid,xu2024integrated,hou2023optimal,zheng2025beyond}, also called Bayesian Cram\'er-Rao bound (BCRB) \cite{attiah2026uplink,attiah2026many,liu2025ris}, can be adopted to characterize a lower bound of the MSE when such prior distribution information is exploited. The overall framework for prior PDF based beamforming and sensing is illustrated in Fig. \ref{PCRB_CRB_block_diagram}. From a practical perspective, different from the CRB, PCRB is only dependent on the unknown parameters' prior distribution information instead of their exact values, making it not only a valuable bound for evaluating the performance of sensing algorithms, but also a suitable sensing performance metric for beamforming optimization, where the obtained beamforming design can be directly applied in practice since it is \emph{independent} of the unknown values of the parameters to be sensed. On the other hand, from a theoretical perspective, PCRB has a drastically different form compared to the CRB, as the posterior Fisher information matrix (PFIM) involves complex integrations over the parameters' possible values weighted by their corresponding probability densities as well as additional terms solely from the prior information. Therefore, beamforming optimization problems with PCRB as the sensing performance metric are fundamentally different from their genie-aided counterparts with CRB as the sensing performance metric. For example, in location sensing, the PCRB-minimizing beampattern generally needs to cover all possible highly-probable locations and achieve a novel \emph{probability-dependent power focusing} effect, as revealed in our prior works \cite{xu2023mimo,xu2023mimo1}; while in contrast, the CRB-minimizing beampattern tends to focus power at the target's actual location (which is assumed to be known from a genie).\looseness=-1
\vspace{-4mm}
\subsection{Prior Works and Research Gap}
\vspace{-0mm}
The above differences are evidenced by the initial results on beamforming optimization for sensing or ISAC systems with PCRB as the performance metric \cite{xu2023mimo,xu2023mimo1,yao2024optimal,wang2024hybrid,xu2024integrated,hou2023optimal,zheng2025beyond,attiah2026uplink,attiah2026many,liu2025ris}. For sensing-only MIMO radar systems, \cite{xu2023mimo,xu2023mimo1} characterized the PCRB of the MSE for sensing the angle of a single point target, derived the optimal PCRB-minimizing beamforming, and proved that one beam suffices to achieve the optimal sensing performance; while the optimal beamforming for minimizing the sum PCRB of the MSE in sensing the angles of multiple targets was obtained in \cite{yao2024optimal}. In ISAC systems, the \emph{dual-functional} transmit beamforming optimization with PCRB as the performance metric has been studied in a few works. \cite{xu2023mimo1} investigated the PCRB minimization problem for single-target angle sensing while satisfying the communication rate constraint of a multi-antenna user. It was derived that the rank of the optimal transmit covariance matrix is bounded by the rank of the MIMO communication channel, although the sensing target may have a continuous angle PDF with a large number of possible locations. This work was extended in \cite{wang2024hybrid} considering a hybrid analog-digital architecture at the base station (BS) transmitter.

Although dual-functional beamforming incurs minimum change to the transmit signal structure of conventional communication-only systems, it may not be always optimal, particularly in multi-target sensing. Specifically, with various possible values of each parameter to be sensed, the beamforming design needs to accommodate all such values based on their associated probabilities, which is especially challenging as the number of parameters increases. Moreover, the beamforming design needs to strike an optimal balance among the sensing performance of multiple targets to ensure their fairness, or minimize the total sensing error to maximize the overall sensing efficacy. Hence, \emph{dedicated sensing beams} may be needed to provide \emph{sufficiently large design degrees-of-freedom}, especially when the number of communication beams is limited (e.g., by the numbers of antennas at the communication users). However, such sensing beams will also introduce \emph{additional interference} to communication, which makes their optimal design highly non-trivial. It is also worth noting that the number of extra dedicated sensing beams determines the number of radio frequency (RF) chains needed at the transmitter. Hence, unveiling the \emph{minimum number of dedicated sensing beams needed} for achieving the optimal ISAC performance is of paramount practical importance, especially under large-scale transmit antenna arrays. 

Research along this direction is still in its infancy. For a single-target single-user ISAC system, \cite{xu2024integrated} derived the optimal beamforming with dedicated sensing beams, and analytically proved that at most \emph{one} dedicated sensing beam is needed. 
\cite{hou2023optimal} derived the optimal design of artificial noise beams as dedicated sensing beams for a single-target secure ISAC system. A general bound on the number of sensing beams needed was provided in a recent work \cite{attiah2026many}. However, for general multi-target multi-user multi-antenna ISAC systems, how to design transmit beamforming to achieve an \emph{optimal sensing-communication trade-off} by exploiting the prior distribution information still remains an unaddressed problem, while the \emph{number of sensing beams} needed is still unknown.\looseness=-1

\vspace{-2mm}
\subsection{Main Contributions}
\begin{figure}[t]
	\centering
	\includegraphics[width=8.5cm]{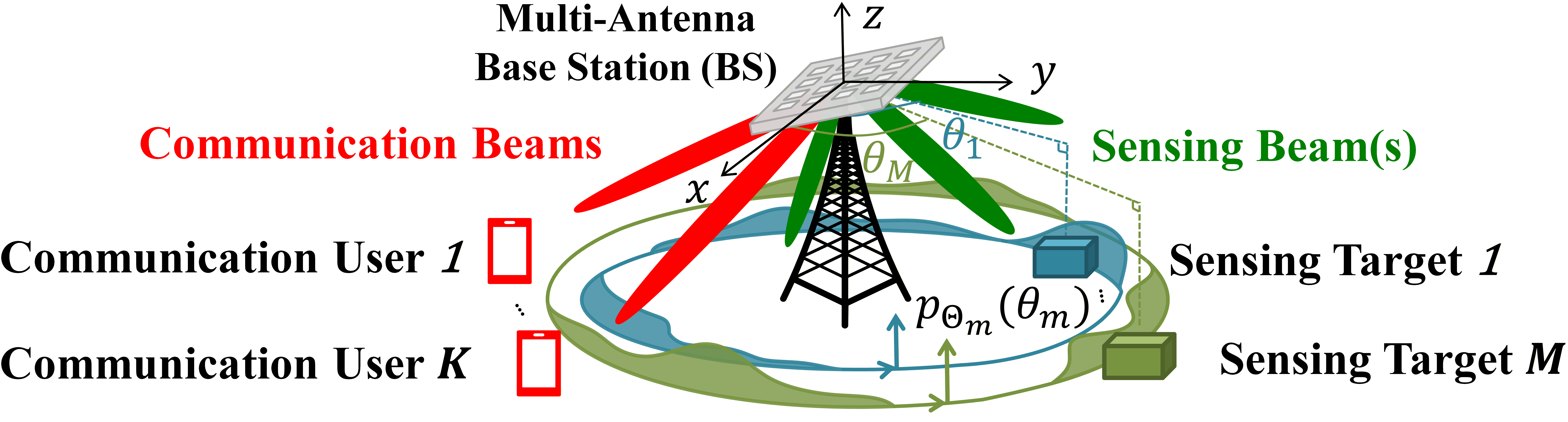}
	\vspace{-4mm}
	\caption{Illustration of a multi-target multi-user ISAC system with prior distribution information on unknown parameters.}
	\vspace{-1mm}
	\label{system_model}
\end{figure}

\vspace{-0mm}
Motivated by the above, this paper studies the transmit beamforming optimization in an ISAC system where a multi-antenna BS communicates with multiple single-antenna users in the downlink and senses the \emph{unknown} and \emph{random} angles of multiple targets based on their reflected signals and prior distribution information, as illustrated in Fig. \ref{system_model}. We focus on a general beamforming structure with one communication beam for each single-antenna user and potentially multiple \emph{dedicated sensing beams}. Our main contributions are summarized below.
\vspace{-5mm}
\begin{itemize}[leftmargin=*]
	\item Firstly, we adopt a novel mean-cyclic error (MCE) metric to quantify the angle sensing error due to its accuracy and suitability in handling \emph{periodic} parameters to be sensed. We derive the \emph{periodic PCRB} as a lower bound of the MCE in multi-target angle sensing, which is an  explicit and tractable function of the transmit beamforming and the prior PDF.
	\item Next, we study the transmit beamforming optimization problem to minimize the \emph{maximum (worst-case) periodic PCRB} among the multiple targets, under individual rate constraints at the multiple communication users. To tackle this non-convex problem, we leverage the semi-definite relaxation (SDR) technique and prove its tightness. Based on this, we propose a \emph{general} construction method for the \emph{optimal solution}, and prove the number of dedicated sensing beams needed is no larger than the \emph{square root of the number of sensing targets} as long as the unknown parameters are independent of each other and the targets' reflection coefficients are of zero mean. Moreover, by further examining the Karush-Kuhn-Tucker (KKT) optimality conditions of the problem after SDR, we unveil the \emph{specific structures} of the optimal solution to the original problem in various practical cases and devise low-complexity methods for obtaining the optimal solution. Furthermore, we derive \emph{tighter bounds on the number of dedicated sensing beams needed}. For instance, \emph{no} sensing beam is needed in the \emph{high-rate regime}, while \emph{at most one} sensing beam is needed when the targets are \emph{homogeneous} with identical distributions.\looseness=-10
	\item Furthermore, we extend the above studies to the transmit beamforming optimization problem to minimize the \emph{sum periodic PCRB} among the multiple targets under individual rate constraints at the multiple communication users. This allows more flexibility in the beamforming design to enhance the overall sensing performance at the cost of potentially sacrificed fairness among targets.
	\item Finally, we provide extensive numerical results which validate the derived general bound and tighter bounds. It is shown that the number of dedicated sensing beams generally increases as the number of targets increases and/or the number of users decreases, and is sensitive to the angle PDF of the multiple targets. Furthermore, our proposed optimal solutions are shown to outperform various benchmark schemes in terms of both periodic PCRB and MCE, and achieve close performance to the sensing-oriented scheme in the low-rate regime.
\end{itemize}

The rest of this paper is organized as follows. Section \ref{section_system_model} presents the system model. Section \ref{section_sensing_performance} characterizes the multi-target sensing performance via deriving the periodic PCRB. Section \ref{section_min_max} formulates and solves the beamforming optimization problem to minimize the maximum (worst-case) periodic PCRB among multiple targets under individual communication user rate constraints. Section \ref{section_min_sum} extends the studies to the sum periodic PCRB minimization problem. Section \ref{numerical_results} provides numerical results. Finally, Section \ref{section_conclusion} concludes this paper.

\emph{Notations:} Boldface lower-case letters denote vectors and boldface upper-case letters denote matrices. $\mathbb{C}^{N\times L}$ and $\mathbb{R}^{N\times L}$ denote the spaces of $N\times L$ complex or real matrices, respectively. $\bm{I}_N$ denotes an $N\times N$ matrix, and $\bm{0}$ denotes an all-zero matrix with appropriate dimension. Imaginary unit is denoted by $j=\sqrt{-1}$. For a complex number, $|\cdot|$ and $\mathfrak{Re}\{\cdot\}$ represent the absolute value and real part, respectively. $(\cdot)^T$ and $(\cdot)^H$ denote transpose and conjugate transpose, respectively. $\otimes$ represents the Kronecker product. For a vector, $\|\cdot\|$ denotes its $l_2$-norm. For an arbitrary-sized matrix, $\mathrm{rank}(\cdot)$, $\lambda_{\max}(\cdot)$, and $[\cdot]_{i,j}$ denote its rank, largest eigenvalue, and $(i,j)$-th element, respectively. For a square matrix, $\mathrm{det}(\cdot)$, $\mathrm{tr}(\cdot)$, and $(\cdot)^{-1}$ represent the determinant, trace, and inverse, respectively. $\bm{X}\succeq \bm{0}$ means $\bm{X}$ is a positive semi-definite (PSD) matrix. $\mathrm{diag}\{x_1,...,x_M\}$ denotes an $M\times M$ matrix with $x_1,...,x_M$ being its diagonal elements. $\mathrm{card}(\mathcal{K})$ represents the cardinality of a set $\mathcal{K}$. $\mathcal{CN}(\bm{0},\bm{\Sigma})$ denotes the distribution of a circularly symmetric complex Gaussian (CSCG) random vector with zero mean and covariance matrix $\bm{\Sigma}$; and $\sim$ means ``distributed as''. $\mathbb{E}[\cdot]$ represents the statistical expectation. $\mathcal{O}(\cdot)$ denotes the standard big-O notation. $\dot{f}(\cdot)$ denotes the derivative of $f(\cdot)$.
\vspace{-2mm}
\section{System Model}\label{section_system_model}
\vspace{-0mm}
Consider a multi-antenna multi-target multi-user ISAC system with $N_t\geq 1$ transmit antennas and $N_r\geq 1$ co-located receive antennas at the BS, as illustrated in Fig. \ref{system_model}. By transmitting beamformed signals in the downlink, the BS aims to sense the \emph{unknown} and \emph{random} azimuth angles of $M\geq 1$ point targets via the echo signals reflected by the targets and received at the BS, as well as to simultaneously communicate with $K\geq 1$ single-antenna communication users. Let $\mathcal{M}=\{1,...,M\}$ and $\mathcal{K}=\{1,...,K\}$ denote the sets of targets and users, respectively. Let $h_\mathrm{B}$ and $h_m$ denote the height of the BS antennas and the $m$-th target, respectively. Let $\theta_m\in[-\pi ,\pi),\ \forall m\in\mathcal{M}$ denote the azimuth angle of each $m$-th target with respect to the BS, as illustrated in Fig. \ref{system_model}. We assume that the angles of different targets are independent of each other and let $p_{\Theta_m}(\theta_m)$ denote the PDF of the angle $\theta_m$ which is assumed to be known \emph{a priori} via historic data or target appearance pattern \cite{yao2024optimal,xu2023mimo1,hou2023optimal,zheng2025beyond,wang2024hybrid,attiah2026uplink,xu2024integrated,attiah2026many,liu2025ris}.\looseness=-1

We consider a quasi-static block-fading frequency-flat channel model between the BS and the communication users, where the channels remain unchanged within each channel coherence block consisting of $L_\mathrm{C}$ symbol intervals. Denote $\bm{h}^H_k\in \mathbb{C}^{1\times N_t}$ as the channel from the BS to the $k$-th user, which is assumed to be perfectly known at both the $k$-th user and the BS. Denote $L$ as the total number of symbol intervals used for sensing $\theta_m$'s, during which the targets remain static. We further focus on the case where $L\leq L_{\mathrm{C}}$, while our studies are readily applicable to the case where $L>L_{\mathrm{C}}$ by designing beamforming separately in each channel coherence block.

We focus on a linear transmit beamforming design within the $L$ symbol intervals during which the communication channels and sensing targets' angles remain unchanged. Specifically, let $\bm{w}_k\in \mathbb{C}^{N_t\times 1}$ denote the linear beamforming vector for each user $k$. We further consider the general beamforming structure with potentially \emph{dedicated sensing beams}, as using $K$ dual-functional beams for both communication and sensing may not be sufficient for high-quality sensing of multiple targets each with a wide range of possible locations. Let $\bm{S}=[\bm{s}_1,...,\bm{s}_{N_t}]\in \mathbb{C} ^{N_t\times N_t}$ denote the sensing beamforming matrix where $\bm{s}_i\in \mathbb{C}^{N_t\times 1}$ denotes the beamforming vector for the $i$-th sensing beam. Note that there are at most $N_t$ sensing beams, and the number of non-zero columns in the optimized $\bm{S}$ will represent the number of dedicated sensing beams needed. Let $\bm{c}_l\sim\mathcal{CN}(\bm{0},\bm{I}_K)$ denote the collection of independent information symbols for the $K$ communication users and $\bm{v}_l\in \mathbb{C}^{N_t\times 1}$ denote the dedicated sensing signals in each $l$-th symbol interval, where the elements in $\bm{c}_l$ and $\bm{v}_l$ all have unit average power and are independent of each other. The baseband equivalent transmit signal vector in each $l$-th symbol interval is thus given by
\vspace{-0.5mm}
\begin{align}
	\bm{x}_l=\sum_{k=1}^K\bm{w}_kc_{l,k}+\bm{S}\bm{v}_l.
\end{align}
\vspace{-2.5mm}%
\par\noindent%
The transmit covariance matrix is given by $\sum_{k=1}^K\bm{w}_k\bm{w}_k^H+\bm{SS}^H$. Denote $P$ as the total transmit power budget across all antennas, which yields $\sum_{k=1}^K \|\bm{w}_k\|^2+\mathrm{tr}(\bm{SS}^H ) \leq P$.
\vspace{-6mm}
\subsection{Multi-User Communication Model and Performance}
\vspace{-2mm}
Based on the above, the received signal at the $k$-th communication user receiver in each $l$-th symbol interval is given by
\vspace{-3mm}
\begin{align}
	y^\mathrm{C}_{l,k}=\bm{h}_k^H\bm{w}_kc_{l,k}+\sum_{j=1,j\neq k}^K\bm{h}_k^H\bm{w}_jc_{l,j}+\bm{h}_k^H\bm{S}\bm{v}_l+n^\mathrm{C}_{l,k},
\end{align}
where $n^\mathrm{C}_{l,k}\sim \mathcal{C}\mathcal{N}(0,\sigma^2_\mathrm{C})$ denotes the CSCG noise at the $k$-th communication user receiver in the $l$-th symbol interval, with $\sigma^2_\mathrm{C}$ denoting the average noise power. We consider a practical and challenging scenario where the dedicated sensing signals are not known at the user receivers, thus introducing extra interference; moreover, the sensing signals follow the CSCG distribution which corresponds to the worst-case interference. The achievable rate for the $k$-th communication user is given by\looseness=-1
\begin{equation}\label{rate}
	R_k=\log_2\left(1+\frac{|\bm{h}_k^H\bm{w}_k|^2}{\sum_{j=1,j\ne k}^K{|\bm{h}_{k}^{H}\bm{w}_j|^2}+\|\bm{h}_k^H\bm{S}\|^2+\sigma _\mathrm{C}^{2}}\right) \! \!\!
\end{equation}
in bits per second per Hertz (bps/Hz).
\vspace{-6.5mm}
\subsection{Multi-Target Sensing Model}
\vspace{-2mm}
Unlike downlink communication which is interfered by the dedicated sensing beams, multi-target sensing can make use of \emph{both} sensing and communication beams, as the sensing signals in $\{\bm{v}_l\}_{l=1}^L$ and the information symbols in $\{\bm{c}_l\}_{l=1}^L$ are both generated and known at the BS. Considering a line-of-sight (LoS) channel model between the BS and each target to draw fundamental insights, the channel from the BS transmitter to the BS receiver via the reflection of each $m$-th target is modeled as $\bm{G}_m(\theta_m) =\alpha_m\bm{b}(\theta_m) \bm{a}^H(\theta_m)$. Specifically, $\alpha_m=\sqrt{\frac{\lambda^2}{(4\pi)^3r_m^4}}e^{-j\frac{4\pi r_m}{\lambda}}\eta_m=\alpha_m^\mathrm{R}+j\alpha_m^\mathrm{I}\in \mathbb{C}$ denotes the overall reflection coefficient for the $m$-th target, where $\lambda$ denotes the wavelength in meter (m), $r_m$ denotes the distance between the BS and the $m$-th target in m, and $\eta_m$ denotes the radar cross-section (RCS) coefficient. The reflection coefficient $\alpha_m$ is generally \emph{unknown} and assumed to follow a known distribution. Moreover, $\bm{b}(\theta_m)\in \mathbb{C}^{N_r\times 1}$ and $\bm{a}^H(\theta_m)\in \mathbb{C}^{1\times N_t}$ denote the array steering vectors at the BS receive antennas and the BS transmit antennas at angle $\theta_m$, respectively, in which $\theta_m$ is the only unknown parameter.\footnote{Note that this is applicable to various practical scenarios, e.g., when both $r_m$'s and $\eta_m$'s are unknown and random with $r_m\gg h_\mathrm{B}-h_m,\ \forall m\in\mathcal{M}$, or when $r_m$'s are known while $\eta_m$'s are unknown and random.} Assuming the $M$ targets are within the same range bin, the overall MIMO channel via the reflections of $M$ targets is modeled as $\sum_{m=1}^M\bm{G}_m(\theta_m)=\sum_{m=1}^M\alpha_m\bm{b}(\theta_m)\bm{a}^H\!(\theta_m)$.\footnote{In this paper, we ignore the effect of multi-reflection links via multiple targets, since their channel gains are generally small and negligible.} The received echo signal vector at the BS receiver in each $l$-th symbol interval is thus given by
\begin{equation}
	\bm{y}_l^\mathrm{S}
	=\sum_{m=1}^M\alpha_m\bm{b}(\theta_m)\bm{a}^H\!(\theta_m)\left(\sum_{k=1}^K\bm{w}_kc_{l,k}+\bm{S}\bm{v}_l\right)+\bm{n}_l^\mathrm{S},	\label{Y}
\end{equation}
\vspace{-1mm}%
\par\noindent%
where $\bm{n}^\mathrm{S}_l\sim \mathcal{C}\mathcal{N}(\mathbf{0},\sigma^2_\mathrm{S}\bm{I}_{N_r})$ denotes the CSCG noise vector at the BS receiver in each $l$-th symbol interval, with $\sigma^2_\mathrm{S}$ denoting the average receiver noise power.

Let $\bm{\alpha}=[\alpha _{1}^{\mathrm{R}},\alpha_1^{\mathrm{I}},...,\alpha_M^{\mathrm{R}},\alpha_M^{\mathrm{I}}]^T$ and $\bm{\theta}=[\theta_1,...,\theta_M]^T$ denote the collections of the reflection coefficients and angles for the $M$ targets, respectively. Since  $\bm{\alpha}$ is also unknown in (\ref{Y}), the sensing of targets' angles in $\bm{\theta}$ need to consider the randomness of $\bm{\alpha}$ based on the received echo signals $\bm{Y}^\mathrm{S}=[\bm{y}^\mathrm{S}_1,...,\bm{y}^\mathrm{S}_L]$ and the prior distribution information of all unknown parameters in $\bm{\zeta }=[\bm{\theta}^T,\bm{\alpha }^{T}] ^T\in \mathbb{R} ^{3M \times 1}$. Note that both the echo signals for sensing in (\ref{Y}) and the communication rate in (\ref{rate}) are critically dependent on the beamforming designs in $\{\bm{w}_k\}_{k=1}^K$ and $\bm{S}$. To optimize beamforming and unveil the number of dedicated sensing beams needed, we will first characterize the multi-target sensing performance for $\bm{\theta}$ exploiting prior \hbox{distribution information.}
\vspace{-4mm}
\section{Multi-Target Sensing Performance Characterization Exploiting Prior Information}\label{section_sensing_performance}
\vspace{-1mm}
Note that each angle in $\bm{\theta}$ to be sensed is a \emph{periodic} parameter with period $2\pi$. Thus, the MSE with directly calculated squared error may not be the most suitable sensing error metric. For instance, if $\theta_m=2\pi-\epsilon$ and its estimate $\hat{\theta}_m=\epsilon$, where $\epsilon$ is a small positive value, the direct error is $2\pi-2\epsilon$, while the actual \emph{cyclic (periodic)} error is only $2\epsilon\ll 2\pi-2\epsilon$. Motivated by this, we adopt the MCE as the sensing error metric, which is given by $\mathrm{MCE}(\hat{\theta}_m) =2-2\mathbb{E}_{\bm{Y}^\mathrm{S},\bm{\zeta}}[\cos(\hat{\theta}_m-\theta_m)]$ \cite{willsky2003fourier,shen2010fundamental1,nitzan2015new}. Note that the MCE is guaranteed to be no larger than the MSE given by $\mathbb{E}_{\bm{Y}^\mathrm{S},\bm{\zeta}}[|\hat{\theta}_m-\theta_m|^2]$ due to exploitation of the periodic property, thus being more suitable for periodic parameters. As the exact value of MCE or the minimum MCE is determined by the specific estimator applied and is difficult to be expressed analytically, we derive the \emph{periodic PCRB} as a lower bound of it for any estimator exploiting prior distribution information.

Let $p_{\mathrm{Z}}(\bm{\zeta})$ and $p_{\alpha _{m}^{\mathrm{R}},\alpha _{m}^{\mathrm{I}}}(\alpha _{m}^{\mathrm{R}},\alpha _{m}^{\mathrm{I}})$ denote the (joint) PDF of $\bm{\zeta}$ and the PDF for each $\alpha_m$, respectively. We consider the following practical and mild conditions for $\bm{\theta}$ and $\bm{\alpha}$:
\vspace{-0mm}
\begin{align}
	p_{\mathrm{Z}}(\bm{\zeta})&\!=\!\prod_{m=1}^M{p_{\Theta_m}(\theta_m)p_{\alpha _{m}^{\mathrm{R}},\alpha _{m}^{\mathrm{I}}}( \alpha _{m}^{\mathrm{R}},\alpha _{m}^{\mathrm{I}} )},\label{mild_condition_1}  \\[-0mm] 
	\mathbb{E}[\alpha_m]&\!=\!\iint{\!(\alpha _{m}^{\mathrm{R}}\!+\!j\alpha _{m}^{\mathrm{I}})p_{\alpha _{m}^{\mathrm{R}},\alpha _{m}^{\mathrm{I}}}( \alpha _{m}^{\mathrm{R}},\alpha _{m}^{\mathrm{I}} ) d\alpha _{m}^{\mathrm{R}}d\alpha _{m}^{\mathrm{I}}}\!=\!0. \label{mild_condition_2} 
\end{align}
\vspace{-3mm}

Note that the first condition means all parameters in $\bm{\theta}$ and $\{\alpha_m\}_{m=1}^M$ are independent of each other. The second condition means each $\alpha_m$ is a zero-mean random variable (RV), which holds for a wide variety of RVs including all proper complex RVs (e.g., CSCG RVs). For example, the RCS coefficient is typically uncorrelated with the target-BS distance, thus $\mathbb{E}[\alpha_m]=0$ if $\mathbb{E}[\eta_m]=0$, which holds under typical RCS models such as the Swerling-I or -II model that considers the target as a large number of small fluctuating reflectors, and the Swerling-0 model that treats the target as a nearly spherical reflector with microscopic structural vibrations \cite{swerling1960probability,aittomaki2010performance}.\footnote{It is worth noting that our framework can also be extended to the general case without the mild conditions, which will lead to more complex PCRB expressions. Due to limited space, we leave the complete development of the beamforming optimization algorithms in such general case as our future work.}\looseness=-10

Define $\bm{A}_m\overset{\Delta}{=}\int\dot{\bm{M}}^H\!(\theta _m) \dot{\bm{M}}(\theta _m) p_{\Theta_m}(\theta_m)d\theta_m\succeq \bm{0}$ with $\bm{M}(\theta_m) \overset{\Delta}{=}\!\bm{b}( \theta _m) \bm{a}^H(\theta _m)$; $\beta _m\!\overset{\Delta}{=}\frac{2Lc_m}{\sigma_\mathrm{S} ^2}>0$ with $c_m\!\overset{\Delta}{=}\!\iint\!({\alpha _{m}^{\mathrm{R}}} ^2+{\alpha _{m}^{\mathrm{I}}} ^2) p_{\alpha _{m}^{\mathrm{R}},\alpha _{m}^{\mathrm{I}}}\!\!( \alpha _{m}^{\mathrm{R}},\alpha _{m}^{\mathrm{I}} ) d\alpha _{m}^{\mathrm{R}}d\alpha _{m}^{\mathrm{I}}$; and $\delta _m\!\!\overset{\Delta}{=}\!\!\big[ \mathbb{E} _{\bm{\zeta }}\big[ \frac{\partial \ln ( \tilde{p}_{\mathrm{Z}}( \tilde{\bm{\zeta}} ) )}{\partial \tilde{\bm{\zeta}}} \big( \frac{\partial \ln ( \tilde{p}_{\mathrm{Z}}(\tilde{\bm{\zeta}} ) )}{\partial \tilde{\bm{\zeta}}} \big) ^T \big] \big] _{m,m}$ where $\tilde{p}_{\mathrm{Z}}(\tilde{\bm{\zeta}})=p_{\mathrm{Z}}(\tilde{\bm{\zeta}}+\bm{\varsigma}(\tilde{\bm{\theta}}))$ is a $2\pi$-periodic extended PDF, $\tilde{\bm{\theta}}=[\tilde{\theta}_1,...,\tilde{\theta}_M]^T\in\mathbb{R}^{M\times 1}$ with $\theta_m=\tilde{\theta}_m-2\pi \lfloor \frac{\tilde{\theta}_m+\pi}{2\pi} \rfloor$, $\tilde{\bm{\zeta}}=[\tilde{\bm{\theta}}^T,\bm{\alpha }^{T}]^T$, and $\bm{\varsigma}(\tilde{\bm{\theta}})=[-2\pi \lfloor \frac{\tilde{\theta}_1+\pi}{2\pi} \rfloor,...,-2\pi \lfloor \frac{\tilde{\theta}_M+\pi}{2\pi} \rfloor,0,...,0]^T$. Let $\bm{F}\!\!\in\!\! \mathbb{R}^{3M\times 3M}$ denote the periodic PFIM of $\bm{\zeta}$ \cite{nitzan2015new}. We then have the following theorem.
\vspace{-0mm}
\begin{theorem}\label{prop_PCRB_expression}
	Under the conditions in (\ref{mild_condition_1}) and (\ref{mild_condition_2}), the periodic PCRB for the MCE in sensing $\theta_m$ is given by
	\begin{align}\label{PCRB_theta_m}
		&\mathrm{PCRB}_{\theta _m}^{\mathrm{P}}=2-2( 1+\left[ \bm{F}^{-1} \right] _{m,m} ) ^{-\frac{1}{2}}\\
		=&2-\frac{2}{\sqrt{1+\frac{1}{\beta _m\mathrm{tr}\left(\boldsymbol{A}_m\left( \sum_{k=1}^K{\boldsymbol{w}_k\boldsymbol{w}_{k}^{H}}+\boldsymbol{SS}^H \right) \right)+\delta _m}}},\forall m\in \mathcal{M}.\nonumber
	\end{align}
\end{theorem}
\begin{IEEEproof}
	The periodic PCRB can be derived based on the periodic PFIM, which contains the periodic PFIMs from observation and prior information. Please refer to Appendix \ref{Appendix_PCRB_expression} for the derivation details.
\end{IEEEproof}
Note that $\beta _m$, $\bm{A}_m$, and $\delta_m$ in $\mathrm{PCRB}_{\theta _m}^{\mathrm{P}}$ can be efficiently obtained \emph{offline} based on the known prior distribution information $p_{\mathrm{Z}}(\bm{\zeta})$ and the system parameters such as $L$ and $\sigma_{\mathrm{S}}^2$.

In this paper, we focus on two practical performance metrics for multi-target sensing: \emph{i)} the maximum (or worst-case) periodic PCRB among all angles in $\bm{\theta}$, $\max_{m\in \mathcal{M}}\ \mathrm{PCRB}_{\theta_m}^{\mathrm{P}}$, to consider \emph{fairness} in multi-target sensing; and \emph{ii)} the sum periodic PCRB for MCE among all targets, $\sum_{m=1}^M\ \mathrm{PCRB}_{\theta_m}^{\mathrm{P}}$, to represent the \emph{overall efficacy} of multi-target sensing. In the following, we will study the transmit beamforming optimization to strike an optimal balance between multi-target sensing and multi-user communication considering either metric above.
\vspace{-6mm}
\section{Min-Max Periodic PCRB under Individual Communication Rate Constraints}\label{section_min_max}
\vspace{-3mm}
\subsection{Problem Formulation}
\vspace{-1.5mm}
In this section, we aim to optimize the transmit beamforming to minimize the \emph{maximum (worst-case) periodic PCRB} among $M$ targets, subject to an individual communication rate constraint at each $k$-th communication user denoted by $\bar{R}_k>0$. Under the general setup potentially with dedicated sensing beams, the optimization problem is formulated as
\begin{align}
	\mbox{(P1)}\
	\min_{\{\bm{w}_k\}_{k=1}^K,\bm{S}}\quad \max_{m\in \mathcal{M}}\quad &\mathrm{PCRB}_{\theta_m}^{\mathrm{P}}\\[-2mm]
	\mathrm{s.t.}\quad & R_k\geq \bar{R}_k,\quad   \forall k\in \mathcal{K}\label{P1_rate}\\[-1.5mm]
	&\sum_{k=1}^K \|\bm{w}_k\|^2+\mathrm{tr}( \bm{SS}^H ) \leq P.\label{P1_power}
\end{align}

The feasibility of (P1) can be checked via solving a convex feasibility problem with (\ref{P1_rate}) equivalently expressed as second-order cone constraints, which is similar to classic quality-of-service constrained feasibility problems for multi-user multiple-input single-output communication. In the sequel, we study (P1) assuming it has been \hbox{verified to be feasible.}

Note that (P1) is a non-convex optimization problem due to the non-convexity of the objective function and the constraints in (\ref{P1_rate}), and is particularly difficult to solve due to the following reasons. From the \emph{sensing} perspective, a common set of sensing and communication beams affects the periodic PCRBs for all the $M$ targets via the multiplications with $\beta_m$'s and $\bm{A}_m$'s (with potentially high ranks), which makes even sensing-oriented beamforming optimization to ensure fairness among multiple targets without the rate constraint challenging. From the \emph{communication} perspective, introducing dedicated sensing beams incurs interference to every communication user, while using only the $K$ communication beams for sensing may also compromise the desired signal power for the communication users due to the wide range of possible locations of multiple targets which generally differ from the user locations. Hence, finding the optimal solution to (P1) and the number of sensing beams that achieve an optimal trade-off between multi-target sensing and multi-user communication is highly non-trivial. In the following, we will devise the optimal solution via SDR and unveil useful insights by exploiting the solution structure.

\vspace{-7mm}
\subsection{Equivalent Transformation and SDR of (P1)}\label{optimal_solution_P1}
\vspace{-1.5mm}
Motivated by the quadratic terms with respect to $\{\bm{w}_k\}_{k=1}^K$ and $\bm{S}$ in both periodic PCRB and rate, we define $\bm{R}_k\overset{\Delta}{=}\bm{w}_k\bm{w}_k^H,\ \forall k\in \mathcal{K}$, $\bm{R}_\mathrm{S}\overset{\Delta}{=}\bm{SS}^H$, and $\gamma _k\!\overset{\Delta}{=}\!2^{\bar{R}_k}\!-\!1>\!0,\ \forall k\in \mathcal{K}$. Then, (P1) can be equivalently transformed into the following problem by introducing an auxiliary variable $t$:
\begin{align}
	\mbox{(P2)}\quad &\nonumber\\[-2.5mm]
	\!\!\!\!\!\!\!\!\max_{\scriptstyle \{\bm{R}_k\}_{k=1}^{K}\atop \scriptstyle \bm{R}_\mathrm{S},t}\ &t  \\[-5.5mm]
	\mathrm{s.t.}\quad  &\beta _m\mathrm{tr}(\bm{A}_m(\sum_{k=1}^K{\bm{R}_k}+\bm{R}_\mathrm{S})) +\delta _m\ge t,\forall m\in \mathcal{M}\label{P2_PCRB} \\[-1.5mm]
	&\bm{h}_{k}^{H}(\bm{R}_k\!-\!\gamma _k(\sum_{j\ne k}{\bm{R}_j+\bm{R}_\mathrm{S}}))\bm{h}_k\ge \gamma _k\sigma _\mathrm{C}^{2},\forall k\in \mathcal{K} \!\!\label{P2_rate} \\[-3mm]
	&\sum_{k=1}^K\mathrm{tr}(\bm{R}_k)+\mathrm{tr}(\bm{R}_\mathrm{S}) \leq P\label{P2_power} \\[-1.5mm]
	&\bm{R}_k\succeq \bm{0},\ \forall k\in \mathcal{K}\label{P2_PSD_C}\\[-1mm]
	&\bm{R}_\mathrm{S}\succeq \bm{0}\label{P2_PSD_S}\\[-1.5mm]
	& \mathrm{rank}(\bm{R}_k)=1,\ \forall k\in \mathcal{K}.\label{rank_k}
\end{align}

Let (P2-R) denote the relaxed version of (P2) via SDR by dropping the rank-one constraints in (\ref{rank_k}). Note that (P2-R) is a convex optimization problem, whose optimal solution can be obtained via the interior-point method or existing software such as CVX \cite{boyd2004convex}. Let $(\{ \bm{R}_{k}^{\star} \} _{k=1}^{K},\bm{R}_\mathrm{S}^\star)$ denote the optimal solution of $(\{\bm{R}_k\}_{k=1}^{K},\bm{R}_\mathrm{S})$ to (P2-R). In the following, we investigate the rank of each $\bm{R}_{k}^{\star}$ to evaluate the tightness of SDR, and unveil useful insights on the number of sensing beams needed by examining $\bm{R}_\mathrm{S}^\star$.
\vspace{-5mm}
\subsection{Properties of Optimal Solutions to (P2-R) and (P1)}\label{subsec_propeties_P2}
\vspace{-1.5mm}
Since (P2-R) satisfies the Slater's condition, strong duality holds for (P2-R). Thus, the properties of the optimal solution can be analyzed via the Lagrange duality theory. Denote $\bm{\psi}=[\psi_1,...,\psi_M]^T\succeq \bm{0}$, $\bm{\nu}=[\nu_1,...,\nu_K]^T \succeq \bm{0}$, $\mu \ge 0$, $\bm{\varPsi}_k\succeq \bm{0},\ \forall k
\in\mathcal{K}$, and $\bm{\varPsi}_\mathrm{S}\succeq \bm{0}$ as the dual variables associated with the constraints in (\ref{P2_PCRB}), (\ref{P2_rate}), (\ref{P2_power}), (\ref{P2_PSD_C}), and (\ref{P2_PSD_S}), respectively. The Lagrangian of $\mbox{(P2-R)}$ is given by $\mathcal{L} ( \{ \bm{R}_{k} \} _{k=1}^{K},\bm{R}_\mathrm{S},t,\bm{\psi },\bm{\nu},\mu, \{\bm{\varPsi}_k\}_{k=1}^{K},\bm{\varPsi}_\mathrm{S})
=t+\sum_{m=1}^M{\psi_m( \beta _m\mathrm{tr}(\bm{A}_m(\sum_{k=1}^K{\bm{R}_k}+\bm{R}_\mathrm{S}))+\delta _m-t )}
+\sum_{k=1}^K{\nu _k}(\boldsymbol{h}_{k}^{H}( \boldsymbol{R}_k-\gamma _k( \sum_{j\ne k}{\boldsymbol{R}_j}+\boldsymbol{R}_{\mathrm{S}} ) ) \boldsymbol{h}_k-\gamma _k\sigma _{\mathrm{C}}^{2} )
+\mu ( P\!-\!\mathrm{tr}(\sum_{k=1}^K{\bm{R}_k})\!-\!\mathrm{tr}(\bm{R}_\mathrm{S}) )\!+\!\sum_{k=1}^K\mathrm{tr}(\bm{\varPsi }_k\bm{R}_k)\!+\!\mathrm{tr}(\bm{\varPsi }_\mathrm{S}\bm{R}_\mathrm{S})$.

Denote $\{\bm{R}_{k}^{\star}\}_{k=1}^{K},\bm{R}_\mathrm{S}^\star,t^\star,\bm{\psi}^\star,\bm{\nu}^\star,\mu^\star, \{ \bm{\varPsi}_k^\star \} _{k=1}^{K}$, and $\bm{\varPsi}_\mathrm{S}^\star$ as the optimal primal and dual variables for $\mbox{(P2-R)}$. The KKT optimality conditions consist of primal and dual feasibility constraints, first-order optimality conditions given by
\begin{align}
	1-\sum_{m=1}^M\psi_m^\star=0\label{KKT_1}\\[-1mm]
	\sum_{m=1}^M{\psi _{m}^{\star}\beta _m\boldsymbol{A}_m}+\nu _{k}^{\star}(\gamma_k+1)\boldsymbol{h}_k\boldsymbol{h}_{k}^{H}-\sum_{j=1}^K{\gamma_j\nu _{j}^{\star}}\boldsymbol{h}_j\boldsymbol{h}_{j}^{H}\nonumber\\[-2mm]
	+\boldsymbol{\varPsi }_{k}^{\star}
	-\mu ^{\star}\boldsymbol{I}_{N_t}=\bm{0}, \forall k\in \mathcal{K}\label{KKT_2}\\[-4mm]
	\sum_{m=1}^M{\psi_m^\star\beta _m\bm{A}_{m}}-\sum_{k=1}^K{\gamma _k\nu _{k}^{\star}\bm{h}_k\bm{h}_{k}^{H}}+\bm{\varPsi }_\mathrm{S}^\star-\mu^\star \bm{I}_{N_t}=\mathbf{0},\label{KKT_3}
\end{align}
and complementary slackness conditions given by $\psi _m^\star( \beta _m\mathrm{tr}(\bm{A}_m(\sum_{k=1}^K\bm{R}_k^\star\!+\!\bm{R}_\mathrm{S}^\star))\!+\!\delta _m-t^\star )\!=\!0$, $\nu_{k}^{\star}(\boldsymbol{h}_{k}^{H}( \boldsymbol{R}_{k}^{\star}\!\\ -\!\gamma _k( \sum_{j\ne k}{\boldsymbol{R}_{j}^{\star}}\!+\!\boldsymbol{R}_{\mathrm{S}}^{\star} ) ) \boldsymbol{h}_k\!-\!\gamma _k\sigma _{\mathrm{C}}^{2})\!=\!0$, $\mathrm{tr}(\bm{\varPsi}_k^\star\bm{R}_k^\star)\!=\!0$, $\mu^\star (P\!-\! \mathrm{tr}( \sum_{k=1}^K\bm{R}_k^\star)\!-\!\mathrm{tr}(\bm{R}_\mathrm{S}^\star))\! =\!0$, and $\mathrm{tr}(\bm{\varPsi}_\mathrm{S}^\star\bm{R}_\mathrm{S}^\star)\!=\!0$.

We denote $\bm{U}_\mathrm{M}^\star=\sum_{m=1}^M\psi_m^\star\beta_m\bm{A}_m$ as it appears in various KKT conditions. Since $\psi_m^\star\geq 0$, $\beta_m>0$, and $\bm{A}_m\succeq \bm{0},\forall m\in \mathcal{M}$, $\bm{U}_\mathrm{M}^\star$ is a PSD matrix. Thus, the eigenvalue decomposition (EVD) of $\bm{U}_\mathrm{M}^\star$ can be expressed as $\bm{U}_\mathrm{M}^\star=\bm{Q\varLambda Q}^H$ where $\bm{\varLambda}=\mathrm{diag}\{\lambda_1,...,\lambda_{N_t}\}$ with $\lambda_1=\lambda_2=...=\lambda_N>\lambda_{N+1}\geq ...\geq \lambda_{N_t}\geq 0$ and $\bm{Q}=[\bm{q}_1,...,\bm{q}_{N_t}]$, with $N$ denoting the number of equally largest eigenvalues in $\bm{U}_\mathrm{M}^\star$.

Moreover, under the constraints in (\ref{P2_rate}), $\bm{R}_k^\star\ne \bm{0}$ must hold, thus we have $\mathrm{det}(\bm{\varPsi }_k^\star)=0$ since $\mathrm{tr}(\bm{\varPsi}_k^\star\bm{R}_k^\star)=0$, $\bm{\varPsi}_k^\star\succeq \bm{0}$, and $\bm{R}_k^\star\succeq \bm{0},\ \forall k\in\mathcal{K}$. According to (\ref{KKT_2}) and (\ref{KKT_3}), we further have $\mu ^{\star}=\lambda _{\max}( \bm{U}_\mathrm{M}^\star-\sum_{i=1}^K{\gamma _i\nu _{i}^{\star}\bm{h}_i\bm{h}_{i}^{H}}+\nu_k^\star(\gamma_k+1)\bm{h}_k\bm{h}_k^H)\ge \lambda _{\max}( \bm{U}_\mathrm{M}^\star-\sum_{i=1}^K{\gamma _i\nu _{i}^{\star}\bm{h}_i\bm{h}_{i}^{H}} ), \forall k\in \mathcal{K}$. By examining the structure of (P2), we have the following proposition.
\begin{proposition}[General Method to Construct Optimal Solution to (P1)]\label{prop_case_I}
	Given any optimal $(\tilde{t}^\star,\{\tilde{\bm{R}}_k^\star\}_{k=1}^K,\tilde{\bm{R}}_\mathrm{S}^\star)$ to (P2-R), the following solution is optimal to (P2-R) and (P2):
	\begin{align}
		{\bm{R}}_k^\star&=\frac{(\tilde{\bm{R}}_{k}^{\star}\bm{h}_k)(\tilde{\bm{R}}_{k}^{\star}\bm{h}_k)^H}{\bm{h}_{k}^{H}\tilde{\bm{R}}_{k}^{\star}\bm{h}_k},\quad \forall k\in \mathcal{K}\label{R_k_opt}\\[-1mm]
		 {\bm{R}}_\mathrm{S}^\star&=\sum_{k=1}^K{\tilde{\bm{R}}_k^\star}+\tilde{\bm{R}}_\mathrm{S}^\star-\sum_{k=1}^K{{\bm{R}}_k^\star}.\label{R_S_opt}
	\end{align}
	Thus, the SDR from (P2) to (P2-R) is \emph{tight}. An optimal solution to (P1) can be thus obtained via $\bm{w}_k^\star=(\bm{h}_{k}^{H}\tilde{\bm{R}}_{k}^{\star}\bm{h}_k)^{-\frac{1}{2}}\\ \tilde{\bm{R}}_{k}^{\star}\bm{h}_k,\forall  k\! \in\!\! \mathcal{K}$ and Cholesky decomposition of ${\bm{R}}_\mathrm{S}^\star=\bm{S}^\star{\bm{S}^\star}^H$.
\end{proposition}
\begin{IEEEproof}
	Denote $t^\star=\tilde{t}^\star$. Due to (\ref{R_k_opt}), constraints in (\ref{P2_rate}), (\ref{P2_PSD_C}), and (\ref{rank_k}) hold for $(t^\star,\{\bm{R}_k^\star\}_{k=1}^K,\bm{R}_\mathrm{S}^\star)$. Based on (\ref{R_S_opt}), $(t^\star,\{\bm{R}_k^\star\}_{k=1}^K,\bm{R}_\mathrm{S}^\star)$ achieves the optimal value of (P2-R) and (P2), and satisfies the constraints in (\ref{P2_PCRB}), (\ref{P2_power}), and (\ref{P2_PSD_S}). Thus, $(\{\bm{R}_k^\star\}_{k=1}^K,\bm{R}_\mathrm{S}^\star)$ is optimal to (P2-R) and (P2). Please refer to Appendix \ref{proof_prop_case_I} for details.
\end{IEEEproof}

Moreover, a bound can be derived on the number of sensing beams needed for (P2) and consequently (P1).

\begin{theorem}[General Bound on 
Number of Sensing Beams]\label{prop_rank}
	There exists an optimal solution to (P2) and (P1) where the number of sensing beams is no larger than $\sqrt{M}$,\footnote{When writing this paper, we became aware that a similar bound was independently derived in a very recent work \cite{attiah2026many}. Besides differences in the proof details, this paper is also different from \cite{attiah2026many} as we derive various new tighter bounds by exploiting the specific problem structures and provide explicit expressions/algorithms for the optimal beamforming to the min-max or min-sum periodic PCRB problems under individual communication rate constraints.\looseness=-1} i.e.,
	\begin{equation}
		\mathrm{rank}(\bm{S}^{\star}{\bm{S}^{\star}}^H )=\mathrm{rank}( \bm{R}_\mathrm{S}^{\star}) \le\sqrt{M}.\label{rank_Rs}
	\end{equation}
\end{theorem}

\begin{IEEEproof}
	Given optimal solution to (P2) obtained via Proposition \ref{prop_case_I}, we can always construct $(\{\bm{R}_k^\star\}_{k=1}^K,\bm{R}_\mathrm{S}^\star)$ that satisfies (\ref{rank_Rs}) and is optimal to (P2) via rank reduction \cite{huang2009rank}. Please refer to Appendix \ref{proof_prop_rank} for details.
\end{IEEEproof}

Algorithm \ref{alg_general_construction} presents the complete procedure for obtaining an optimal solution to (P1) with no more than $\sqrt{M}$ dedicated sensing beams, which is based on a rank reduction method in Steps 3-10 detailed in the proof of Theorem \ref{prop_rank}. The complexity of Algorithm \ref{alg_general_construction} is analyzed as follows. The complexity for solving (P2-R) via the interior-point method is $\mathcal{O}(4N_t^{6.5}K^{4.5}+2N_t^2KM^{3.5})$ \cite{boyd2004convex}. The complexity of executing (\ref{R_k_opt}) and (\ref{R_S_opt}) is $\mathcal{O}(2N_t^2K)$. The worst-case complexity for the rank reduction method in Steps 3-9 is $\mathcal{O}(N_t^7+N_tK^3)$. The decompositions of $\{\bm{R}_k^\star\}_{k=1}^K$ and $\bm{R}^\star_\mathrm{S}$ for obtaining $\{\bm{w}_k\}_{k=1}^K$ and $\bm{S}$ have a worst-case complexity of $\mathcal{O}(N_t^3(K+1))$. Thus, the overall worst-case complexity for Algorithm \ref{alg_general_construction} is $\mathcal{O}(N_t^7+4N_t^{6.5}K^{4.5}+2N_t^2KM^{3.5})$ by reserving the highest-order terms.\looseness=-3

Besides the general construction method for the optimal solution to (P1) and the general bound on the number of sensing beams needed shown in Proposition \ref{prop_case_I} and Theorem \ref{prop_rank}, respectively, we further examine the KKT conditions to derive \emph{tighter bounds} on the number of sensing beams needed and unveil \emph{specific structures} of the optimal solution to (P1) which provide new insights and enable further reduction of the computational complexity. 
\begin{algorithm}[t]
	\renewcommand{\algorithmicrequire}{\textbf{Input:}}
	\renewcommand{\algorithmicensure}{\textbf{Output:}}
	\caption{Proposed Algorithm for Obtaining an Optimal Solution to (P1) with No More Than $\sqrt{M}$ Sensing Beams}
	\label{alg_general_construction}
	\begin{algorithmic}[1]
		\STATE Obtain the optimal solution $(\{\tilde{\bm{R}}_k^\star\}_{k=1}^K,\tilde{\bm{R}}_\mathrm{S}^\star)$ to (P2-R).
		\STATE Construct another optimal solution $(\{\bm{R}_k^\star \} _{k=1}^{K},\bm{R}_\mathrm{S}^\star)$ to (P2) via (\ref{R_k_opt}) and (\ref{R_S_opt}).
		\WHILE{$\mathrm{rank}(\bm{R}_\mathrm{S}^\star)>\sqrt{M}$}
		\STATE Decompose $\bm{R}_k^\star=\bm{v}_k\bm{v}_k^H,\ \forall k\in\mathcal{K}$ and $\bm{R}_\mathrm{S}^\star=\bm{V}_\mathrm{S}\bm{V}_\mathrm{S}^H$. Denote $N_\mathrm{S}=\mathrm{rank}(\bm{R}_\mathrm{S}^\star)$.
		\STATE Find a non-zero solution $(\{\Delta_k\}_{k=1}^K,\bm{\Delta}_\mathrm{S})$ to the system of linear equations: $\sum_{k=1}^K\Delta_k\bm{v}_k^H\bm{A}_m\bm{v}_k+\mathrm{tr}( \bm{V}_\mathrm{S}^H\bm{A}_m\bm{V}_\mathrm{S}\bm{\Delta }_\mathrm{S})=0,\forall m\in\mathcal{M}$, and $\Delta _k|\bm{h}_{k}^{H}\bm{v}_k|^2-\gamma _k(\sum_{i\ne k}\Delta _i|\bm{h}_{k}^{H}\bm{v}_i|^2+\bm{h}_{k}^{H}\bm{V}_\mathrm{S}\mathbf{\Delta }_\mathrm{S}\bm{V}_{\mathrm{S}}^{H} \bm{h}_k)=0,\ \forall k\!\in\!\mathcal{K}$.
		\STATE Obtain $\{\xi_{i}\}_{i=1}^{N_\mathrm{S}}$ as the eigenvalues of $\bm{\Delta }_\mathrm{S}$.
		\STATE Obtain $\xi_0=\arg\max_{\{\Delta_k\}_{k=1}^K,\{\xi_{i}\}_{i=1}^{N_\mathrm{S}}}\{|\Delta_k|,k=1,...,K,|\xi_i|,i=1,...,N_\mathrm{S}\}$.
		\STATE Construct new optimal solution to (P2) via ${\bm{R}}_k^\star=\bm{v}_k(1\!-\!\frac{\Delta_k}{\xi _0})\bm{v}_k^H,\ \forall k\in \mathcal{K}$ and ${\bm{R}}_\mathrm{S}^\star\!=\!\bm{V}_\mathrm{S}(\bm{I}_{N_\mathrm{S}}-\frac{\bm{\Delta }_\mathrm{S}}{\xi _0})\bm{V}_\mathrm{S}^H$.
		\ENDWHILE
		\STATE  $\bm{w}_k^\star=\sqrt{1-\frac{\Delta_k}{\xi _0}}\bm{v}_k,\forall k\in \mathcal{K}$, $\bm{S}^\star=\bm{V}_\mathrm{S}(\bm{I}_{N_\mathrm{S}}-\frac{\bm{\Delta }_\mathrm{S}}{\xi _0})^{\frac{1}{2}}$.
	\end{algorithmic}
	\vspace{-1.7mm}
\end{algorithm}
\vspace{-8mm}
\subsection{Optimal Solution to (P1) and Number of Sensing Beams Needed in Different Cases}\label{opti_solution_P1}
\vspace{-0.5mm}
Note that as the communication constraints in (\ref{P2_rate}) are critical to the design flexibility of the sensing beams, we first denote $\mathcal{K}_{\mathrm{A}}=\{k|\nu_k^\star>0,\, \forall k\in\mathcal{K}\}$ as the set of users whose rate constraints are \emph{active}, and $\mathcal{K}_{\mathrm{I}}=\{k|\nu_k^\star=0,\, \forall k\in\mathcal{K}\}$ as the set of users whose rate constraints are \emph{inactive}, with $\mathcal{K}_{\mathrm{A}}\bigcup \mathcal{K}_{\mathrm{I}}=\mathcal{K}$. We then analyze $\bm{R}_{k}^{\star}$'s and  $\bm{R}_\mathrm{S}^\star$ in three cases.

\emph{1) Case I: $\mathrm{card}(\mathcal{K}_{\mathrm{A}})\!=\!0$ (low-rate regime).} In this case, all rate constraints in (\ref{P2_rate}) are \emph{inactive}, which corresponds to \emph{low rate requirements} $\bar{R}_k$'s for all users. We have following result.\looseness=-1

\begin{proposition}\label{lemma_case_I}
	In Case I, if $N=1$ (i.e., there is only one largest eigenvalue in $\bm{U}_\mathrm{M}^\star$), \emph{no} dedicated sensing beam is needed. The optimal solution to (P2-R) and (P2) can be expressed as  $\bm{R}_k^\star=P_{k,1}^\mathrm{C}\bm{q}_1\bm{q}_{1}^{H},\forall k\in \mathcal{K}$, $\bm{R}_\mathrm{S}^\star=P_{1}^\mathrm{S}\bm{q}_1\bm{q}_{1}^{H}$ with $\sum_{k=1}^K{P_{k,1}^\mathrm{C}}+{P_{1}^\mathrm{S}}=P$, based on which another optimal solution to (P2) can be constructed as $\tilde{\bm{R}}_{k}^{\star}=( P_{k,1}^\mathrm{C}+P_{1,k}^\mathrm{S} ) \bm{q}_1\bm{q}_{1}^{H},\forall k\in \mathcal{K}$, $\tilde{\bm{R}}_\mathrm{S}^\star=\bm{0}$ with $\sum_{k=1}^K{P_{1,k}^\mathrm{S}}=P_{1}^\mathrm{S}$. An optimal solution to (P1) is thus given by
	\begin{equation}\label{CaseI}
		\bm{w}_{k}^{\star}=(P_{k,1}^\mathrm{C}+P_{1,k}^\mathrm{S})^{\frac{1}{2}}\bm{q}_1,\forall k\in \mathcal{K},\quad \bm{S}^\star=\bm{0}.
	\end{equation}
\end{proposition}
\begin{IEEEproof}
	Based on (\ref{KKT_2}) and (\ref{KKT_3}), if $N\!=\!1$, $\bm{q}_1$ is the orthogonal basis of the null space of both $\{\bm{\varPsi }_k^\star\}_{k=1}^K$ and $\bm{\varPsi }_\mathrm{S}^\star$. Thus, the optimal solution $(\{\tilde{\bm{R}}_k^\star\}_{k=1}^K,\tilde{\bm{R}}_\mathrm{S}^\star=\bm{0})$ with $\mathrm{rank}(\tilde{\bm{R}}_k^\star)\!\!=\!\!1,\, \forall k\!\!\in\!\!\mathcal{K}$ to (P2) can be constructed. Please refer to Appendix \ref{proof_lemma_case_I} for details.
\end{IEEEproof}

Proposition \ref{lemma_case_I} shows that when the rate requirements are low and $N=1$ (which is very probable in practice as verified by extensive numerical results), communication beams suffice to support optimal sensing while satisfying the rate constraints.

\begin{table}[t]
	\caption{Bounds on Number of Sensing Beams Needed for Min-Max Problem (P1) and Worst-Case Complexity 
    }
	\vspace{-3mm}
	\centering
	\scalebox{0.49}{
	\resizebox{\textwidth}{!}{	\begin{tabular}{|c|c|c|}
			\hline
			&\makecell{Number of \\Sensing Beams\\ Needed} &\makecell{Worst-Case Complexity for Obtaining\\ Optimal Solution to (P1) with\\ Bounded Number of Sensing Beams}\\
			\hline
			Case I, $N=1$ (Low-Rate Regime)& 0&\makecell[c]{$\mathcal{O}(4N_t^{6.5}K^{4.5}+2N_t^2KM^{3.5})$}\\
			\hline
			Case II, $\tilde{N}=1,\mathrm{card}(\mathcal{K}_{\mathrm{A}})<K$& $0$&\makecell[c]{$\mathcal{O}(4N_t^{6.5}K^{4.5}+2N_t^2KM^{3.5})$}\\
			\hline
			Case II, $\tilde{N}=1,\mathrm{card}(\mathcal{K}_{\mathrm{A}})=K$& $\leq 1$&\makecell[c]{$\mathcal{O}(4N_t^{6.5}K^{4.5}+2N_t^2KM^{3.5})$}\\
			\hline
			Case III (High-Rate Regime)& $0$ &\makecell[c]{$\mathcal{O}(4N_t^{6.5}K^{4.5}+2N_t^2KM^{3.5})$}\\
			\hline
			Homogeneous Targets, $K=1$&0&$\mathcal{O}(N_t^3)$ or $\mathcal{O}(N_t^7)$\\
			\hline
			Homogeneous Targets, $K>1$&$\leq1$&$\mathcal{O}(N_t^7+4N_t^{6.5}K^{4.5})$\\
			\hline
			General Case  & $\leq \sqrt{M}$ & $\mathcal{O}(N_t^7+4N_t^{6.5}K^{4.5}+2N_t^2KM^{3.5})$\\
			\hline		
	\end{tabular}}
    }
	\vspace{-1mm}	\label{table_minmax}
\end{table}
\emph{2) Case II: $\mathrm{card}(\mathcal{K}_{\mathrm{A}}\!)\!\!> \!\! 0$, $\mu^\star\!\!\!=\!\!\lambda_{\max}(\!\bm{U}_\mathrm{M}^\star\!+\!\nu_k^\star(\gamma_k\!+\!1)\bm{h}_k\bm{h}_k^H \\-\!\!\sum_{i=1}^{\mathrm{card}(\mathcal{K}_{\mathrm{A}})}{\!\gamma _i\nu _{i}^{\star}\bm{h}_i\bm{h}_{i}^{H}})\!=\!\lambda_{\max}\!( \bm{U}_\mathrm{M}^\star\!-\!\sum_{i=1}^{\mathrm{card}(\mathcal{K}_{\mathrm{A}})}{\!\gamma _i\nu _{i}^{\star}\bm{h}_i\bm{h}_{i}^{H}} ),\\\forall k\!\!\in \!\!\mathcal{K}_\mathrm{A}$ (moderate-rate regime).} In this case, the rate constraints for some users are active, which are assumed to be indexed as $\mathcal{K}_{\mathrm{A}}=\{1,...,\mathrm{card}(\mathcal{K}_{\mathrm{A}})\}$, while the rate requirements are \emph{moderate}.\looseness=-1

Denote $\tilde{\bm{U}}_\mathrm{M}^{\star}=\bm{U}_\mathrm{M}^\star-\sum_{i=1}^{\mathrm{card}(\mathcal{K}_{\mathrm{A}}\!)}{\gamma _i\nu _{i}^{\star}\bm{h}_i\bm{h}_{i}^{H}}$, $\tilde{\lambda}_1$ as the largest eigenvalue of $\tilde{\bm{U}}_\mathrm{M}^{\star}$, and $\tilde{N}$ as the number of largest eigenvalues in $\tilde{\bm{U}}_\mathrm{M}^\star$. Denote the collection of eigenvectors corresponding to $\tilde{\lambda}_1$ as $\tilde{\bm{J}}=[\tilde{\bm{q}}_1,...,\tilde{\bm{q}}_{\tilde{N}}]$. The following proposition provides an explicit expression of the optimal solution to (P1).
\begin{proposition}\label{prop_case_II}
	In Case II, if $\tilde{N}\!\!\!\!=\!\!\!\!1$, the optimal $(\{\bm{R}_k^\star\}_{k=1}^K,\bm{R}_\mathrm{S}^\star)$ to (P2-R) can be expressed as $\bm{R}_\mathrm{S}^\star={P_{1}^\mathrm{S}\tilde{\bm{q}}_1\tilde{\bm{q}}_1^H}$ and $\bm{R}_k^\star=\begin{cases}
		P_{f,k}\bm{f}_k\bm{f}^H_k+P_{k,1}^\mathrm{C}\tilde{\bm{q}}_1\tilde{\bm{q}}_1^H,\ \forall k\in \mathcal{K}_\mathrm{A} \\
		P_{k,1}^\mathrm{C}\tilde{\bm{q}}_1\tilde{\bm{q}}_1^H,\qquad\qquad\qquad \forall k\in \mathcal{K}_\mathrm{I}
	\end{cases}$,
	where $\sum_{k=1}^{\mathrm{card}(\mathcal{K}_{\mathrm{A}})}{P_{f,k}}\!+\!\sum_{k=1}^{K}{P_{k,1}^\mathrm{C}}\!+\!P_1^\mathrm{S}=P$, $P_1^\mathrm{S}\geq0$, $P_{k,1}^\mathrm{C}\ge 0,\\ \forall k\in\mathcal{K}$; $P_{f,k}> 0$, $\bm{f}_k^H\tilde{\bm{J}}\!=\!\bm{0}$, $\bm{f}^H_k\bm{h}_k\!\ne\!0$, and $\|\bm{f}_k\| ^2=1,\ \forall k\!\!\in\!\! \mathcal{K}_\mathrm{A}$. The optimal solution to (P1) can be constructed as
	\vspace{-1mm}
	\begin{itemize}
		\item If $\mathrm{card}(\mathcal{K}_{\mathrm{A}})<K$: ${\bm{S}}^\star=\bm{0}$,
		\begin{equation}\label{CaseII_1}
			\hspace*{-0.5cm}
			\!\!\!\!\!\!\!\!	{\bm{w}}_k^\star=\!\begin{cases}
				\sqrt{P_{k,f}}\bm{f}_k,\quad  & k\!\in\! \mathcal{K}_\mathrm{A}\\
				\big(\!\sum_{i=1}^{\mathrm{card}(\!\mathcal{K}_{\mathrm{A}}\!) \!+\!1}{P_{i,1}^\mathrm{C}}\!\!+\!\!P_1^\mathrm{S}\big)^{\frac{1}{2}}\tilde{\bm{q}}_1,\!\!\!\!\!\! &k\!=\!\mathrm{card}(\mathcal{K}_{\mathrm{A}}\!)+1\\
				\sqrt{P_{k,1}^\mathrm{C}}\tilde{\bm{q}}_1,\quad & k\!\in\! \mathcal{K} _\mathrm{I}\backslash \{\mathrm{card}(\mathcal{K}_{\mathrm{A}}\!)\!\!+\!\!1\}.
			\end{cases} \!\!\!\!\!\!\!
		\end{equation}
		\item If $\mathrm{card}(\mathcal{K}_{\mathrm{A}})=K$:
		\vspace{-3mm}
		\begin{align}\label{CaseII_2}
			\vspace{-5mm}
			\!\!\!\!\!\!	{\bm{w}}_k^\star=\sqrt{P_{k,f}}\bm{f}_k,\forall k\in \mathcal{K}, {\bm{S}}^\star=\big(\sum_{i=1}^{K}{P_{i,1}^\mathrm{C}}+P_1^\mathrm{S}\big)^{\frac{1}{2}}\tilde{\bm{q}}_1.\!\!
		\end{align}
	\end{itemize}
\end{proposition}
\vspace{-1mm}
\begin{IEEEproof}
	Based on (\ref{KKT_2}) and (\ref{KKT_3}), $[\tilde{\bm{J}},\bm{f}_k]$ is the orthogonal basis of the null space of $\bm{\varPsi}_k^\star,\,\forall k\in\!\mathcal{K}_{\mathrm{A}}$, and $\tilde{\bm{J}}$ is the orthogonal basis of the null space of $\bm{\varPsi}_k^\star,\forall k\!\in\! \mathcal{K}_\mathrm{I}$ and $\bm{\varPsi}_\mathrm{S}^\star$. Since $\tilde{N}\!=\!1$, $\bm{w}_k^\star$'s in (\ref{CaseII_1}) and $\bm{S}^\star\!=\!\bm{0}$ are optimal to (P1) if $\mathrm{card}(\mathcal{K}_{\mathrm{A}}\!)\!\!<\!\!K$, and (\ref{CaseII_2}) is optimal to (P1) if $\mathrm{card}(\mathcal{K}_{\mathrm{A}})\!\!=\!\!K$. Please refer to Appendix \ref{proof_prop_case_II} for details.\looseness=-1
\end{IEEEproof}
Note that when $\tilde{N}=1$, i.e., there is only one largest eigenvalue in $\tilde{\bm{U}}_\mathrm{M}^\star$, \emph{no sensing beam} or at most \emph{one sensing beam} is needed; while the bound in (\ref{rank_Rs}) still holds otherwise.

\emph{3) Case III: $\mathrm{card}(\mathcal{K}_{\mathrm{A}}\!)\!\!>\!\!0$, there exists at least one $k\!\!\in\!\! \mathcal{K}_{\mathrm{A}}$ such that $\lambda _{\max}(\bm{U}_\mathrm{M}^\star-\sum_{i=1}^{\mathrm{card}(\mathcal{K}_{\mathrm{A}})}{\gamma _i\nu _{i}^{\star}\bm{h}_i\bm{h}_{i}^{H}}+\nu_k^\star(\gamma_k+1)\\ \bm{h}_k\bm{h}_k^H )\!\!>\!\! \lambda _{\max}(\bm{U}_\mathrm{M}^\star\!-\!\!\sum_{i=1}^{\mathrm{card}(\mathcal{K}_{\mathrm{A}}\!)}{\gamma _i\nu _{i}^{\star}\bm{h}_i\bm{h}_{i}^{H}})$ (high-rate regime).} This case tends to happen when some rate requirements $\bar{R}_k$'s and consequently their corresponding $\gamma_k$'s are \emph{high}.\looseness=-1
\begin{proposition}\label{prop_case_III}
	In Case III, \emph{no} dedicated sensing beam is needed. Any optimal $\{\bm{R}_k^\star\}_{k=1}^K$ and $\bm{R}_\mathrm{S}^{\star}$ to (P2-R) satisfies $\mathrm{rank}(\bm{R}_k^\star)=1,\forall k\in \mathcal{K}$ and $\bm{R}_\mathrm{S}^{\star}=\bm{0}$. The optimal solution to (P1) can be obtained via $\bm{R}_k^\star=\bm{w}_k^\star{\bm{w}_k^\star}^H,\forall k\in \mathcal{K}$ and $\bm{S}^\star=\bm{0}$.
\end{proposition}
\begin{IEEEproof}
	Due to (\ref{KKT_2}) and (\ref{KKT_3}), we have $\bm{R}_\mathrm{S}^\star=\bm{0}$. Moreover, $\bm{\varPsi}_k^\star\!\!=\!\!\bm{\varPsi}_\mathrm{S}^\star\!\!-\!\!\nu_k^\star(\gamma_k\!+\!1)\bm{h}_k\bm{h}_k^H$ yields $\mathrm{rank}(\bm{R}_k^\star)\!\!=\!\!1,\forall k\!\!\in \!\!\mathcal{K}$. Please refer to Appendix \ref{proof_prop_case_III} for details.\looseness=-2
\end{IEEEproof}

Proposition \ref{prop_case_III} indicates that no sensing beam is needed when the communication rate constraints become stringent, since no interference from dedicated sensing beams can be tolerated.

\begin{algorithm}[t]
	\renewcommand{\algorithmicrequire}{\textbf{Input:}}
	\renewcommand{\algorithmicensure}{\textbf{Output:}}
	\caption{Proposed Algorithm for Obtaining an Optimal Solution to (P1) with Tightly Bounded $\#$ of Sensing Beams}
	\label{alg_P1}
	\begin{algorithmic}[1]
		\STATE Obtain the optimal solution $(\{\bm{R}_k^\star \} _{k=1}^{K},\bm{R}_\mathrm{S}^\star)$ to (P2-R).
		\IF{$\mathrm{card}(\mathcal{K}_{\mathrm{A}})\!=\!0$ and $N=1$}
		\STATE Obtain $\{\bm{w}_k^\star\}_{k=1}^K$ and $\bm{S}^\star$ via Proposition \ref{lemma_case_I} (\emph{no sensing beam is needed}).
		\ELSE \IF{$\lambda_{\max}(\bm{U}_\mathrm{M}^\star\!-\!\sum_{i=1}^{\mathrm{card}(\mathcal{K}_{\mathrm{A}})}{\gamma _i\nu _{i}^{\star}\bm{h}_i\bm{h}_{i}^{H}}\!+\!\nu_k^\star(\gamma_k\!+\!1)\bm{h}_k\bm{h}_k^H )\!=\!\lambda_{\max}\!( \bm{U}_\mathrm{M}^\star\!-\!\sum_{i=1}^{\mathrm{card}(\mathcal{K}_{\mathrm{A}})}{\gamma _i\nu _{i}^{\star}\bm{h}_i\bm{h}_{i}^{H}} ),\forall k\in \mathcal{K}_\mathrm{A}$ and $\tilde{N}=1$}
		\STATE Obtain $\{\bm{w}_{k}^{\star}\}_{k=1}^K$ and $\bm{S}^\star$ via Proposition \ref{prop_case_II} (\emph{at most one sensing beam is needed}).
		\ELSE \IF{there exists at least one $k\in \mathcal{K}_{\mathrm{A}}$ such that $\lambda_{\max}(\bm{U}_\mathrm{M}^\star\!-\!\sum_{i=1}^{\mathrm{card}(\mathcal{K}_{\mathrm{A}}\!)}{\gamma _i\nu _{i}^{\star}\bm{h}_i\bm{h}_{i}^{H}}\!+\!\nu_k^\star(\gamma_k\!+\!1)\bm{h}_k\bm{h}_k^H )\!>\!\lambda_{\max}\!( \bm{U}_\mathrm{M}^\star\!-\!\sum_{i=1}^{\mathrm{card}(\mathcal{K}_{\mathrm{A}}\!)}{\gamma _i\nu _{i}^{\star}\bm{h}_i\bm{h}_{i}^{H}} )$}
		\STATE
		Obtain $\{\bm{w}_k^\star\}_{k=1}^K$ and $\bm{S}^\star$ via Proposition \ref{prop_case_III} (\emph{no sensing beam is needed}).
		\ELSE
		\STATE Obtain $\{\bm{w}_k^\star\}_{k=1}^K$ and $\bm{S}^\star$ via Steps 2-10 in Algorithm \ref{alg_general_construction} (\emph{at most $\sqrt{M}$ sensing beams are needed}).
		\ENDIF
		\ENDIF
		\ENDIF
	\end{algorithmic}
	\vspace{-1mm}
\end{algorithm}

Algorithm \ref{alg_P1} summarizes the above results and presents an alternative algorithm for finding an optimal solution to (P1), where the number of sensing beams can be further reduced in various cases compared to the general case in Algorithm \ref{alg_general_construction}. The complexity of obtaining the optimal solution in each case is summarized in Table \ref{table_minmax}, which is lower than that of Algorithm \ref{alg_general_construction} due to our derivation of the specific beamforming structures via judicious exploitation of the interplay between communication and sensing.

\vspace{-6mm}
\subsection{Special Case with Homogeneous Targets}\label{special_case_P1}
\vspace{-2mm}
To obtain further insights, we consider a case with \emph{homogeneous targets} where the marginal PDFs for all $\theta_m$'s are the same, and the (marginal) PDFs for all $\alpha_m$'s are the same. This is highly practical due to the typically identical appearance pattern and properties of targets within the region covered by the same BS. In this case,  $\bm{A}_1=\bm{A}_m,\delta_1=\delta_m,\beta_1=\beta_m,\forall m\in \mathcal{M}$ hold. We then have the following bounds.
\vspace{-0mm}
\begin{lemma}\label{lemma_P1_1}
	With homogeneous targets, \emph{at most one} sensing beam is needed in the optimal solution to (P1).
\end{lemma}
\begin{IEEEproof}
	In this case, (P2) is equivalent to itself with $M=1$ target. Based on Theorem \ref{prop_rank}, we have $\mathrm{rank}(\bm{R}_\mathrm{S}^\star)\leq \sqrt{M}=1$, i.e., at most one sensing beam is needed.
\end{IEEEproof}
The optimal solution can be obtained via Algorithm 1. Note that in this case, the auxiliary variable $t$ in (P2) is not needed by using the left-hand side (LHS) of (\ref{P2_PCRB}) as the objective function and removing the constraints in (\ref{P2_PCRB}), which saves the complexity in solving (P2-R) as shown in Table \ref{table_minmax}.

Moreover, when $K=1$, a tighter bound can be obtained.
\vspace{-0mm}
\begin{lemma}\label{lemma_P1_2}
	With homogeneous targets and $K=1$ communication user, \emph{no} sensing beam is needed.
\end{lemma}	
\begin{IEEEproof}
	The problem can be shown to be equivalent to a problem with only one dual-functional communication beam, which indicates that \emph{no} sensing beam is needed. Please refer to Appendix \ref{Appendix_lemma_P1_2} for details.	
\end{IEEEproof}
Define $\bm{q}^\prime\!\!\overset{\Delta}{=}\!\!\mathrm{arg}\max_{\{\bm{q}^\prime_n\}_{n=1}^{N^\prime}}| \boldsymbol{h}_{1}^{H}\boldsymbol{q}_{n}^{\prime} |^2$, where $N^\prime$ and $\{\bm{q}^\prime_n\}_{n=1}^{N^\prime}$ denote the number of eigenvectors corresponding to the strongest eigenvalue of $\bm{A}_1$ and these eigenvectors, respectively. In this case, the optimal solution to (P1) can be obtained via solving the following problem:
\begin{align}
	\mbox{(P2-HS)}\quad\max_{\scriptstyle \boldsymbol{R}_1^\mathrm{C}\succeq \mathbf{0}:\atop \scriptstyle \mathrm{tr}( \boldsymbol{R}_1^\mathrm{C}) \leq P,\mathrm{rank}(\bm{R}_1^\mathrm{C})=1}\quad&\mathrm{tr}( \boldsymbol{A}_1\boldsymbol{R}_1^\mathrm{C}) \\[-2.5mm]
	\mathrm{s.t.}\qquad\qquad& \boldsymbol{h}_1^H\boldsymbol{R}_1^\mathrm{C}\boldsymbol{h}_1\ge \gamma_1\sigma _{\mathrm{C}}^{2}.\label{P2-HS_rate}
\end{align}
\vspace{-5mm}%
\par\noindent%
The optimal solution to (P2-HS) without the rate constraint in (\ref{P2-HS_rate}) can be shown to be given by ${\bm{R}_1^\mathrm{C}}^\star=P\bm{q}'\bm{q}'^H$, which yields a communication rate of $\log_2(1\!\!+\!\!P\sigma_{\mathrm{C}}^{-2}|\bm{h}_{1}^{H}\bm{q}^{\prime}|^2)$. Thus, if $\bar{R}_1\!\!\leq\!\! \log_2(1+ P\sigma_{\mathrm{C}}^{-2}|\bm{h}_{1}^{H}\bm{q}^{\prime}|^2)$, the optimal solution to (P1) lies in Case I and is given by $\bm{w}_1^\star\!=\!\sqrt{P}\bm{q}^\prime$, $\bm{S}^\star\!\!=\!\!\bm{0}$. Otherwise, an optimal solution of ${\bm{R}_1^\mathrm{C}}^\star$ with rank one can be found by first solving (P2-HS) without the rank-one constraint on $\bm{R}_1^\mathrm{C}$ and then performing rank reduction in a similar manner as Steps 3-9 in Algorithm \ref{alg_general_construction}, since the SDR is guaranteed to be tight \cite{huang2009rank}. Optimal solution to (P1) is obtained via ${\bm{R}_1^\mathrm{C}}^\star=\bm{w}_1^\star{\bm{w}_1^\star}^H$ and $\bm{S}^\star\!\!\!=\!\!\bm{0}$. Table \ref{table_minmax} summarizes the complexity of the above solution.\looseness=-1

\vspace{-6mm}
\subsection{Summary}
\vspace{-1.5mm}
To summarize, the SDR is always \emph{tight}, and an optimal solution to (P1) with no more than $\sqrt{M}$ dedicated sensing beams can be found via Algorithm \ref{alg_general_construction}. Moreover, in various cases, we revealed that \emph{no or at most one} dedicated sensing beam is needed, for which the optimal solution to (P1) can be obtained with lower complexity, as summarized in Table \ref{table_minmax}.
\vspace{-5mm}
\section{Min-Sum Periodic PCRB under Individual Communication Rate Constraints}\label{section_min_sum}
\vspace{-2mm}
\subsection{Problem Formulation}
\vspace{-1mm}
In this section, we employ the \emph{sum periodic PCRB} of the MCE for sensing the $M$ targets, which bounds the sum MCE in multi-target sensing. Note that this allows more flexibility in improving the overall sensing performance, at the cost of potentially sacrificed sensing performance for certain targets (e.g., those located far from the BS and/or with significantly different PDF from others). We aim to optimize the transmit beamforming potentially with dedicated sensing beams to minimize the sum periodic PCRB subject to individual rate constraints at each $k$-th communication user denoted by $\bar{R}_k>0$. The optimization problem is formulated as
\begin{align}
	\mbox{(P3)}\quad \min_{\scriptstyle \{\bm{w}_k\}_{k=1}^K\atop \scriptstyle \bm{S}: (\ref{P1_rate})-(\ref{P1_power})}\quad&\sum_{m=1}^M\mathrm{PCRB}_{\theta _m}^{\mathrm{P}}.
\end{align}
The feasibility of (P3) can be checked via solving the same convex feasibility problem as (P1). In the following, we study (P3) assuming it has been verified to be feasible.

Problem (P3) is a non-convex optimization problem due to the non-convexity of the objective function and the constraints in (P3). Moreover, compared with (P1) where the objective function can be directly expressed as a set of constraints on each individual periodic PCRB, the objective function of (P3) involves the summation of multiple complex PCRB functions which are difficult to be decoupled. In the following, we obtain the optimal solution to (P3) by performing equivalent transformations, leveraging SDR, and proving its tightness.
\vspace{-5mm}
\subsection{Equivalent Transformation and SDR of Problem (P3)}\label{proposed_alg_P3}
\vspace{-1.5mm}
By defining $\bm{R}_k\overset{\Delta}{=}\bm{w}_k\bm{w}_k^H,\ \forall k\in \mathcal{K}$ and $\bm{R}_\mathrm{S}\overset{\Delta}{=}\bm{SS}^H$, (P3) can be equivalently expressed as the following problem:
\vspace{-1mm}
\begin{align}
	\mbox{(P3')}\qquad&\nonumber\\[-4mm]
	\max_{\{\bm{R}_k\} _{k=1}^{K},\,\bm{R}_\mathrm{S}}\,&\sum_{m=1}^M\left(1\!\!+\!\!\frac{1}{\beta _m\mathrm{tr}( \bm{A}_m(\sum_{k=1}^K{\bm{R}_k} \!\!+\!\!\bm{R}_\mathrm{S})) \!+\!\delta _m} \right) ^{-\frac{1}{2}}\!\!\!\!\!\label{P1ob_min_sum}\\[-2.5mm]
	\mathrm{s.t.}\quad & \bm{h}_{k}^{H}(\bm{R}_k-\gamma_k(\sum_{j\ne k}{\bm{R}_j+\bm{R}_\mathrm{S}}))\bm{h}_k\ge \gamma _k\sigma _\mathrm{C}^{2},\nonumber\\[-2.5mm]
	&\qquad\qquad\qquad\qquad\qquad \forall k\in\mathcal{K} \label{P3_R'_2}
\end{align}
\begin{align}
	&\sum_{k=1}^K\mathrm{tr}(\bm{R}_k) +\mathrm{tr}(\bm{R}_\mathrm{S}) \leq P\label{P3_R'_3}\\[-1mm]
	&\bm{R}_k\succeq \mathbf{0},\quad \forall k\in\mathcal{K}\label{P3_R'_4}\\[-1mm]
	&\bm{R}_\mathrm{S}\succeq \mathbf{0}\label{P3_R'_5}\\[-1.5mm]
	&\mathrm{rank}(\bm{R}_k)=1,\quad \forall k\in\mathcal{K}.\label{P3_R'_rank}
\end{align}
\vspace{-5mm}%
\par\noindent%
By introducing an auxiliary vector $\bm{y}\!=\![y_1,...,y_M]^T\in \mathbb{R}^{M\times 1}$, Problems (P3') and (P3) are equivalent to the problem below:
\begin{align}
	\mbox{(P4)}\!\!\!\!
	\max_{\scriptstyle \{\bm{R}_k\}_{k=1}^{K}\atop \scriptstyle \bm{R}_\mathrm{S},\bm{y}:(\ref{P3_R'_2})-(\ref{P3_R'_rank})}\,&\sum_{m=1}^M\sqrt{1-y_m^2}\\[-3mm]
	\mathrm{s.t.}\quad &\frac{1}{y_{m}^{2}}-\beta _m\mathrm{tr}(\bm{A}_m(\sum_{k=1}^K{\bm{R}_k}+\bm{R}_\mathrm{S}))-\delta _m\le 1,\nonumber\\[-3mm]
	&\qquad\qquad\qquad\qquad\qquad \forall m\in\mathcal{M}.\label{P3_R'_1}
\end{align}
The proof of the equivalence between (P3) and (P4) can be found in Appendix \ref{proof_equivalence}. Let (P4-R) denote the relaxed version of Problem (P4) by removing the constraints in (\ref{P3_R'_rank}). Note that (P4-R) is a convex optimization problem, whose optimal solution can be obtained via the interior-point method or CVX \cite{boyd2004convex}. In the following, we unveil useful properties of the optimal solution to (P4-R) which enable us to \hbox{find the optimal solution to (P3).}
\vspace{-5mm}
\subsection{Properties of Optimal Solutions to (P4-R) and (P3)}
\vspace{-1mm}
As (P4-R) satisfies the Slater's condition, strong duality holds for (P4-R). In the following, we analyze the KKT optimality conditions of (P4-R). Denote $\bar{\bm{\nu}}=[\bar{\nu}_1,...,\bar{\nu}_K]^T \succeq \bm{0}$, $\bar{\mu}\ge 0$, $\bar{\bm{\varPsi}}_k\succeq \bm{0},\ \forall k\in\mathcal{K}$, $\bar{\bm{\varPsi}}_\mathrm{S}\succeq \bm{0}$, and $\bm{z}=[z_1,...,z_M]\succeq \bm{0}$ as the dual variables associated with the constraints in (\ref{P3_R'_2}), (\ref{P3_R'_3}), (\ref{P3_R'_4}), (\ref{P3_R'_5}), and (\ref{P3_R'_1}), respectively. The Lagrangian of $\mbox{(P4-R)}$ is given by $	\mathcal{L}( \{\bm{R}_k\}_{k=1}^K,\bm{R}_\mathrm{S},\bm{y},\bar{\bm{\nu}},\bar{\mu},\{\bar{\bm{\varPsi }}_k\}_{k=1}^K,\bar{\bm{\varPsi }}_\mathrm{S},\bm{z})
=-\sum_{m=1}^M{z_m(y_{m}^{-2}-\beta _m\mathrm{tr}(\bm{A}_m(\sum_{k=1}^K{\bm{R}_k}+\bm{R}_\mathrm{S}))-\!\delta_m\!\!-\!\!1)}
-\sum_{k=1}^K\bar{\nu}_k(\gamma _k(\sum_{j\ne k}{\bm{h}_{k}^{H}\bm{R}_j\bm{h}_k}+\bm{h}_k^H\bm{R}_\mathrm{S}\bm{h}_k\!+\!\sigma _\mathrm{C}^{2})\!-\!\bm{h}_k^H\bm{R}_k\bm{h}_k)
+\bar{\mu}(P-\mathrm{tr}(\sum_{k=1}^K{\bm{R}_k}\!+\!\bm{R}_\mathrm{S}))\!+\!\sum_{k=1}
^K{\mathrm{tr}( \bar{\bm{\varPsi}_k}\bm{R}_k)}+\mathrm{tr}( \bar{\bm{\varPsi}_\mathrm{S}}\bm{R}_\mathrm{S})
+\sum_{m=1}^M{\sqrt{1-y_m^2}}$.
Let $\{\bm{R}^\star_k\}_{k=1}^K,\bm{R}_\mathrm{S}^\star,\bm{y}^\star, \bar{\bm{\nu}}^\star,\bar{\mu}^\star,\{\bar{\bm{\varPsi}}_k^\star\}_{k=1}^K,\bar{\bm{\varPsi}}_\mathrm{S}^\star$, and $\bm{z}^\star$ denote the optimal primal and dual variables for (P4-R). The KKT optimality conditions consist of primal and dual feasibility constraints, first-order optimality conditions given by
\begin{align}
	-{{y_{m}^{\star}}\big/{\sqrt{1-y_m^{\star2}}}}+{{z_m^\star}\big/{y_{m}^{\star3}}}=0\\[-1mm]
	\sum_{m=1}^M{z_{m}^{\star}\beta _m\boldsymbol{A}_m}+\bar{\nu}_{k}^{\star}(\gamma_k+1)\boldsymbol{h}_k\boldsymbol{h}_{k}^{H}-\sum_{j=1}^K{\gamma _j\bar{\nu}_{j}^{\star}}\boldsymbol{h}_j\boldsymbol{h}_{j}^{H}\nonumber\\[-1.5mm]
	+\bar{\boldsymbol{\varPsi}}_{k}^{\star}-\bar{\mu}^{\star}\boldsymbol{I}_{N_t}=\bm{0},\ \forall k\in\mathcal{K}\label{KKT_2_2}
\end{align}
\begin{align}
	\sum_{m=1}^M{z^\star_{m}\beta _m\bm{A}_{m}}-\sum_{k=1}^K{\gamma _k\bar{\nu} _{k}^{\star}\bm{h}_k\bm{h}_{k}^{H}}+\bar{\bm{\varPsi }}_\mathrm{S}^\star-\bar{\mu}^\star \bm{I}_{N_t}\!=\!\mathbf{0},\label{KKT_2_3}
\end{align}
and complementary slackness conditions given by $z_m^\star(y_{m}^{\star-2}\!\!-\\ \!\!\beta _m\mathrm{tr}(\bm{A}_m(\sum_{k=1}^K{\bm{R}_k^\star}+\bm{R}_\mathrm{S}^\star))-\delta_m -1)=0$, $\bar{\nu}_{k}^{\star}(\boldsymbol{h}_{k}^{H}(\gamma_k\times \\(\sum_{j\ne k}{\!\boldsymbol{R}_{j}^{\star}}+\boldsymbol{R}_{\mathrm{S}}^{\star})-\boldsymbol{R}_{k}^{\star}) \boldsymbol{h}_k+\gamma_k\sigma _{\mathrm{C}}^{2})=0$, $\mathrm{tr}(\bar{\bm{\varPsi}}_k^\star\bm{R}_k^\star)=0$, $\bar{\mu}^\star(\mathrm{tr}( \sum_{k=1}^K\bm{R}_k^\star)+\mathrm{tr}(\bm{R}_\mathrm{S}^\star)-P)=0$, and $\mathrm{tr}(\bar{\bm{\varPsi}}_\mathrm{S}^\star\bm{R}_\mathrm{S}^\star)=0$.

Based on the structure of (P4), we have following results.\looseness=-10
\begin{proposition}[General Method to Construct Optimal Solution to (P3)]\label{prop_P3_optimal}
	Given any optimal $(\tilde{\bm{y}}^\star,\{\tilde{\bm{R}}_k^\star\}_{k=1}^K,\tilde{\bm{R}}_\mathrm{S}^\star)$ to (P4-R), the following solution is optimal to (P4-R) and (P4):
	\begin{align}
		{\bm{R}}_k^\star=&\frac{(\tilde{\bm{R}}_{k}^{\star}\bm{h}_k)(\tilde{\bm{R}}_{k}^{\star}\bm{h}_k)^H}{\bm{h}_{k}^{H}\tilde{\bm{R}}_{k}^{\star}\bm{h}_k},\quad \forall k\in \mathcal{K}\label{R_k_opt_sum}\\[-2mm]
		{\bm{R}}_\mathrm{S}^\star=&\sum_{k=1}^K{\tilde{\bm{R}}_k^\star}+\tilde{\bm{R}}_\mathrm{S}^\star-\sum_{k=1}^K{{\bm{R}}_k^\star}.\label{R_S_opt_sum}
	\end{align}
	Thus, the SDR from (P4) to (P4-R) is \emph{tight}. An optimal solution to (P3) can be thus obtained via $\bm{w}_k^\star=(\bm{h}_{k}^{H}\tilde{\bm{R}}_{k}^{\star}\bm{h}_k)^{-\frac{1}{2}}\\ \tilde{\bm{R}}_{k}^{\star}\bm{h}_k,\forall k\!\in\!\! \mathcal{K}$ and Cholesky decomposition of ${\bm{R}}_\mathrm{S}^\star=\bm{S}^\star{\bm{S}^\star}^H$.
\end{proposition}
\begin{IEEEproof}
	Denote $\bm{y}^\star=\tilde{\bm{y}}^\star$. Due to (\ref{R_k_opt_sum}), constraints in (\ref{P3_R'_2}), (\ref{P3_R'_4}), and (\ref{P3_R'_rank}) hold for $(\bm{y}^\star,\{\bm{R}_k^\star\}_{k=1}^K,\bm{R}_\mathrm{S}^\star)$. Based on (\ref{R_S_opt_sum}), $(\bm{y}^\star,\{\bm{R}_k^\star\}_{k=1}^K,\bm{R}_\mathrm{S}^\star)$ achieves the optimal value of (P4-R) and (P4), and satisfies the constraints in (\ref{P3_R'_3}), (\ref{P3_R'_5}), and (\ref{P3_R'_1}). Thus, $(\{\bm{R}_k^\star\}_{k=1}^K,\bm{R}_\mathrm{S}^\star)$ is optimal to (P4-R) and (P4). Other details are similar to the proof of Proposition \ref{prop_case_I}.
\end{IEEEproof}

Moreover, we have the following bound on the number of sensing beams needed for (P4) and consequently (P3).
\begin{theorem}[General Bound on 
Number of Sensing Beams]\label{prop_rank_sum}
	There exists an optimal solution to (P4) and (P3) where the number of sensing beams is no larger than $\sqrt{M}$, i.e.,
	\begin{equation}
		\mathrm{rank}(\bm{S}^{\star}{\bm{S}^{\star}}^H )=\mathrm{rank}( \bm{R}_\mathrm{S}^{\star}) \le\sqrt{M}.\label{rank_Rs_sum}
	\end{equation}
\end{theorem}

\begin{IEEEproof}
	Given optimal solution to (P4) obtained via Proposition \ref{prop_P3_optimal}, we can always construct $(\{\bm{R}_k^\star\}_{k=1}^K,\bm{R}_\mathrm{S}^\star)$ that satisfies (\ref{rank_Rs_sum}) and is optimal to (P4) via rank reduction \cite{huang2009rank}. Other details are similar to the proof of Theorem \ref{prop_rank}.
\end{IEEEproof}

An optimal solution to (P4) and (P3) with $\mathrm{rank}(\bm{S}^{\star}{\bm{S}^{\star}}^H )=\mathrm{rank}( \bm{R}_\mathrm{S}^{\star}) \le\sqrt{M}$ can be obtained via first solving (P4-R) and obtaining the optimal solution to (P4) via (\ref{R_k_opt_sum}) and (\ref{R_S_opt_sum}), and then applying Steps 3-10 in Algorithm \ref{alg_general_construction}. The complexity for obtaining the optimal solution to (P3) via this general method with no more than $\sqrt{M}$ sensing beams is summarized in Table \ref{table_minsum}. Specifically, the complexity for solving (P4-R) via the interior-point method is $\mathcal{O}(4N_t^{6.5}K^{4.5}+2N_t^2KM^{3.5}+M^{4.5})$. Since the rank reduction method has a worst-case complexity of $\mathcal{O}(N_t^7+N_tK^3)$, the overall worst-case complexity for this general method is $\mathcal{O}(N_t^7+4N_t^{6.5}K^{4.5}+2N_t^2KM^{3.5}+M^{4.5})$ by reserving the highest-order terms. Next, we further examine the KKT optimality conditions to obtain specific structures of the optimal solution to (P3) and tighter bounds on the number of sensing beams needed.
\vspace{-3mm}
\subsection{Optimal Solution to Problem (P3) and Number of Sensing Beams Needed in Different Cases}\label{opti_solution_P3}
\vspace{-0mm}
Denote $\bm{U}_\mathrm{S}^\star=\sum_{m=1}^M{z^\star_{m}\beta _m\bm{A}_{m}}$. The EVD of $\bm{U}_\mathrm{S}^\star$ can be expressed as $\bm{U}_\mathrm{S}^\star=\bar{\bm{Q}}\bar{\bm{\varLambda }}\bar{\bm{Q}}^H$ where $\bar{\bm{\varLambda}}=\mathrm{diag}\{\bar{\lambda}_1,...,\bar{\lambda}_{N_t}\}$ with $\bar{\lambda}_1=...= \bar{\lambda}_{\bar{N}}>\bar{\lambda}_{\bar{N}+1}\ge...\ge\bar{\lambda}_{N_t}$ and $\bar{N}$ denotes the number of equally largest eigenvalues in $\bm{U}_\mathrm{S}^\star$. Define $\bar{\bm{Q}}=[\bar{\bm{q}}_1,...,\bar{\bm{q}}_{N_t}]$. Similar to the analysis in Section \ref{opti_solution_P1}, we split the user set $\mathcal{K} $ into two subsets: $\bar{\mathcal{K}}_\mathrm{A}=\{k|\bar{\nu}_k^\star>0,\, \forall k\in\mathcal{K}\}=\{1,...,\mathrm{card}(\bar{\mathcal{K}}_\mathrm{A})\}$ for ease of exposition consisting of users with active rate constraints, and $\bar{\mathcal{K}}_\mathrm{I}=\{k| \bar{\nu}_k^\star=0,\, \forall k\in\mathcal{K}\}$ consisting of users with inactive rate constraints. Denote $\bar{\lambda}_1^\prime=\bar{\lambda}_2^\prime=...=\bar{\lambda}_{\bar{N}^\prime}^\prime$ as the largest eigenvalues of $\bar{\bm{U}}^\star_\mathrm{S}=\bm{U}_\mathrm{S}^\star-\sum_{i=1}^{\mathrm{card}(\bar{\mathcal{K}}_\mathrm{A})}{\gamma _i\bar{\nu} _{i}^{\star}\bm{h}_i\bm{h}_{i}^{H}}$ with $\bar{N}^\prime$ being the number of largest eigenvalues. Denote the collection of the corresponding eigenvectors as $\bar{\bm{J}}^\prime=[\bar{\bm{q}}_1^\prime,...,\bar{\bm{q}}_{\bar{N}^\prime}^\prime]$. Then, we have the following proposition.
\vspace{-0mm}
\begin{proposition}\label{prop_P3_bounds}
	In the following cases, the optimal solution to (P3) has a specific structure with a tightly bounded number of dedicated sensing beams needed:
	\vspace{-1mm}
	\begin{itemize}[leftmargin=*]
		\item \emph{Case I: $\mathrm{card}(\bar{\mathcal{K}}_\mathrm{A})=0$ (low-rate regime). If $\bar{N}=1$, no sensing beam is needed.} The optimal solution to (P4-R) and (P4) can be written as $\bm{R}_k^\star=\bar{P}_{k,1}^\mathrm{C}\bar{\bm{q}}_1\bar{\bm{q}}_{1}^{H},\forall k\in \mathcal{K}$, $\bm{R}_\mathrm{S}^\star=\bar{P}_{1}^\mathrm{S}\bar{\bm{q}}_1\bar{\bm{q}}_{1}^{H}$ with $\sum_{k=1}^K{\bar{P}_{k,1}^\mathrm{C}}+{\bar{P}_{1}^\mathrm{S}}=P$, based on which the optimal solution to (P3) with \emph{no} sensing beam can be obtained as $\bm{w}_{k}^{\star}=(\bar{P}_{k,1}^\mathrm{C}+\bar{P}_{1,k}^\mathrm{S})^{\frac{1}{2}}\bar{\bm{q}}_1,\forall k\in \mathcal{K},\ \bm{S}^\star=\bm{0}$ with $\sum_{k=1}^K{\bar{P}_{1,k}^\mathrm{S}}=\bar{P}_{1}^\mathrm{S}$.
		\item \emph{Case II: $\mathrm{card}(\bar{\mathcal{K}}_\mathrm{A})>0$, $\lambda_{\max}(\bar{\bm{U}}_\mathrm{S}^\star\!+\!\bar{\nu}_k^\star(\gamma_k\!+\!1)\bm{h}_k\bm{h}_k^H )\!=\!\lambda_{\max}\!(\bar{\bm{U}}_\mathrm{S}^\star),\forall k\in\bar{\mathcal{K}}_\mathrm{A}$ (moderate-rate regime). If $\bar{N}^\prime=1$, no or at most one sensing beam is needed.} The optimal solution $(\{\bm{R}_k^\star\}_{k=1}^K,\bm{R}_\mathrm{S}^\star)$ to (P4-R) can be expressed as
		\begin{align}
			\bm{R}_k^\star&=\bar{P}_{f,k}\bar{\bm{f}}_k\bar{\bm{f}}_k^H+\bar{P}_{k,1}^\mathrm{C}\bar{\bm{q}}^\prime_1\bar{\bm{q}}_1^{\prime H},\,k\in \bar{\mathcal{K}}_\mathrm{A}\\
			\bm{R}_k^\star&=\bar{P}_{k,1}^\mathrm{C}\bar{\bm{q}}^\prime_1\bar{\bm{q}}_1^{\prime H},\,k\in \bar{\mathcal{K}}_\mathrm{I}\\
			\bm{R}_\mathrm{S}^\star&=\bar{P}_1^\mathrm{S}\bar{\bm{q}}_1^\prime\bar{\bm{q}}_1^{\prime H},
		\end{align}
		with $\sum_{k=1}^{\mathrm{card}(\bar{\mathcal{K}}_\mathrm{A}\!)}{\!\!\bar{P}_{f,k}}\!\!+\!\!\sum_{k=1}^{K}{\!\!\bar{P}_{k,1}^\mathrm{C}}\!\!+\!\!\bar{P}_1^\mathrm{S}\!\!=\!\!P$, $\bar{P}_1^\mathrm{S}\!\!\geq\!\!0$, $\bar{P}_{k,1}^\mathrm{C}\ge 0,\ \forall k\in\mathcal{K}$; $\bar{P}_{f,k}> 0$, $\bar{\bm{f}}_k^H\bar{\bm{J}}^\prime=\bm{0}$, $\bar{\bm{f}}^H_k\bm{h}_k\ne0$, and $\|\bar{\bm{f}}_k\|^2=1,\ \forall k\in \bar{\mathcal{K}}_\mathrm{A}$. Specifically, when $\mathrm{card}(\bar{\mathcal{K}}_\mathrm{A})\!\!<\!\!K$, the optimal solution to (P3) can be given by $\bm{S}^\star=\bm{0}$ and \looseness=-1 \\${\bm{w}}_k^\star\!\!=\!\!\!\begin{cases}
			\sqrt{\bar{P}_{k,f}}\bar{\bm{f}}_k,\quad  & k\!\in\!\bar{\mathcal{K}}_\mathrm{A}\\
			\!\big(\!\sum_{n=1}^{\mathrm{card}(\bar{\mathcal{K}}_\mathrm{A}\!) +\!1}{\!\!\bar{P}_{n,1}^\mathrm{C}}\!+\!\bar{P}_1^\mathrm{S}\big)^{\!\frac{1}{2}}\bar{\bm{q}}^\prime_1,\!\!\!\!\!\! &k\!=\!\mathrm{card}(\bar{\mathcal{K}}_\mathrm{A}\!)\!+\!1\\
			\sqrt{\bar{P}_{k,1}^\mathrm{C}}\bar{\bm{q}}^\prime_1,\quad & k\!\in\!\bar{\mathcal{K}}_\mathrm{I}\backslash \{\mathrm{card}(\bar{\mathcal{K}}_\mathrm{A}\!)\!+\!\!1\}.
		\end{cases}$
		Moreover, with $\mathrm{card}(\bar{\mathcal{K}}_\mathrm{A})=K$, the optimal solution to (P3) can be expressed as ${\bm{w}}_k^\star=\sqrt{\bar{P}_{k,f}}\bar{\bm{f}}_k,\forall k\in \mathcal{K}$, ${\bm{S}}^\star=(\sum_{i=1}^{K}{\bar{P}_{i,1}^\mathrm{C}}+\bar{P}_1^\mathrm{S})^{\frac{1}{2}}\bar{\bm{q}}^\prime_1$.
		\item \emph{Case III: $\mathrm{card}(\bar{\mathcal{K}}_\mathrm{A})\!>\!0$, there exists at least one $k\in \bar{\mathcal{K}}_{\mathrm{A}}$ such that $\lambda_{\max}( \bar{\bm{U}}_\mathrm{S}^\star+\bar{\nu}_k^\star(\gamma_k+1)\bm{h}_k\bm{h}_k^H)>\lambda _{\max}(\bar{\bm{U}}_\mathrm{S}^\star)$ (high-rate regime), no sensing beam is needed.} Any optimal $\{\bm{R}_k^\star\}_{k=1}^K$ and $\bm{R}_\mathrm{S}^\star$ to (P4-R) satisfies $\mathrm{rank}(\bm{R}_k^\star)=1,\ \forall k\in\mathcal{K}$ and $\bm{R}_\mathrm{S}^\star=\bm{0}$. The optimal solution to (P3) can be obtained via $\bm{R}_k^\star=\bm{w}_k^\star{\bm{w}_k^\star}^H,\forall k\in \mathcal{K}$ and $\bm{S}^\star=\bm{0}$.
	\end{itemize}
\end{proposition}
\vspace{-0mm}
\begin{IEEEproof}
	In Case I with $\bar{N}=1$, based on (\ref{KKT_2_2}) and (\ref{KKT_2_3}), $\bar{\bm{q}}_1$ is the orthogonal basis of the null space of both $\{\bar{\bm{\varPsi}}_k^\star\}_{k=1}^K$ and $\bar{\bm{\varPsi }}_\mathrm{S}^\star$. Thus, the optimal solution $(\{\tilde{\bm{R}}_k^\star\}_{k=1}^K,\tilde{\bm{R}}_\mathrm{S}^\star=\bm{0})$ with $\mathrm{rank}(\tilde{\bm{R}}_k^\star)=1,\ \forall k\in\mathcal{K}$ to (P4) can be constructed. In Case II, based on (\ref{KKT_2_2}) and (\ref{KKT_2_3}), $[\bar{\bm{J}}^\prime,\bar{\bm{f}}_k]$ is the orthogonal basis of the null space of $\bar{\bm{\varPsi}}_k^\star,\forall k\in \bar{\mathcal{K}}_\mathrm{A}$, and $\bar{\bm{J}}^\prime$ is the orthogonal basis of the null space of both $\bar{\bm{\varPsi}}_k^\star,\forall k\in \bar{\mathcal{K}}_\mathrm{I}$ and $\bar{\bm{\varPsi}}_\mathrm{S}^\star$. Since $\bar{N}^\prime=1$, an optimal solution to (P3) with $\bm{S}^\star=\bm{0}$ and ${\bm{S}}^\star=(\sum_{i=1}^{K}{\bar{P}_{i,1}^\mathrm{C}}+\bar{P}_1^\mathrm{S})^{\frac{1}{2}}\bar{\bm{q}}^\prime_1$ can be constructed when $\mathrm{card}(\bar{\mathcal{K}}_\mathrm{A})<K$ and $\mathrm{card}(\bar{\mathcal{K}}_\mathrm{A})=K$, respectively. In Case III, based on (\ref{KKT_2_2}) and (\ref{KKT_2_3}), we have $\bm{R}_\mathrm{S}^\star=\bm{0}$. Moreover, $\bar{\bm{\varPsi}}_k^\star=\bar{\bm{\varPsi}}_\mathrm{S}^\star-\bar{\nu}_k^\star(\gamma_k+1)\bm{h}_k\bm{h}_k^H, \forall k\in\mathcal{K}$ yields $\mathrm{rank}(\bm{R}_k^\star)=1,\forall k\in \mathcal{K}$. Other details are similar to the proofs of Propositions \ref{lemma_case_I}, \ref{prop_case_II}, and \ref{prop_case_III}.
\end{IEEEproof}
\begin{table}[t]
	\vspace{-2mm}
	\caption{Bounds on Number of Sensing Beams Needed for Min-Sum Problem (P3) and Worst-Case Complexity 
    }
	\vspace{-3mm}
	\centering
	\scalebox{0.48}{
	\resizebox{\textwidth}{!}{	\begin{tabular}{|c|c|c|}
			\hline
			&\makecell{Number of\\ Sensing Beams\\ Needed} &\makecell{Worst-Case Complexity for Obtaining\\ Optimal Solution to (P3) with\\ Bounded Number of Sensing Beams}\\
			\hline
			Case I, $\bar{N}=1$ (Low-Rate Regime)& 0&\makecell[c]{$\mathcal{O}(4N_t^{6.5}K^{4.5}+2N_t^2KM^{3.5}+M^{4.5})$}\\
			\hline
			Case II, $\bar{N}^\prime=1,\mathrm{card}(\bar{\mathcal{K}}_\mathrm{A})<K$& $0$&\makecell[c]{$\mathcal{O}(4N_t^{6.5}K^{4.5}+2N_t^2KM^{3.5}+M^{4.5})$}\\
			\hline
			Case II, $\bar{N}^\prime=1,\mathrm{card}(\bar{\mathcal{K}}_\mathrm{A})=K$& $\leq 1$&\makecell[c]{$\mathcal{O}(4N_t^{6.5}K^{4.5}+2N_t^2KM^{3.5}+M^{4.5})$}\\
			\hline
			Case III (High-Rate Regime)& $0$ &\makecell[c]{$\mathcal{O}(4N_t^{6.5}K^{4.5}+2N_t^2KM^{3.5}+M^{4.5})$}\\
			\hline
			Homogeneous Targets, $K=1$&0&$\mathcal{O}(N_t^3)$ or $\mathcal{O}(N_t^7)$\\
			\hline
			Homogeneous Targets, $K>1$&$\leq1$&$\mathcal{O}(N_t^7+4N_t^{6.5}K^{4.5})$\\
			\hline
			General Case & $\leq \sqrt{M}$ & $\mathcal{O}(N_t^7+4N_t^{6.5}K^{4.5}+2N_t^2KM^{3.5}+M^{4.5})$\\
			\hline
	\end{tabular}}}
	\label{table_minsum}
	\vspace{-1mm}
\end{table}

The tight bounds on the number of sensing beams needed presented in Proposition \ref{prop_P3_bounds} are summarized in Table \ref{table_minsum}, together with the complexities for finding the optimal solution to (P3) in different cases, which are observed to be lower than that for the general case. Note that the optimal beamforming design solutions in Propositions \ref{prop_case_I}-\ref{prop_case_III} and Propositions \ref{prop_P3_optimal}-\ref{prop_P3_bounds} for the min-max and min-sum periodic PCRB problems (P1) and (P3) are generally different due to the drastically different objective functions. Specifically, the beamforming design for the min-max problem tends to cover the possible angles of every target based on their probabilities, to avoid excessively poor sensing performance for any target. In contrast, the beamforming design for the min-sum problem may sacrifice targets with drastically different PDFs or long distances to the BS, and focus power over common highly-probable locations for all targets to enhance the overall sensing performance. Moreover, even under similar bounds on the number of sensing beams needed shown in Tables \ref{table_minmax} and \ref{table_minsum}, the exact number of sensing beams needed in the min-max and min-sum cases may still be different, which will be observed from Section \ref{numerical_results}.
\vspace{-5mm}
\subsection{Special Case with Homogeneous Targets}\label{special_case_P3}
\vspace{-1mm}
With homogeneous targets defined in Section \ref{special_case_P1}, we have $\bm{A}_1=\bm{A}_m,\beta_1=\beta_m,\delta_1=\delta_m$, and consequently $\mathrm{PCRB}_{\theta _1}^{\mathrm{P}}=\mathrm{PCRB}_{\theta _m}^{\mathrm{P}},\forall m\in\mathcal{M}$. Thus, (P3) is equivalent to (P1). The following lemma directly follows from Lemmas \ref{lemma_P1_1} and \ref{lemma_P1_2}.
\begin{lemma}\label{lemma_P3}
	With homogeneous targets, \emph{at most one} sensing beam is needed in the optimal solution to (P3). With $K=1$, \emph{no} sensing beam is needed.
\end{lemma}
\vspace{-5mm}
\subsection{Summary}
\vspace{-1mm}
To summarize, the SDR is always \emph{tight}, and an optimal solution to (P3) with no more than $\sqrt{M}$ dedicated sensing beams can be found via the similar procedure as in Algorithm \ref{alg_general_construction}. Moreover, in various cases, \emph{no or at most one} sensing beam is needed, and the optimal solution to (P3) can be obtained with lower complexity, as summarized in Table \ref{table_minsum}.

\vspace{-2mm}
\section{Numerical Results}\label{numerical_results}
\vspace{-1mm}
In this section, we present numerical results to validate our analytical results and evaluate the performance of our optimal beamforming solutions. Consider a uniform planar array (UPA) configuration for the BS antennas with half-wavelength antenna spacing, where $N_t\!\!=\!\!T_x\!\!\times \!\! T_y\!\!=\!\!3\!\!\times \!\!3$ and $N_r\!\!=R_x\!\!\times \!\! R_y\!\!=\!\!3\!\times \!4$. We set $L\!=\!25$, $M\!\!=\!\!30$, $K\!\!=\!\!2$, $P\!\!=\!\!30$ dBm, $\sigma ^2_\mathrm{C}\!\!=\!\!-90$ dBm, and $\sigma^2_\mathrm{S}\!\!=\!\!-90$ dBm unless otherwise specified. Let $h_{\mathrm{B}}\!=\!11$ m and $h_m\!=\!1$ m, $\forall m\in\mathcal{M}$. The array steering vectors are given by $\bm{a}(\theta_m)=\bm{a}_1(\theta_m)\otimes \bm{a}_2(\theta_m)$ and $\bm{b}(\theta_m)\!=\!\bm{b}_1(\theta_m)\otimes \bm{b}_2(\theta_m)$, where $\bm{a}_1(\theta_m)=[ e^{\frac{-\!j\pi \cos\phi_m( T_x\!-\!1 )\!\cos\! \theta _m}{2}},... ,\!e^{\frac{j\pi \cos\phi_m( T_x\!-\!1 )\!\cos\! \theta _m}{2}} ] ^T$, $\bm{a}_2(\theta_m)=[ e^{\frac{-\!j\pi \cos\phi_m( T_y\!-\!1 )\!\sin\! \theta _m}{2}},... ,\!e^{\frac{j\pi \cos\phi_m( T_y\!-\!1 )\!\sin\! \theta _m}{2}} ] ^T$, $\bm{b}_1(\theta_m)=[ e^{\frac{-\!j\pi \cos\phi_m( R_x\!-\!1 )\!\cos\! \theta _m}{2}},... ,\!e^{\frac{j\pi \cos\phi_m( R_x\!-\!1 )\!\cos\! \theta _m}{2}} ] ^T$, and $\bm{b}_2(\theta_m)=[ e^{\frac{-\!j\pi \cos\phi_m( R_y\!-\!1 )\!\sin\! \theta _m}{2}},...,\!e^{\frac{j\pi \cos\phi_m( R_y\!-\!1 )\!\sin\! \theta _m}{2}} ] ^T$ with $\phi_m=\mathrm{arcsin}(-\frac{h_\mathrm{B}-h_m}{r_m})$ denoting the elevation angle of target $m$ and $r_m$ denoting the distance between the BS and target $m$. We further assume $r_m=100$ m, $\forall m\in \mathcal{M}$, while the overall reflection coefficient for each $m$-th target follows an independent CSCG distribution of $\alpha_m\sim\mathcal{CN}(0,2\times 10^{-13})\,\forall m\in\mathcal{M}$. 

\vspace{-1mm}
Motivated by practical scenarios where each target's angle PDF is typically concentrated around several highly-probable angles, we consider a \emph{von-Mises mixture model} for the PDF of each angle $\theta_m\in [-\pi,\pi)$, which is the periodic version of the more well-known Gaussian mixture model \cite{xu2023mimo1};\footnote{Note that the periodic PCRB is also applicable to any other non-periodic angle PDF by transforming it to a $2\pi$-periodic extended PDF as introduced before Theorem \ref{prop_PCRB_expression} and Appendix \ref{Appendix_PCRB_expression}.} while different $\theta_m$'s are assumed to be independent of each other. The PDF of each target is expressed as $p_{\Theta _m}( \theta _m ) =\sum_{v=1}^{V_m}{p_{m,v}\frac{e^{\kappa _{m,v}\cos ( \theta _m-\theta _{m,v})}}{2\pi I_0( \kappa _{m,v} )}},\,\forall m\in\mathcal{M}$, where $V_m$ denotes the number of von-Mises PDF components of the $m$-th target; $p_{m,v}$, $\kappa _{m,v}$, and $\theta _{m,v}$ represent the weight, concentration, and mean of the $(m,v)$-th von-Mises PDF, respectively; $I_n(\kappa _{m,v})=\frac{1}{2\pi}\int_{-\pi}^{\pi}{e^{\kappa _{m,v}\cos t}\cos ( nt ) dt}$ is the modified Bessel function of order $n$. The communication channel from the BS to each $k$-th user is assumed to be given by $\bm{h}_k^H=({\beta^\mathrm{C}_k}/{(K_\mathrm{C}+1)})^{\frac{1}{2}}(\sqrt{K_\mathrm{C}}{\bm{h}_k^{\mathrm{LoS}}}^H+{\bm{h}_k^{\mathrm{NLoS}}}^H)$ under the Rician fading model, where $K_\mathrm{C}=70$ dB denotes the Rician factor, $\beta^\mathrm{C}_k$ denotes the path power gain given by $\beta^\mathrm{C}_k=10^{-3}/r_{\mathrm{U}_k}^{\alpha_\mathrm{C}}$ with $r_{\mathrm{U}_k}$ denoting the distance from the BS to the $k$-th user and $\alpha_\mathrm{C}=3$ denoting the path loss exponent. The LoS component is given by $\bm{h}_k^{\mathrm{LoS}}=\bm{h}_k^{(1)}\otimes\bm{h}_k^{(2)},\,\forall k\in\mathcal{K}$, where $\bm{h}_k^{(1)}=[ e^{\frac{-\!j\pi \cos\phi_{\mathrm{U}_k}( T_x\!-\!1 )\!\cos\! \theta _{\mathrm{U}_k}}{2}},... ,\!e^{\frac{j\pi \cos\phi_{\mathrm{U}_k}( T_x\!-\!1 )\!\cos\! \theta _{\mathrm{U}_k}}{2}} ] ^T$, and $\bm{h}_k^{(2)}=[ e^{\frac{-\!j\pi \cos\phi_{\mathrm{U}_k}( T_y\!-\!1 )\!\sin\! \theta _{\mathrm{U}_k}}{2}},... ,\!e^{\frac{j\pi \cos\phi_{\mathrm{U}_k}( T_y\!-\!1 )\!\sin\! \theta _{\mathrm{U}_k}}{2}} ] ^T$, with $\phi_{\mathrm{U}_k}\!\!\!=\!\mathrm{arcsin}(\!-\!\frac{h_\mathrm{B}-h_{\mathrm{U}_k}}{r_{\mathrm{U}_k}}\!\!)$ denoting the elevation angle of the $k$-th user and $h_{\mathrm{U}_k}$ denoting the height of the $k$-th user. The elements in the non-LoS (NLoS) component ${\bm{h}^\mathrm{NLoS}_k}^H$ follow independent CSCG distributions with $[{\bm{h}_{k}^{\mathrm{NLoS}}}^H]_i\!\!\sim\!\mathcal{CN}(0,1), \forall i$. We further assume that the two users are located at height $h_{\mathrm{U}_1}\!=\!h_{\mathrm{U}_2}\!=\!1$ m and distance $r_{\mathrm{U}_1}\!\!=\!r_{\mathrm{U}_2}\!\!=\!500$ m, with azimuth angles $\theta_{\mathrm{U}_1}\!\!=\!\!0.5$, $\theta_{\mathrm{U}_2}\!\!=\!\!-2$, unless specified otherwise.\looseness=-1

\vspace{-5mm}
\subsection{Optimal Beamforming Designs: Number of Dedicated Sensing Beams and Radiated Power Patterns}
\vspace{-2mm}
First, we illustrate the number of sensing beams needed in the optimal beamforming designs (equivalently $\mathrm{rank}(\bm{R}_{\mathrm{S}}^\star)$ for (P2) and (P4)) and their radiated power patterns at $50$ m. 
\begin{figure}[t]
	\centering		
	\subfigtopskip=2pt
	\subfigbottomskip=2pt
	\subfigcapskip=-5pt
	\subfigure[Solution obtained with only the general rank reduction method.]{
		\label{regime_max_cvx}
		\includegraphics[width=4.2cm]{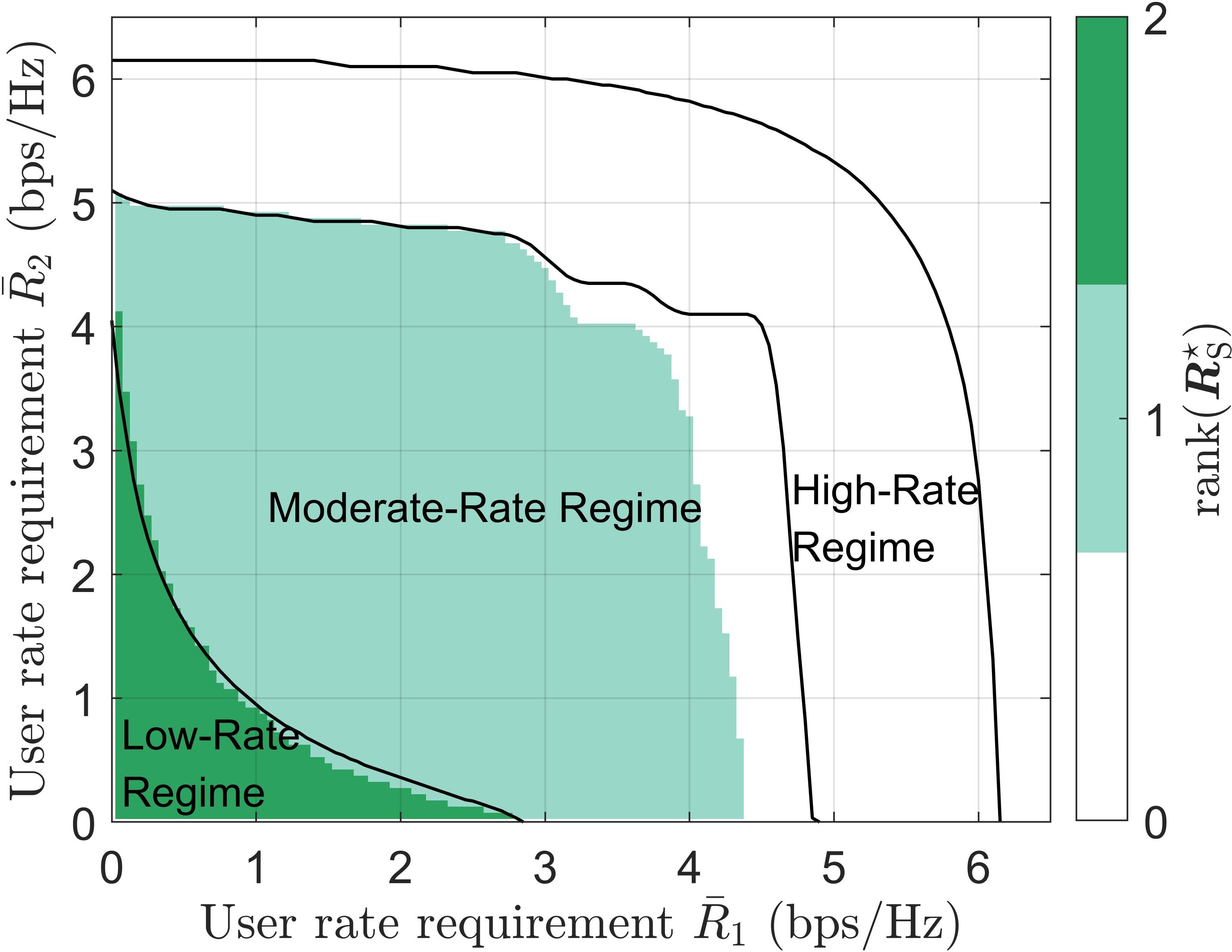}}
	\subfigure[Solution obtained with discussion of the three cases in different rate regimes.]{
		\label{regime_max_proposed}
		\includegraphics[width=4.2cm]{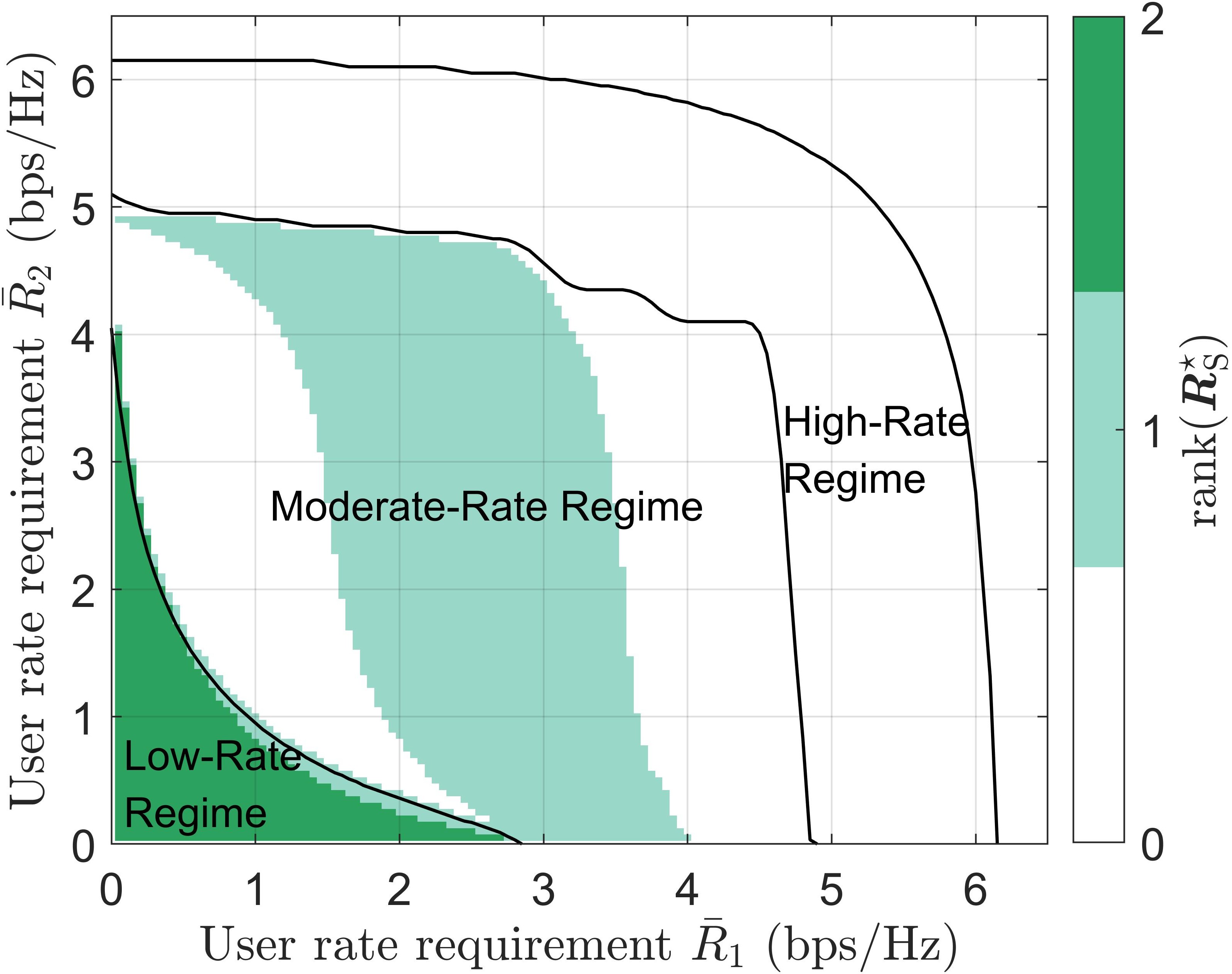}}
	\vspace{-4mm}
	\caption{Illustration of the number of sensing beams needed for Problem (P1).}
	\label{regime_max_all}
	\vspace{-1mm}
\end{figure}

\subsubsection{Optimal Beamforming Designs under General Bound and Tighter Bounds}
To demonstrate the effectiveness of the tighter bounds and beamforming designs obtained by discussing the three cases in Propositions \ref{lemma_case_I}-\ref{prop_case_III} and Proposition \ref{prop_P3_bounds}, Fig. \ref{regime_max_all} shows the number of sensing beams needed under the concentrated angle PDF for the min-max problem (P1) either with only the general rank reduction method or with discussion of the three cases. It is observed that the latter under tighter bounds requires fewer beams over a large rate region. Moreover, Table IV in Appendix \ref{Appendix_numerical_results} shows that the cases where tighter bounds exist occupy a large portion of random system setup realizations for both the min-max and min-sum problems. These results validate the effectiveness of discussing the tighter bounds in Propositions \ref{lemma_case_I}-\ref{prop_case_III} and Proposition \ref{prop_P3_bounds}.\looseness=-5
\vspace{-0.5mm}
\subsubsection{Optimal Beamforming Designs with Different Numbers of Targets/Users}
\begin{figure*}[t]
	\centering
	\vspace{-0.2cm}
	\subfigtopskip=2pt
	\subfigbottomskip=2pt
	\subfigcapskip=-5pt
	\subfigure[$M=30$ targets, $K=2$ users.]{
		\label{rank_beamforming_base}
		\includegraphics[width=0.32\linewidth]{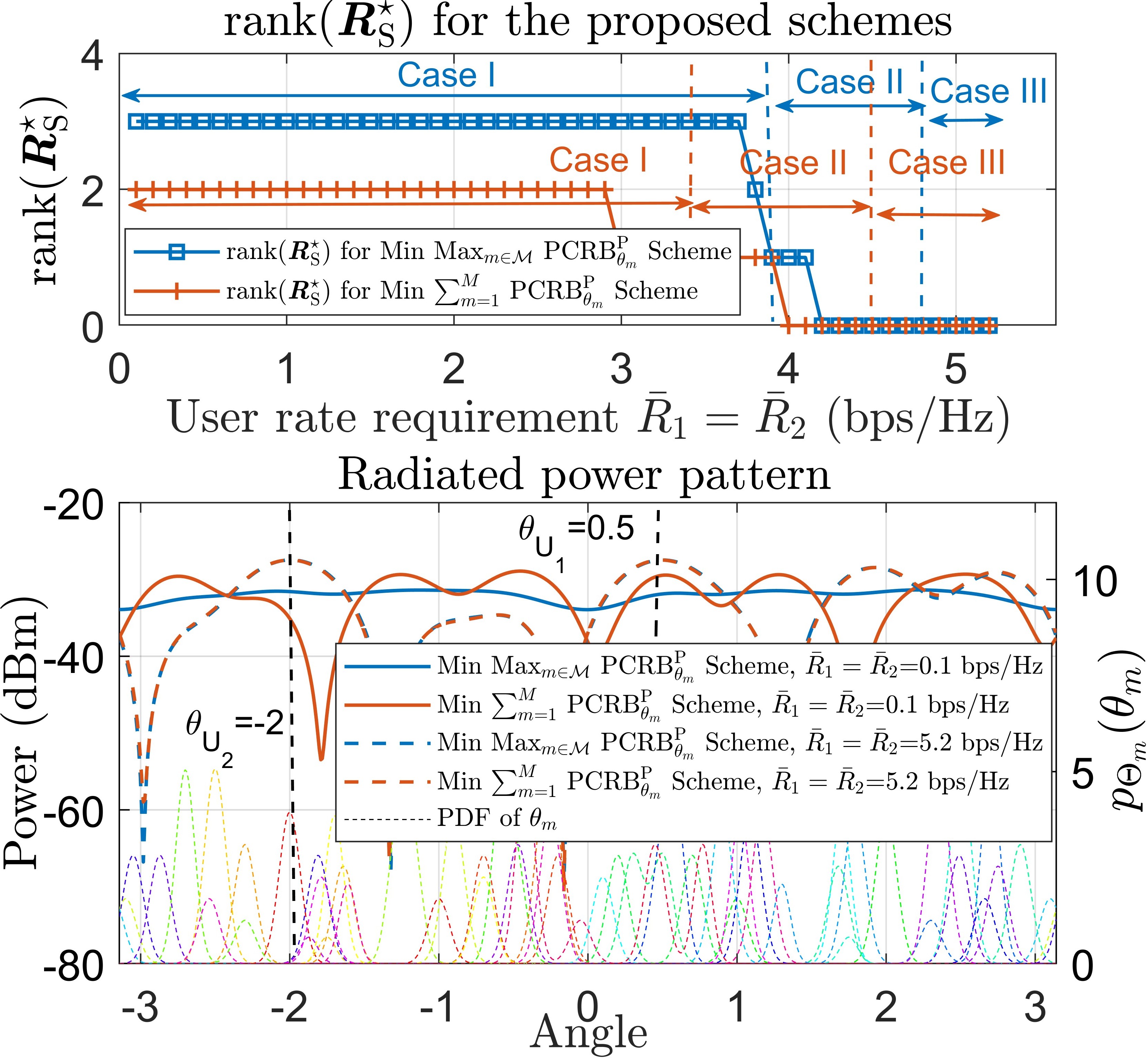}}
	\subfigure[$M=20$ targets, $K=2$ users.]{
		\label{rank_beamforming_M10}
		\includegraphics[width=0.32\linewidth]{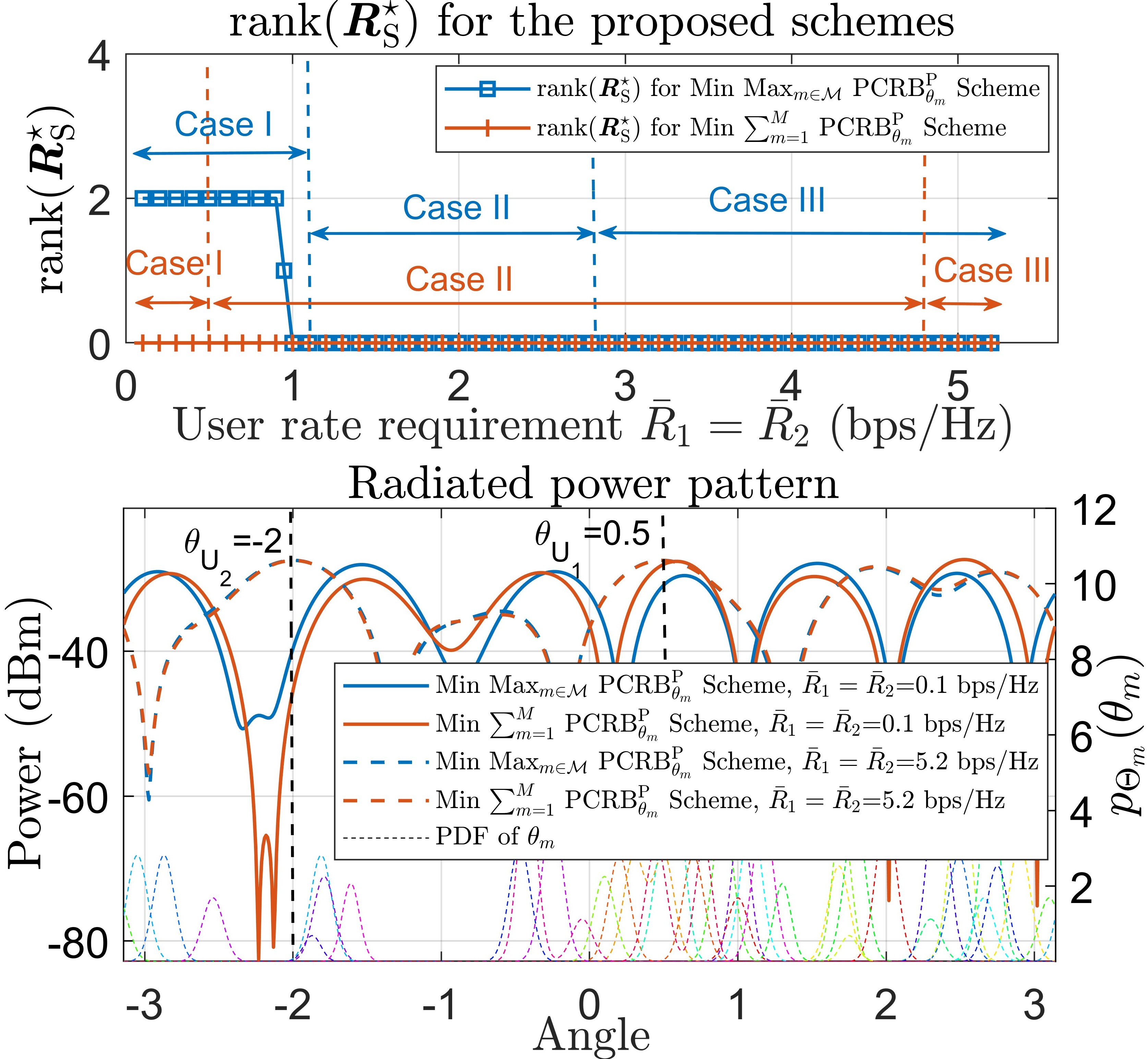}}
	\subfigure[$M=30$ targets, $K=1$ user.]{
		\label{rank_beamforming_user1}
		\includegraphics[width=0.32\linewidth]{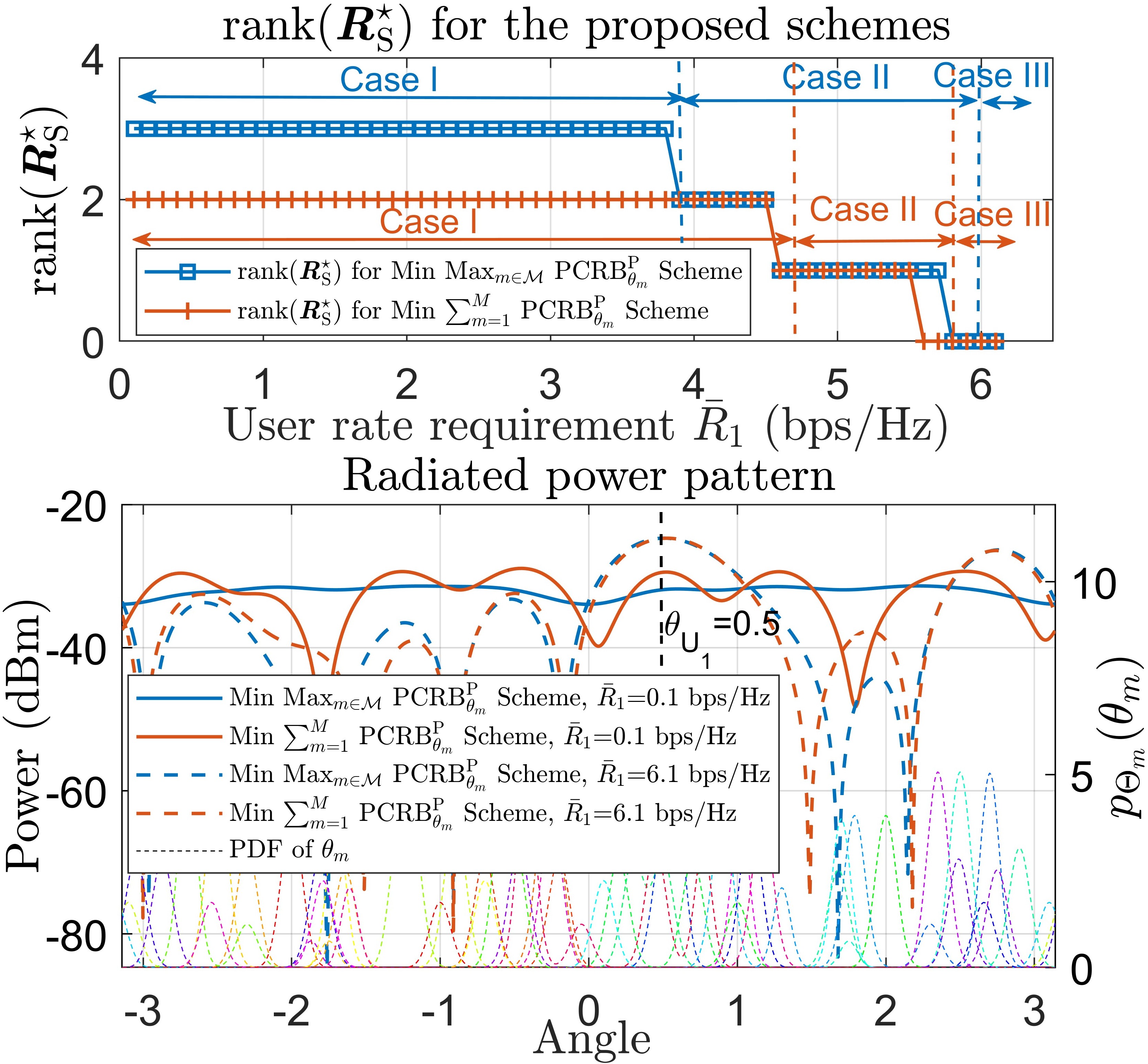}}
	\vspace{-2mm}
	\caption{Illustration of the number of sensing beams and radiated power pattern under optimal solutions to (P1) and (P3) with different numbers of targets/users.}
	\label{rank_beamforming}
	\vspace{-4.5mm}
\end{figure*}
\begin{figure*}[t]   
	\centering
		\subfigtopskip=2pt
	\subfigbottomskip=2pt
	\subfigcapskip=-5pt
	\begin{minipage}{0.65\textwidth}
		\centering
		\subfigure[Concentrated angle PDF.]{
			\label{rank_beamforming_gathered}
			\includegraphics[width=0.48\linewidth]{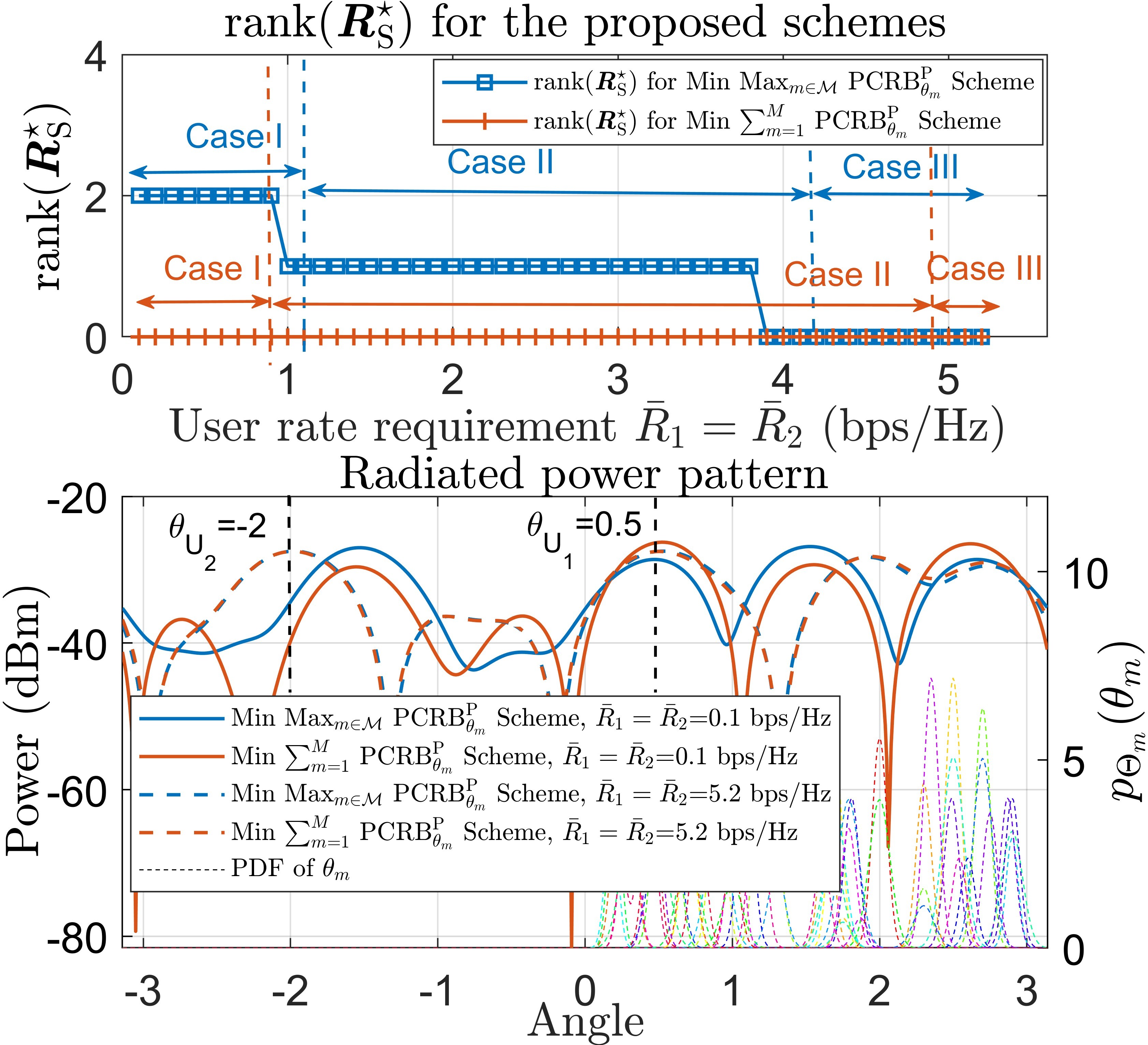}}
		\subfigure[Uniform (dispersed) angle PDF.]{
			\label{rank_beamforming_uniform}
			\includegraphics[width=0.48\linewidth]{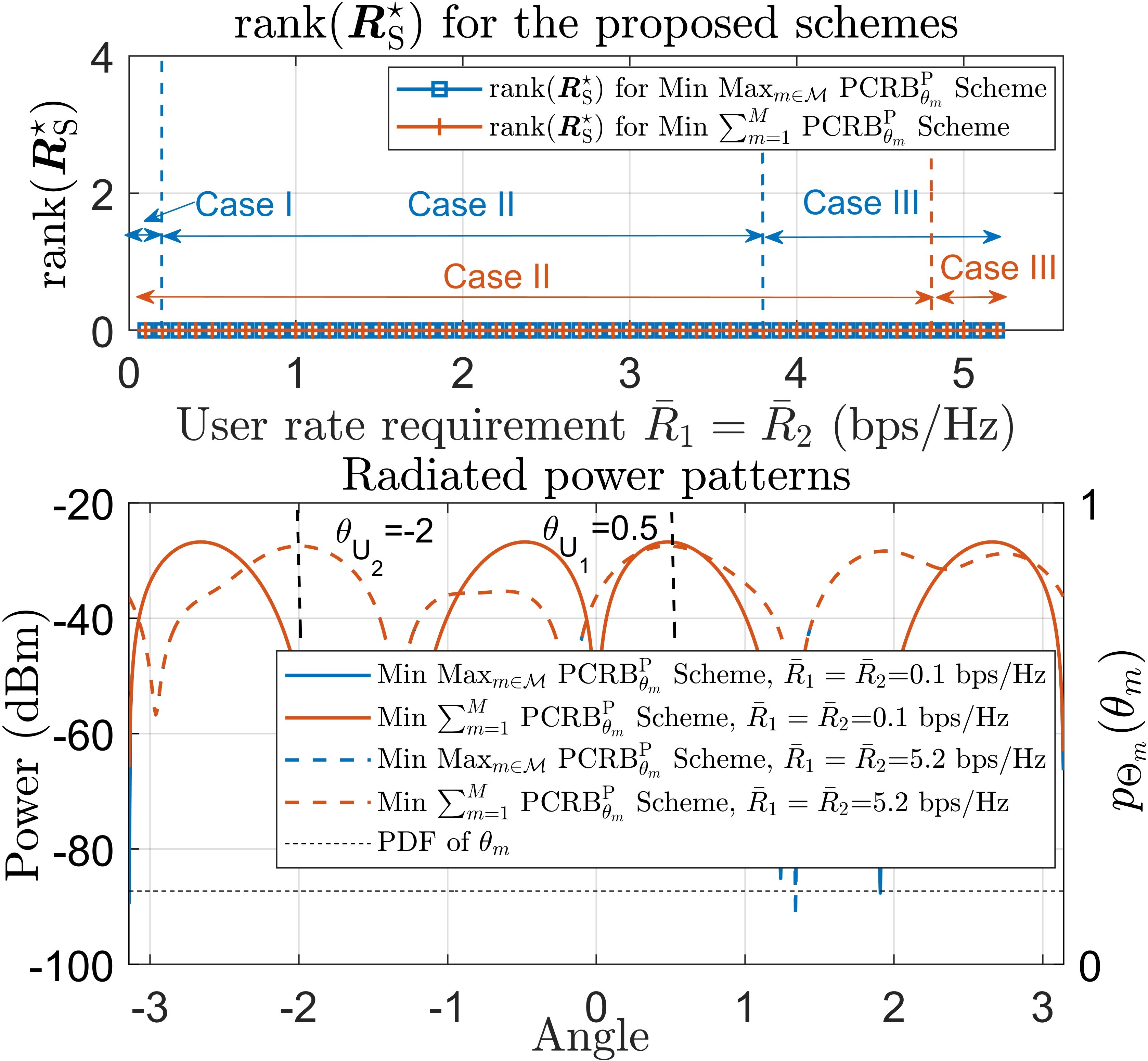}}
		\vspace{-2.5mm}
		\caption{Illustration of the number of sensing beams and radiated power pattern under optimal solutions to (P1) and (P3) with different prior PDFs under $M=30$ targets and $K=2$ users.}
	\end{minipage}
	\hfill  
	\begin{minipage}{0.32\textwidth}
		\includegraphics[width=0.98\linewidth]{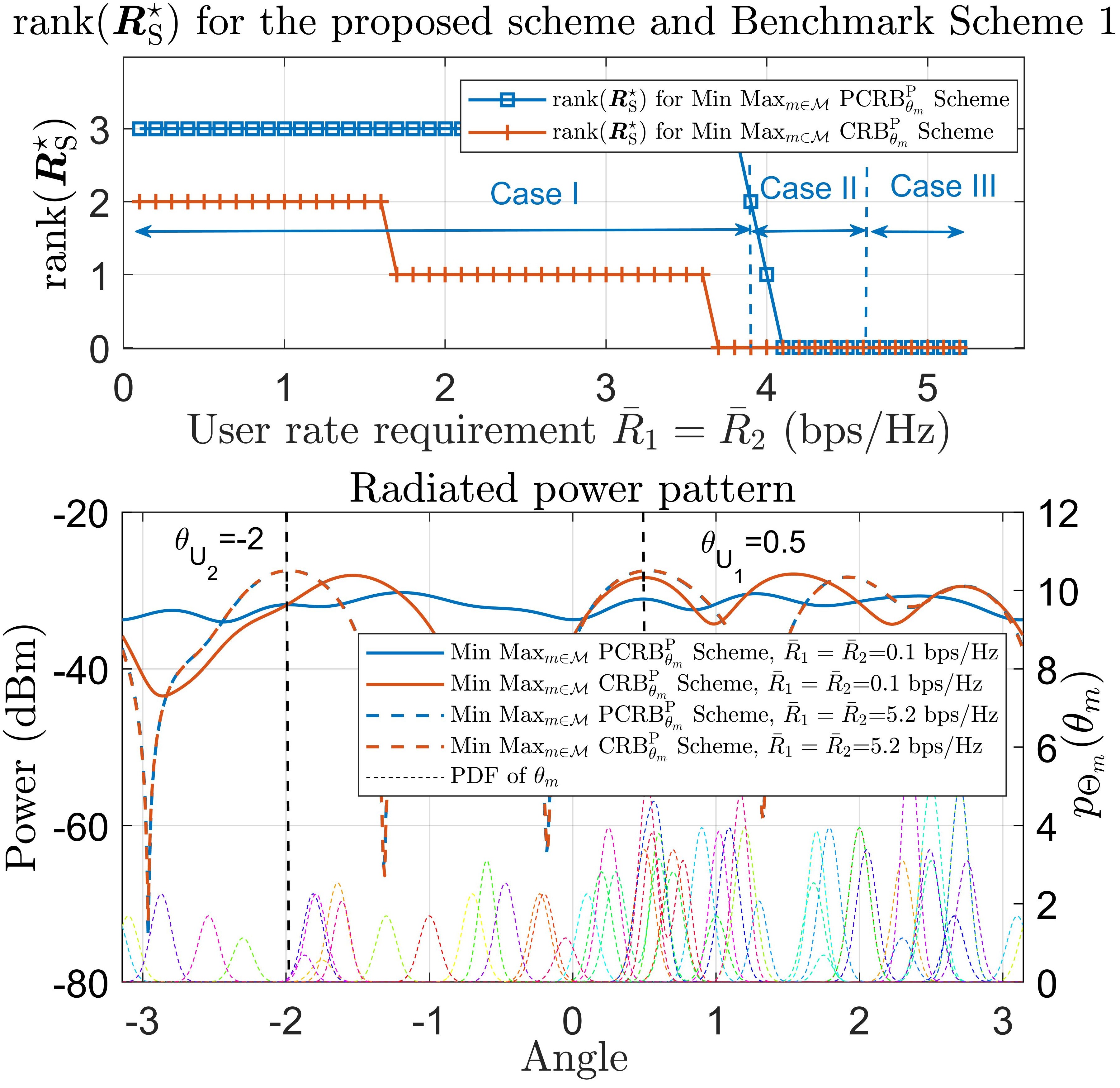}
		\vspace{-2mm}
		\caption{Comparison of the optimal solution to (P1) and Benchmark Scheme 1.}
		\label{rank_beamforming_CRB}
	\end{minipage}
	\vspace{-7mm}
\end{figure*}
Fig. \ref{rank_beamforming} compares the beamforming designs under different numbers of targets and/or users with different rate requirements, where the PDFs of the targets are illustrated as fine dashed curves in the bottom figure of each subfigure with different colors representing different targets. With $M=30$ targets and $K=2$ users, it is observed from Fig. \ref{rank_beamforming_base} that the number of sensing beams is always limited by the general bound $\sqrt{M}$ derived in Theorem \ref{prop_rank} and Theorem \ref{prop_rank_sum}; while \emph{one} or \emph{no} sensing beam is needed in various rate regimes, which are consistent with our analysis of Case I, Case II, and Case III in Sections \ref{opti_solution_P1} and \ref{opti_solution_P3}. Moreover, compared to the optimal beamforming design for the min-max periodic PCRB problem, its counterpart for the min-sum problem generally requires fewer sensing beams, because the former needs to cater to the possible angles of every target so as to optimize the worst-case performance among the multiple targets. This is also reflected on the radiated power pattern in the low-rate regime, where the min-max beamforming yields more evenly distributed power patterns, while the min-sum beamforming radiates less power over various angles with low probability densities and higher power over highly-probable angles. Furthermore, as the user rate requirement becomes more stringent, the number of dedicated sensing beams becomes smaller since less interference from sensing can be tolerated, while the radiated power towards the user locations is increased. With a high rate requirement (e.g., $5.2$ bps/Hz for both users), the system will be communication-limited, thus the min-max and min-sum beamforming solutions tend to yield the same radiated power pattern despite their different sensing performance metrics.

\vspace{-1.5mm}
Fig. \ref{rank_beamforming_M10} shows the optimal beamforming designs with the first $M=20$ targets and $K=2$ users. By comparing it with Fig. \ref{rank_beamforming_base}, it is observed that the number of dedicated sensing beams becomes smaller as the number of targets becomes smaller, as fewer angles need to be covered. It can be observed from the radiated power pattern that with fewer targets to consider, there is more flexibility in the beamforming design, which enables power saving over an enlarged range of angles. Fig. \ref{rank_beamforming_user1} shows the optimal beamforming designs with $M=30$ targets and $K=1$ user (user 1). Compared with Fig. \ref{rank_beamforming_base}, it is observed that the reduction in the number of users allows for an increased number of dedicated sensing beams, as it becomes easier to control the interference generated by sensing to communication. \looseness=-1

\vspace{-0mm}
\subsubsection{Optimal Beamforming Designs with Different PDFs}
In Fig. \ref{rank_beamforming_gathered}, we consider a more concentrated target angle PDF where all targets' angles are gathered together with means $\theta_{m,v}\in [0, \pi )$. It is observed that compared with Fig. \ref{rank_beamforming_base}, fewer dedicated sensing beams are needed in general. Moreover, the radiated power pattern becomes more concentrated to reduce unnecessary power waste on angles with small probability densities. On the other hand, Fig. \ref{rank_beamforming_uniform} considers the most dispersed PDF of the target's angles, i.e., the uniform distribution with $p_{\Theta_m}(\theta_m)=\frac{1}{2\pi},\forall m\in \mathcal{M}$, where the targets are \emph{homogeneous}. It is observed that \emph{no} dedicated sensing beam is needed; moreover, the two schemes achieve the same radiated power patterns,\footnote{The minor difference is due to numerical calculation inaccuracies in CVX.} which are consistent with our analytical results in Sections \ref{opti_solution_P1} and \ref{opti_solution_P3}. In addition, compared to Fig. \ref{rank_beamforming_uniform}, the optimal beamforming for Fig. \ref{rank_beamforming_base} strengthens the power at various angles based on knowledge of their high probability densities. \looseness=-1

\vspace{-5mm}
\subsection{Performance of Proposed Optimal Beamforming Designs Versus Benchmark Schemes}
\vspace{-2mm}
In this subsection, we compare the performance of our proposed optimal solutions versus three benchmark schemes.
\begin{itemize}[leftmargin=*]
	\item \textbf{Benchmark Scheme 1: CRB-based beamforming.} Similar to the sensing performance metric considered in \cite{liu2021cramer}, we optimize the beamforming to minimize the maximum or sum periodic CRB corresponding to the most probable angles of the multiple targets under individual rate constraints.\normalcolor
	\item \textbf{Benchmark Scheme 2: Dual-functional beamforming.} We only optimize dual-functional communication beamforming vectors $\{\bm{w}_k\}_{k=1}^K$ for (P1) or (P3) with $\bm{S}=\bm{0}$.
	\item \textbf{Benchmark Scheme 3: Sensing-oriented beamforming.} We only optimize the sensing beamforming vectors in $\bm{S}$ to minimize the maximum or sum periodic PCRB.
	\vspace{-1mm}
\end{itemize}
\begin{figure}[t]
	\centering
	\vspace{-0.4cm}
	\subfigtopskip=2pt
	\subfigbottomskip=2pt
	\subfigcapskip=-5pt
	\subfigure[Maximum periodic PCRB versus user rate requirement.]{
		\label{PCRB_rate_max}
		\includegraphics[width=8.5cm]{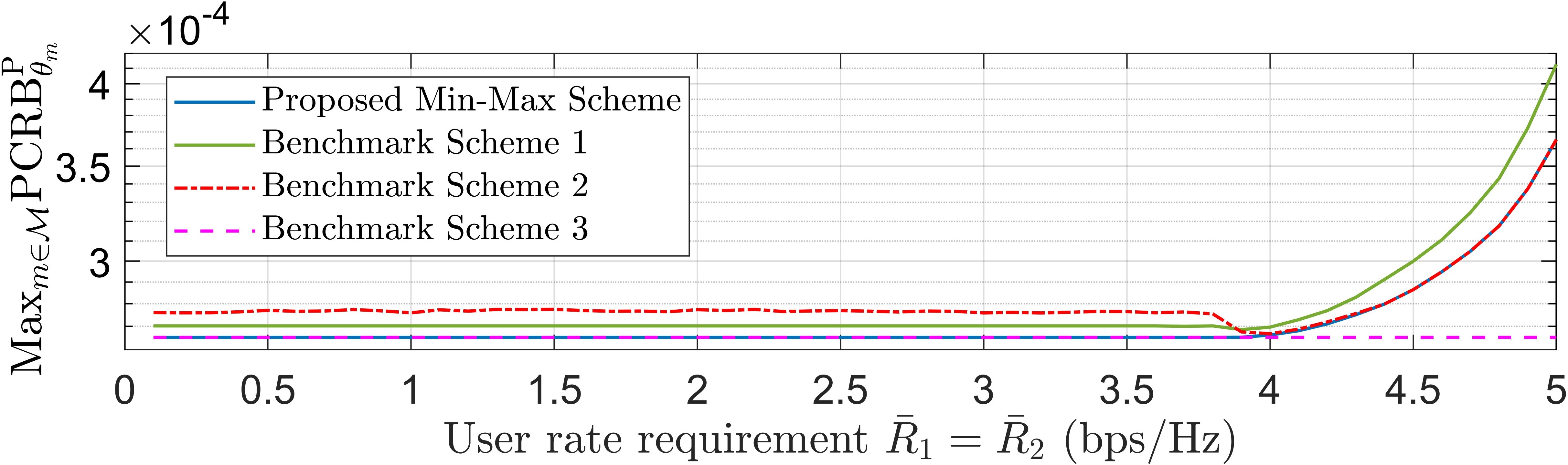}}
	\subfigure[Sum periodic PCRB versus user rate requirement.]{
		\label{PCRB_rate_sum}
		\includegraphics[width=8.5cm]{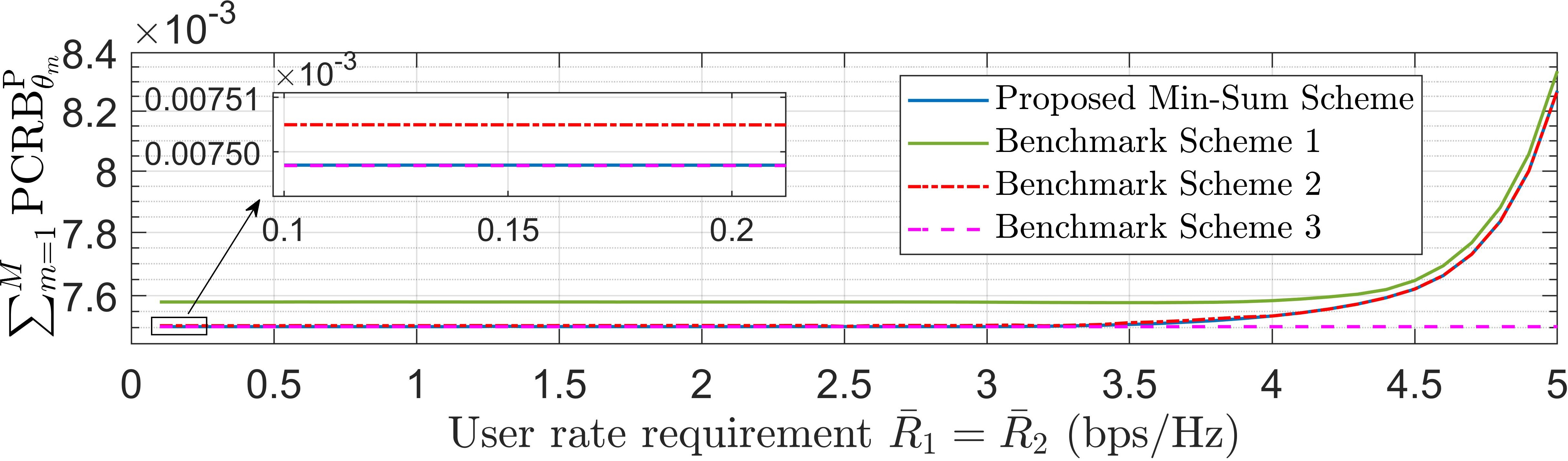}}
	\vspace{-2mm}
	\caption{Maximum or sum periodic PCRB versus user rate requirement.}
	\label{PCRB_rate_all}
	\vspace{-0mm}
\end{figure}

\vspace{-0mm}
Fig. \ref{rank_beamforming_CRB} shows that our proposed min-max periodic PCRB scheme generally requires more dedicated sensing beams than Benchmark Scheme 1 as it considers a broader range of angles with high probability densities instead of focusing only on the most probable ones. Moreover, the radiated power patterns also show that our proposed design focuses sufficient power across wider ranges of angles with high probability densities. Fig. \ref{PCRB_rate_all} shows the trade-off between the multi-target sensing performance characterized by the maximum or sum periodic PCRB and the communication rate requirement $\bar{R}_1=\bar{R}_2$, under the same setup as Fig. \ref{rank_beamforming_base}. It is observed that our proposed optimal designs outperform Benchmark Scheme 1 in both cases as they make full use of the prior distribution information. Moreover, both proposed designs outperform Benchmark Scheme 2 in low-to-moderate rate regimes, which demonstrates the need of dedicated sensing beams; while in the high-rate regime, their performances overlap, which verifies our analytical results in Sections \ref{section_min_max} and \ref{section_min_sum} that no sensing beam is needed. In addition, we show the maximum or sum periodic PCRB achieved by Benchmark Scheme 3 as a lower bound, although they are not feasible for (P1) and (P3) due to the absence of communication beams. It is observed that our proposed designs can achieve the same sensing performance as Benchmark Scheme 3 in the low-rate regime, which demonstrates their effectiveness.

\begin{figure}[t]
	\centering
		\vspace{-0.5cm}
		\subfigtopskip=2pt
		\subfigbottomskip=4pt
		\subfigcapskip=-4pt
	\subfigure[Maximum MCE/periodic PCRB in sensing $\bm{\theta}$ versus transmit power.]{
		\label{PCRB_MCE_max}
		\includegraphics[width=8.4cm]{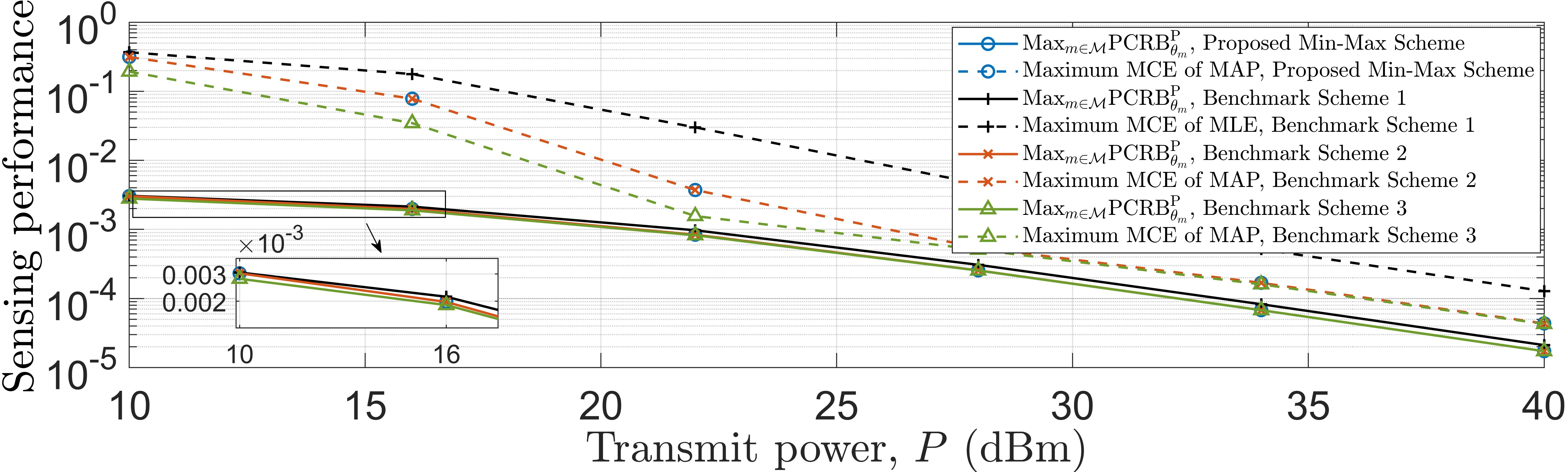}}
	\subfigure[Sum MCE/periodic PCRB in sensing $\bm{\theta}$ versus transmit power.]{
		\label{PCRB_MCE_sum}
		\includegraphics[width=8.4cm]{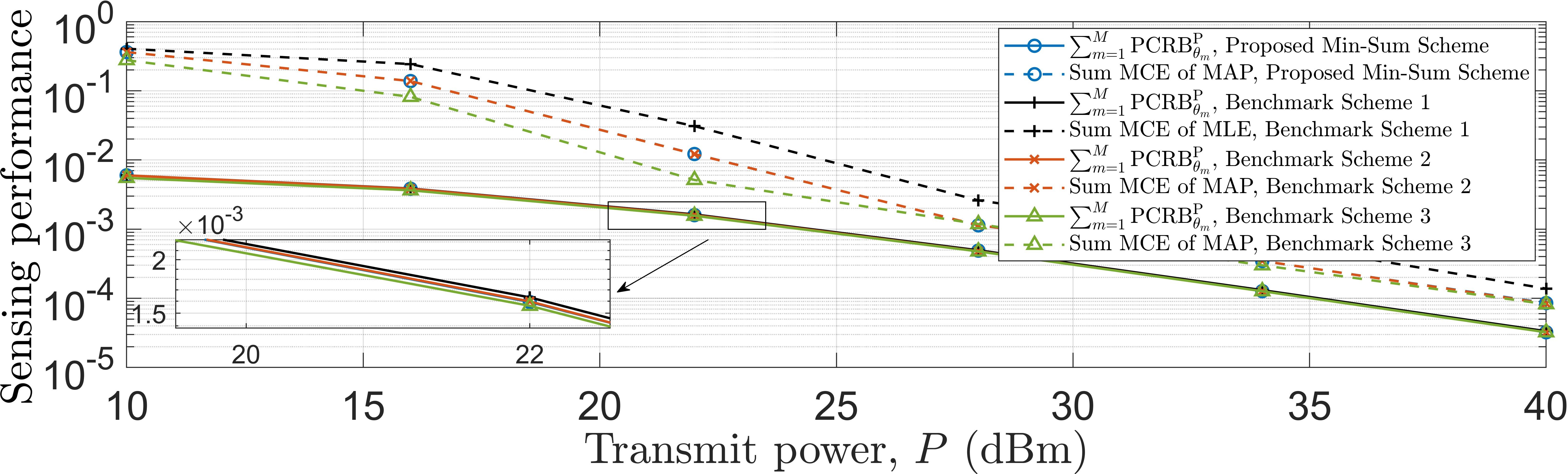}}
	\vspace{-3mm}
	\caption{Maximum or sum MCE/periodic PCRB versus transmit power.}
	\label{PCRB_MCE_all}
	\vspace{-5mm}
\end{figure}

Fig. \ref{PCRB_MCE_all} evaluates the MCE of various schemes under $M=2$, $K=2$, and $\bar{R}_1=\bar{R}_2=4.1$ bps/Hz. Denote $f( \bm{Y}^{\mathrm{S}}|\boldsymbol{\zeta })$ as the likelihood function of $\bm{\zeta}$ given $\bm{Y}^\mathrm{S}$. For Benchmark Scheme 1, the maximum likelihood estimation (MLE) method is adopted to estimate $\bm{\zeta}$ with $\hat{\boldsymbol{\zeta}}_{\mathrm{MLE}}=\mathrm{arg} \underset{\boldsymbol{\zeta }}{\max}\,\ln f( \bm{Y}^{\mathrm{S}}|\boldsymbol{\zeta } ) $. For the proposed schemes as well as Benchmark Schemes 2 and 3, the estimation of $\bm{\zeta}$ is obtained via the maximum a posteriori (MAP) estimation method with $\hat{\boldsymbol{\zeta}}_{\mathrm{MAP}}\!\!=\!\! \mathrm{arg} \underset{\boldsymbol{\zeta }}{\max}\,\ln f( \bm{Y}^{\mathrm{S}}|\boldsymbol{\zeta })\! +\!\ln p_{\mathrm{Z}}( \boldsymbol{\zeta } ) $. We perform a six-dimensional exhaustive search to obtain the MAP or MLE estimates. It is observed that our proposed schemes outperform Benchmark Scheme 1 and Benchmark Scheme 2, despite the small number of targets under which at most one dedicated sensing beam is needed; moreover, they achieve close performance to the sensing-optimal Benchmark Scheme 3 in moderate-to-high power regimes, which further validates their effectiveness. Lastly, it is observed that the MCE for each scheme approaches the periodic PCRB as the transmit power increases, which validates the efficacy of adopting the periodic PCRB as a tractable sensing performance metric.\looseness=-5
\vspace{-6mm}
\subsection{Effect of Communication User Locations}
\vspace{-2mm}
\begin{figure}[t]
	\centering
	\vspace{3pt}
	\includegraphics[width=8.5cm]{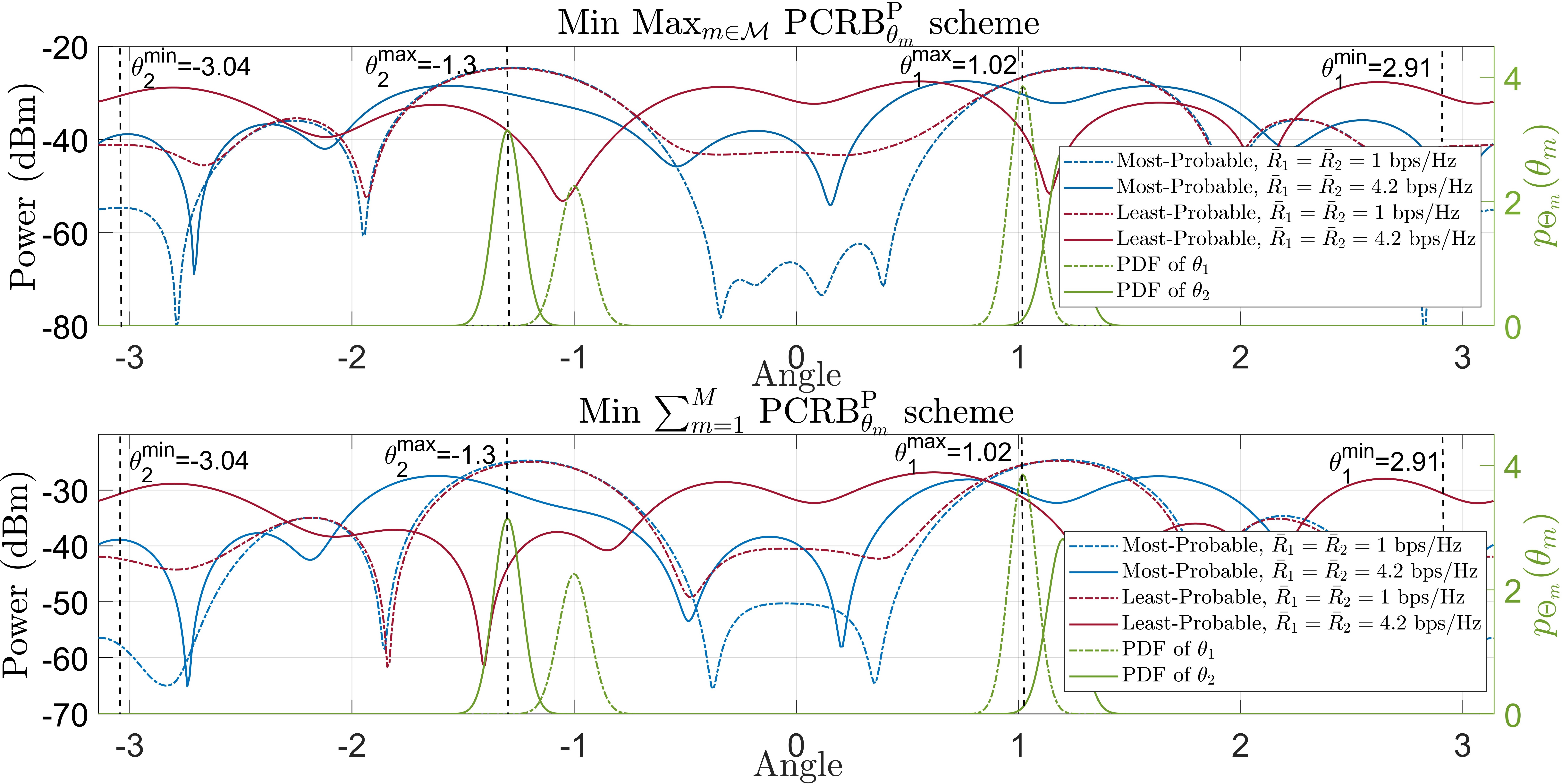}
	\vspace{-4mm}
	\caption{Radiated power pattern with different communication user locations.}
	\label{most_least_probable_beamforming}
	\vspace{-1.5mm}
\end{figure}

Finally, we explore the effect of communication user locations on the optimal beamforming design. Fig. \ref{most_least_probable_beamforming} shows the radiated power pattern when the users are located at the \emph{most-probable} or \emph{least-probable} angles of the targets, respectively, under $M=2$ and $K=2$. Specifically, user $k$ is set to be located at either the most-probable angle $\theta_k^{\max}={\arg\max}_{\theta_k\in [-\pi,\pi)}p_{\Theta_k}(\theta_k)$ or the least-probable angle $\theta_k^{\min}={\arg\min}_{\theta_k\in [-\pi,\pi)}p_{\Theta_k}(\theta_k)$ of target $k$. It is observed that when the users are located at least-probable angles, more power is radiated to these angles compared to the most-probable case, which, however, can mainly be used for communication. This effect is particularly severe if the communication rate requirement is high. Moreover, we show in Table \ref{Table_fig3} the sensing performance with different user locations and rate requirements. It is observed that when the users move from most-probable angles to least-probable angles, the sensing performance deteriorates significantly, especially under stringent user rate requirements. This is because when users are located at most-probable angles, the radiated power along these angles can be efficiently reused for both sensing and communication, which also alleviates the need of dedicated sensing beams as the communication beams pointed towards these angles can already provide sufficient power for sensing. This unveils that by exploiting the prior distribution information, each BS can select its associated users and targets based on their location relationships to further enhance the performance.
\vspace{-3mm}
\section{Conclusions}\label{section_conclusion}
\vspace{-0mm}
This paper studied the beamforming optimization for a multi-target multi-user multi-antenna ISAC system where the unknown and random angles of multiple targets need to be sensed by exploiting their joint distribution information. We analytically quantified the multi-target sensing performance by deriving the periodic PCRB. Then, we studied the beamforming optimization problems to minimize the maximum (worst-case) periodic PCRB or sum periodic PCRB among multiple sensing targets under individual communication rate constraints at the users. By leveraging the SDR technique on these non-convex and challenging problems, we proposed general algorithms to obtain the optimal solutions to both problems, and general bounds on the number of sensing beams needed. Moreover, by judiciously examining the KKT conditions, we devised more explicit forms of the optimal solutions in various practical cases and proposed tighter bounds on the number of sensing beams needed. Numerical results demonstrated the efficacy of our proposed solutions and validated our analysis.

\begin{table}[t]
	\caption{Max or Sum Periodic PCRB with Different User Locations}
	\vspace{-3mm}
	\centering
	\scalebox{0.48}{
			\resizebox{\textwidth}{!}{
			\begin{tabular}{|c|c|c|c|}
				\hline
				\multicolumn{2}{|c|}{} &\makecell[c]{Most-Probable Target Angles,  $\theta_{\mathrm{U}_k}=\theta_k^\mathrm{max}$ }& \makecell[c]{Least-Probable Target Angles,   $\theta_{\mathrm{U}_k}=\theta_k^\mathrm{min}$ }\\
				\hline
				\multirow{2}*{\raisebox{-0\height}{$\underset{m\in\mathcal{M}}{\mathrm{Max}}\mathrm{PCRB}_{\theta_m}^\mathrm{P}$}}&\makecell{$\bar{R}_1=\bar{R}_2=1$ bps/Hz}&$1.2916\times10^{-4}$&$1.3326\times10^{-4}$\\
				\cline{2-4}
				&\makecell{$\bar{R}_1=\bar{R}_2=4.2$ bps/Hz}&$1.8995\times10^{-4}$&$5.0757\times10^{-4}$\\
				\hline
				\multirow{2}*{\raisebox{-0\height}{$\sum_{m=1}^{M}\mathrm{PCRB}_{\theta_m}^\mathrm{P}$}}&\makecell{$\bar{R}_1=\bar{R}_2=1$ bps/Hz}&$2.5426\times10^{-4}$&$2.6088\times10^{-4}$\\
				\cline{2-4}
				&\makecell{$\bar{R}_1=\bar{R}_2=4.2$ bps/Hz}&$3.6082\times10^{-4}$&$8.8158\times10^{-4}$\\
				\hline
			\end{tabular}
		}
			}
	\label{Table_fig3}
\end{table}

\vspace{-4mm}
\bibliographystyle{IEEEtran}
\bibliography{reference}

\appendices
\section{Proof of Theorem 1}\label{Appendix_PCRB_expression}
Due to the periodic property of $\bm{\theta}$, we introduce $\tilde{\bm{\theta}}$ and $\tilde{\bm{\zeta}}$ as the $2\pi$-periodic extended versions of $\bm{\theta}$ and $\bm{\zeta}$, respectively. Note that $f(\bm{Y}^\mathrm{S}|\bm{\zeta})$ is the likelihood function of $\bm{\zeta}$ given $\bm{Y}^\mathrm{S}$, the $2\pi$-periodic extended functions can be thus written as $\tilde{f}(\bm{Y}^\mathrm{S}|\tilde{\bm{\zeta}})=f(\bm{Y}^\mathrm{S}|\tilde{\bm{\zeta}}+\bm{\varsigma}(\tilde{\bm{\theta}}))$ and $\tilde{p}_{\mathrm{Z}}(\tilde{\bm{\zeta}})=p_{\mathrm{Z}}(\tilde{\bm{\zeta}}+\bm{\varsigma}(\tilde{\bm{\theta}})) $.

When the prior information is exploited, the periodic PFIM is given by $\bm{F}=\bm{F}_{\mathrm{O}}+\bm{F}_{\mathrm{P}}$ \cite{shen2010fundamental1} with $\bm{F}_{\mathrm{O}}\in \mathbb{R} ^{3M \times 3M}$ and $\bm{F}_{\mathrm{P}}\in \mathbb{R} ^{3M \times 3M}$ being the periodic PFIMs from observation (i.e., the received signals $\bm{Y}^\mathrm{S}$ shown in (\ref{Y})) and prior information (i.e., $\tilde{p}_{\mathrm{Z}}(\tilde{\bm{\zeta}})$), respectively. In the following, we first derive $\bm{F}_{\mathrm{O}}$ and $\bm{F}_{\mathrm{P}}$, based on which $\mathrm{PCRB}_{\theta _m}^{\mathrm{P}}$ can be further derived.

Firstly, $\bm{F}_{\mathrm{O}}$ can be expressed as
\begin{align}
	\bm{F}_{\mathrm{O}}=&\mathbb{E}_{\bm{Y}^\mathrm{S},\bm{\zeta}}\Big[ \frac{\partial \ln ( \tilde{f}( \bm{Y}^\mathrm{S}| \tilde{\bm{\zeta}}))}{\partial \tilde{\bm{\zeta}}}\Big(\frac{\partial\ln(\tilde{f}( \bm{Y}^\mathrm{S}| \tilde{\bm{\zeta}}))}{\partial \tilde{\bm{\zeta}}}\Big)^T\Big]\nonumber\\
	=&[\bm{F}^{\mathrm{\bm{\theta\theta}}}_{\mathrm{O}},\bm{F}^{\mathrm{\bm{\theta\alpha}}}_{\mathrm{O}};	{\bm{F}^{\mathrm{\bm{\theta\alpha}}}_{\mathrm{O}}}^T,\bm{F}^{\mathrm{\bm{\alpha\alpha}}}_{\mathrm{O}}],
\end{align}
where $\bm{F}^{\mathrm{\bm{\theta\theta}}}_{\mathrm{O}}\in \mathbb{R}^{M\times M}$, $\bm{F}^{\mathrm{\bm{\theta\alpha}}}_{\mathrm{O}}\in \mathbb{R}^{M\times2M}$, and $\bm{F}^{\mathrm{\bm{\alpha\alpha}}}_{\mathrm{O}}\in \mathbb{R}^{2M\times 2M}$. Due to the periodic characteristic of trigonometric functions, we have $\bm{M}(\tilde{\theta}_m)=\bm{M}(\tilde{\theta}_m-2\pi \lfloor \frac{\tilde{\theta} _m+\pi}{2\pi} \rfloor)=\bm{M}(\theta_m)$ and $\dot{\bm{M}}(\tilde{\theta}_m)=\dot{\bm{M}}(\tilde{\theta}_m-2\pi \lfloor \frac{\tilde{\theta}_m+\pi}{2\pi} \rfloor)=\dot{\bm{M}}(\theta_m)$. Moreover, under the practical and mild conditions for $\bm{\theta}$ and $\bm{\alpha}$ in (\ref{mild_condition_1}) and (\ref{mild_condition_2}), the $(m,n)$-th element of $\bm{F}^{\mathrm{\bm{\theta\theta}}}_{\mathrm{O}}$ with $m\neq n$ can be derived as
\begin{align}
	&[\bm{F}^{\mathrm{\bm{\theta\theta}}}_{\mathrm{O}}]_{m,n}\nonumber\\[-2mm]
	=&\frac{2L}{\sigma_\mathrm{S}^2}\mathfrak{Re}\Big\{\mathbb{E}_{\bm{\zeta}}\Big[\alpha _{m}^{*}\alpha_n\mathrm{tr}(\dot{\bm{M}}^H(\theta_m)\dot{\bm{M}}(\theta_n)(\sum_{k=1}^K\bm{w}_k\bm{w}_k^H\nonumber\\[-2mm]
	&+\bm{SS}^H))\Big]\Big\}=0,
\end{align}
and the $(m,m)$-th element of $\bm{F}^{\mathrm{\bm{\theta\theta}}}_{\mathrm{O}}$ can be derived as
\begin{align}
	&[\bm{F}^{\mathrm{\bm{\theta\theta}}}_{\mathrm{O}}]_{m,m}\nonumber\\
	=&\frac{2L}{\sigma_\mathrm{S}^2}\mathfrak{Re}\Big\{\mathbb{E}_{\bm{\zeta}}\Big[({\alpha_{m}^{\mathrm{R}}}^2+{\alpha_{m}^{\mathrm{I}}}^2)\nonumber\\[-2mm]
	&\times \mathrm{tr}(\dot{\bm{M}}^H(\theta_m)\dot{\bm{M}}(\theta_m)(\sum_{k=1}^K\bm{w}_k\bm{w}_k^H+\bm{SS}^H))\Big]\Big\}\nonumber\\
	=&\frac{2L}{\sigma_\mathrm{S}^2}\mathfrak{Re}\Big\{(\iint{({\alpha _{m}^{\mathrm{R}}} ^2+{\alpha _{m}^{\mathrm{I}}} ^2)p_{\alpha _{m}^{\mathrm{R}},\alpha _{m}^{\mathrm{I}}}( \alpha _{m}^{\mathrm{R}},\alpha _{m}^{\mathrm{I}} ) d\alpha _{m}^{\mathrm{R}}d\alpha _{m}^{\mathrm{I}}})\nonumber\\[-2mm]
	&\times \mathrm{tr}(\mathbb{E}_{\bm{\theta}}[\dot{\bm{M}}^H(\theta_m) \dot{\bm{M}}( \theta _m)] (\sum_{k=1}^K\bm{w}_k\bm{w}_k^H+\bm{SS}^H))\Big\}\nonumber\\[-2mm]
	=&\frac{2Lc_m}{\sigma_\mathrm{S} ^2}\mathrm{tr}(\bm{A}_m(\sum_{k=1}^K\bm{w}_k\bm{w}_k^H+\bm{SS}^H)).
\end{align}

Furthermore, each $(m,n)$-th $1\times 2$ block of $\bm{F}^{\mathrm{\bm{\theta\alpha}}}_{\mathrm{O}}$ \hbox{is given by}
\begin{align}
	&[\bm{F}^{\mathrm{\bm{\theta\alpha}}}_{\mathrm{O}}]_{m,2n-1:2n}\nonumber\\[-2mm]
	=&\frac{2L}{\sigma_\mathrm{S} ^2}\mathfrak{Re}\Big\{\mathbb{E}_{\bm{\zeta}}\Big[\alpha _{m}^{*}\mathrm{tr}( \dot{\bm{M}}^H(\theta _m) \bm{M}( \theta _n)(\sum_{k=1}^K\bm{w}_k\bm{w}_k^H\nonumber\\[-2mm]
	&+\bm{SS}^H))[ 1,j]\Big]\Big\} \nonumber\\[-2mm]
	=&\frac{2L}{\sigma_\mathrm{S} ^2}\mathfrak{Re} \Big\{ \iint{\alpha _{m}^{*}}p_{\alpha _{m}^{\mathrm{R}},\alpha _{m}^{\mathrm{I}}}( \alpha _{m}^{\mathrm{R}},\alpha _{m}^{\mathrm{I}} ) d\alpha _{m}^{\mathrm{R}}d\alpha _{m}^{\mathrm{I}} \nonumber\\[-2mm]
	&\times\mathbb{E}_{\bm{\theta}}\Big[\!\mathrm{tr}(\!\dot{\bm{M}}^H(\theta _m)\bm{M}(\theta _n)(\sum_{k=1}^K\bm{w}_k\bm{w}_k^H+\bm{SS}^H))[1,j]\Big]\!\Big\}\nonumber\\[-2mm]
	=&\mathbf{0},\quad\forall m,n.
\end{align}

For $\bm{F}^{\mathrm{\bm{\alpha\alpha}}}_{\mathrm{O}}$, each $(m,n)$-th $2\times 2$ block is given by
\begin{align}
	&[\bm{F}^{\mathrm{\bm{\alpha\alpha}}}_{\mathrm{O}}]_{2m-1:2m,2n-1,2n}=\frac{2L}{\sigma_\mathrm{S} ^2}\mathfrak{Re}\Big\{\Big[ \begin{matrix}
		1&		j\\
		-j&		1\\
	\end{matrix} \Big]\\[-1mm]
	\times& \mathrm{tr}(\mathbb{E}_{\bm{\theta}}[\bm{M}^H\!(\theta _m) \bm{M}( \theta _n )](\sum_{k=1}^K\bm{w}_k\bm{w}_k^H+\bm{SS}^H)) \!\Big\},\forall m,n.\nonumber
\end{align}

On the other hand, based on the prior information $\tilde{p}_{\mathrm{Z}}(\tilde{\bm{\zeta}})$, $\bm{F}_{\mathrm{P}}$ can be calculated offline as
\begin{align}
	\label{equ11}
	\bm{F}_{\mathrm{P}}=\mathbb{E}_{\bm{\zeta}}\Big[ \frac{\partial \ln ( \tilde{p}_{\mathrm{Z}}( \tilde{\bm{\zeta}} ) )}{\partial \tilde{\bm{\zeta}}}\Big(\frac{\partial \ln ( \tilde{p}_{\mathrm{Z}}( \tilde{\bm{\zeta}} ) )}{\partial \tilde{\bm{\zeta}}}\Big)^H\Big].
\end{align}

Denote $\bm{\zeta}^{( 1 )}=[\zeta_1,...,\zeta_{m-1},\zeta_{m+1},...,\zeta_{3M}]^T $, $\bm{\zeta}^{( 2 )}=[\zeta_1,...,\zeta_{n-1},\zeta_{n+1},...,\zeta_{3M}]^T $, and $\bm{\zeta}^{( 3 )}=[\zeta_1,...,\zeta_{m-1},\zeta_{m+1},...,\zeta_{n-1},\zeta_{n+1},...,\zeta_{3M}]^T $. Moreover, denote $\tilde{\bm{\zeta}}^{( 1 )}=[\tilde{\zeta}_1,...,\tilde{\zeta}_{m-1},\tilde{\zeta}_{m+1},...,\tilde{\zeta}_{3M}]^T $, $\tilde{\bm{\zeta}}^{( 2 )}=[\tilde{\zeta}_1,...,\tilde{\zeta}_{n-1},\tilde{\zeta}_{n+1},...,\tilde{\zeta}_{3M}]^T $, and $\tilde{\bm{\zeta}}^{( 3 )}=[\tilde{\zeta}_1,...,\tilde{\zeta}_{m-1},\tilde{\zeta}_{m+1},...,\tilde{\zeta}_{n-1},\tilde{\zeta}_{n+1},...,\tilde{\zeta}_{3M}]^T $. Since for any $m\neq n$, we have
\begin{align}
	&[ \boldsymbol{F}_{\mathrm{P}} ] _{m,n}\\[-2mm]
	=&\int\Big( \iint{\frac{\partial \ln\mathrm{(}\tilde{p}_{\mathrm{Z}}(\tilde{\bm{\zeta}}^{\left( 1 \right)},\tilde{\zeta}_m))}{\partial \tilde{\zeta}_m}}\frac{\partial \ln\mathrm{(}\tilde{p}_{\mathrm{Z}}(\tilde{\bm{\zeta}}^{\left( 2 \right)},\tilde{\zeta}_n))}{\partial \tilde{\zeta} _n}\nonumber\\[-2mm]
	& \times\tilde{p}_{\mathrm{Z}}(\tilde{\bm{\zeta}}^{\left( 3 \right)},\tilde{\zeta}_m,\tilde{\zeta}_n)d\zeta_md\zeta_n \Big)d\bm{\zeta}^{\left( 3 \right)}\nonumber
	\\[-1mm]
	=&\iint{\Big( \int{\frac{1}{\tilde{p}_{\mathrm{Z}}( \tilde{\bm{\zeta}}^{\left( 1 \right)},\tilde{\zeta}_m )}\frac{\partial \tilde{p}_{\mathrm{Z}}(\tilde{\bm{\zeta}}^{\left( 1 \right)},\tilde{\zeta}_m)}{\partial \tilde{\zeta}_m}\tilde{p}_{\mathrm{Z}}(\tilde{\bm{\zeta}}^{\left( 1 \right)},\tilde{\zeta}_m)d\zeta_m} \Big)} \nonumber\\[-1mm]
	&\times\frac{\partial \ln\mathrm{(}\tilde{p}_{\mathrm{Z}}(\tilde{\bm{\zeta}}^{\left( 2 \right)},\tilde{\zeta}_n))}{\partial \tilde{\zeta}_n}d\zeta_nd\bm{\zeta}^{\left( 3 \right)}\nonumber
	\\[-2mm]
	=&\begin{cases}
		\iint{\tilde{p}_{\mathrm{Z}}(\tilde{\boldsymbol{\zeta}}^{\left( 1 \right)},\tilde{\zeta}_m)\mid _{-\pi}^{\pi}}\frac{\partial \ln\mathrm{(}\tilde{p}_{\mathrm{Z}}(\tilde{\boldsymbol{\zeta}}^{\left( 2 \right)},\tilde{\zeta}_n))}{\partial \tilde{\zeta}_n}d\zeta _nd\boldsymbol{\zeta }^{\left( 3 \right)}, 1\le m\le M\\
		\iint{\tilde{p}_{\mathrm{Z}}(\tilde{\boldsymbol{\zeta}}^{\left( 1 \right)},\tilde{\zeta}_m)\mid _{-\infty}^{+\infty}}\frac{\partial \ln\mathrm{(}\tilde{p}_{\mathrm{Z}}(\tilde{\boldsymbol{\zeta}}^{\left( 2 \right)},\tilde{\zeta}_n))}{\partial \tilde{\zeta}_n}d\zeta _nd\boldsymbol{\zeta }^{\left( 3 \right)},
		\\\qquad\qquad\qquad\qquad\qquad\qquad\qquad\quad M+1\le m\le 3M\\
	\end{cases}\nonumber\\
	=&0,\nonumber
\end{align}
indicating $\bm{F}_{\mathrm{P}}$ is a diagonal matrix.

Finally, based on the above, $\bm{F}^{-1}$ is given by
\begin{equation}
	\bm{F}^{-1}=\Big[ \begin{matrix}
		\big( \bm{F}^{\mathrm{\bm{\theta\theta}}}_{\mathrm{O}}+\left[ \bm{F}_{\mathrm{P}}\right] _{1:M,1:M} \big) ^{-1}&		\bm{C}\\
		\bm{C}^T&		\bm{D}\\
	\end{matrix} \Big]
\end{equation}
with $\bm{C}\in \mathbb{R} ^{M\times 2M}$ and $\bm{D}\in \mathbb{R} ^{2M\times 2M}$ being functions of $\bm{F}$. As the inverse of the overall periodic PFIM, $\bm{F}^{-1}$, is a lower bound of the MCE matrix for sensing $\bm{\zeta}$ which is denoted by $\bm{E}$, (i.e., $\bm{E}\succeq \bm{F}^{-1}$), the periodic PCRB for the MCE in sensing $\theta_m$ can be written as
\begin{align}
	&\mathrm{PCRB}_{\theta _m}^{\mathrm{P}}\\[-2mm]
	=&2-2( 1+[ \bm{F}^{-1} ] _{m,m}) ^{-\frac{1}{2}}\nonumber\\[-2mm]
	=&2-2(1+(\frac{2Lc_m}{\sigma _\mathrm{S}^{2}}\mathrm{tr}(\bm{A}_m(\sum_{k=1}^K\bm{w}_k\bm{w}_k^H+\bm{SS}^H))\nonumber\\[-2mm]
	&+\left[\bm{F}_{\mathrm{P}}\right]_{m,m})^{-1} ) ^{-\frac{1}{2}}\nonumber\\[-2.5mm]
	=&2-2(1+(\beta _m\mathrm{tr}(\bm{A}_m(\sum_{k=1}^K\bm{w}_k\bm{w}_k^H+\bm{SS}^H))+\delta _m)^{-1} ) ^{-\frac{1}{2}}\nonumber\\[-1mm]
	=&2-\frac{2}{\sqrt{1+\frac{1}{\beta _m\mathrm{tr}\left(\boldsymbol{A}_m\left( \sum_{k=1}^K{\boldsymbol{w}_k\boldsymbol{w}_{k}^{H}}+\boldsymbol{SS}^H \right) \right)+\delta _m}}},\forall m\in\mathcal{M}.\nonumber
\end{align}
This thus completes the proof of Theorem 1.

\vspace{-4.5mm}
\section{Proof of Proposition 1}\label{proof_prop_case_I}
\vspace{-1mm}
Since $\bm{R}_\mathrm{S}^\star=\sum_{k=1}^K{\tilde{\bm{R}}_k^\star}+\tilde{\bm{R}}_\mathrm{S}^\star-\sum_{k=1}^K{\bm{R}_k^\star}$, we have $\beta _m\mathrm{tr}\big( \bm{A}_m\big( \sum_{k=1}^K{\tilde{\bm{R}}_{k}^{\star}}+\tilde{\bm{R}}_\mathrm{S}^{\star} \big)\big) +\delta _m=\beta _m\mathrm{tr}\big(\bm{A}_m\big(\sum_{k=1}^K{\bm{R}_{k}^{\star}}+\bm{R}_\mathrm{S}^{\star}\big)\big)+\delta _m,\ \forall m\in\mathcal{M}$.
Suppose $(\tilde{t}^\star,\{\tilde{\bm{R}}_k^\star\}_{k=1}^K,\tilde{\bm{R}}^\star_\mathrm{S})$ is an optimal solution to (P2-R), by setting $t^\star=\tilde{t}^\star$, $(t^\star,\{\bm{R}_k^\star\}_{k=1}^K,\bm{R}^\star_\mathrm{S})$ satisfies the constraints in (\ref{P2_PCRB}), and achieves the same objective value of (P2-R). Moreover, we have
\begin{align}
	\bm{h}_{k}^{H}\bm{R}_{k}^{\star}\bm{h}_k=\bm{h}_{k}^{H}\frac{\tilde{\bm{R}}_{k}^{\star}\bm{h}_k\bm{h}_{k}^{H}\tilde{\bm{R}}_{k}^{\star}}{\bm{h}_{k}^{H}\tilde{\bm{R}}_{k}^{\star}\bm{h}_k}\bm{h}_k=\bm{h}_{k}^{H}\tilde{\bm{R}}_{k}^{\star}\bm{h}_k
\end{align}
and
\begin{align}
	\bm{h}_{k}^{H}(\sum_{j\ne k}{\bm{R}_{j}^{\star}}\!+\!\bm{R}_\mathrm{S}^{\star}) \bm{h}_k&=\bm{h}_{k}^{H}(\sum_{j\ne k}{\tilde{\bm{R}}_{j}^{\star}}\!+\!\tilde{\bm{R}}_\mathrm{S}^{\star}\!+\!\tilde{\bm{R}}_{k}^{\star}-\bm{R}_{k}^{\star})\!\bm{h}_k\nonumber\\
	&=\bm{h}_{k}^{H}(\sum_{j\ne k}{\tilde{\bm{R}}_{j}^{\star}}\!+\!\tilde{\bm{R}}_\mathrm{S}^{\star})\bm{h}_k.
\end{align}
Therefore, (\ref{P2_rate}) holds for $(\{\bm{R}_k^\star\}_{k=1}^K,\bm{R}_\mathrm{S}^\star)$. Since $\mathrm{tr}\big( \sum_{k=1}^K{\tilde{\bm{R}}_{k}^{\star}}+\tilde{\bm{R}}_\mathrm{S}^{\star}\big)=\mathrm{tr}\big( \sum_{k=1}^K{\bm{R}_{k}^{\star}}+\bm{R}_\mathrm{S}^{\star}\big)=P$, (\ref{P2_power}) holds for $(\{\bm{R}_k^\star\}_{k=1}^K,\bm{R}_\mathrm{S}^\star)$. Furthermore, it can be shown that $\bm{R}_{k}^{\star}\succeq \bm{0}$ and thus (\ref{P2_PSD_C}) holds. Finally, for any vector $\bm{\rho }\in \mathbb{C} ^{N_t\times 1}$, we have
\begin{align}
	\bm{\rho }^H\!\!(\tilde{\bm{R}}_{k}^{\star}\!\!-\!\!\bm{R}_{k}^{\star})\bm{\rho }\!\!=\!\!\bm{\rho }^H\tilde{\bm{R}}_{k}^{\star}\bm{\rho}\!\!-\!\!(\bm{h}_{k}^{H}\tilde{\bm{R}}_{k}^{\star}\bm{h}_k) ^{-1}| \bm{\rho }^H\tilde{\bm{R}}_{k}^{\star}\bm{h}_k |^2\!\!\ge\!\! 0,
\end{align}
where the inequality holds due to the Cauchy-Schwartz inequality, i.e., $| \bm{\rho }^H\tilde{\bm{R}}_{k}^{\star}\bm{h}_k |^2\le ( \bm{h}_{k}^{H}\tilde{\bm{R}}_{k}^{\star}\bm{h}_k ) ( \bm{\rho }^H\tilde{\bm{R}}_{k}^{\star}\bm{\rho } ) $. Hence, $\tilde{\bm{R}}_{k}^{\star}-\bm{R}_{k}^{\star}\succeq \bm{0}$ and consequently $\bm{R}_\mathrm{S}^\star=\sum_{k=1}^K{(\tilde{\bm{R}}_k^\star-\bm{R}_k^\star)}+\tilde{\bm{R}}_\mathrm{S}^\star\succeq \bm{0}$ hold. Therefore, $(\{\bm{R}^\star_k\}_{k=1}^K,\bm{R}^\star_\mathrm{S})$ is feasible for (P2-R) and achieves the optimal value, thereby being an optimal solution to (P2-R). 

Due to the fact that $\mathrm{rank}(\bm{R}^\star_k)=1,\ \forall k\in\mathcal{K}$, (\ref{rank_k}) holds for $(\{\bm{R}^\star_k\}_{k=1}^K,\bm{R}^\star_\mathrm{S})$. Thus, $(\{\bm{R}^\star_k\}_{k=1}^K,\bm{R}^\star_\mathrm{S})$ is feasible for (P2) and achieves the optimal value, thereby being an optimal solution to (P2). This thus completes the proof of Proposition 1.

\vspace{-5.5mm}
\section{Proof of Theorem 2}\label{proof_prop_rank}
\vspace{-2mm}
We prove Theorem 2 by showing that for any optimal solution to (P2) and (P2-R) obtained via Proposition 1 denoted by $(\{\tilde{\bm{R}}_{k}^{\star}\} _{k=1}^{K},\tilde{\bm{R}}_\mathrm{S}^\star)$, we can always construct $(\{{\bm{R}}_{k}^{\star}\} _{k=1}^{K},{\bm{R}}_\mathrm{S}^\star)$ which satisfies (\ref{rank_Rs}) and is also an optimal solution to (P2).

Let $N_\mathrm{S}=\mathrm{rank}(\tilde{\bm{R}}_\mathrm{S}^\star)\geq 1$. Note that any optimal $\tilde{\bm{R}}_{k}^\star,\ \forall k\in \mathcal{K}$ to (P2) and (P2-R) obtained by Proposition 1 always has a rank-one structure. By decomposing $\tilde{\bm{R}}^\star_k=\tilde{\bm{v}}_k\tilde{\bm{v}}_k^H$ with $\tilde{\bm{v}}_k\in \mathbb{C}^{N_t \times 1}$ and $\tilde{\bm{R}}^\star_\mathrm{S}=\tilde{\bm{V}}_\mathrm{S}\tilde{\bm{V}}_\mathrm{S}^H$ with $\tilde{\bm{V}}_\mathrm{S}\in \mathbb{C} ^{N_t\times N_\mathrm{S}}$, the constraints in (\ref{P2_PCRB}) and (\ref{P2_rate}) can be expressed as
\begin{align}
	\beta_m(\sum_{k=1}^K \tilde{\bm{v}}_k^H\bm{A}_m\tilde{\bm{v}}_k \!+\!\mathrm{tr}( \tilde{\bm{V}}_\mathrm{S}^H\bm{A}_m\tilde{\bm{V}}_\mathrm{S})) \!+\!\delta _m \geq t,\forall \!m\!\!\in\!\! \mathcal{M}
\end{align}
\begin{align}
	&|\bm{h}_k^H\tilde{\bm{v}}_k|^2-\gamma_k(\sum_{j\ne k}|\bm{h}_k^H\tilde{\bm{v}}_j|^2+\bm{h}_k^H\tilde{\bm{V}}_\mathrm{S}\tilde{\bm{V}}_\mathrm{S}^H\bm{h}_k)\ge \gamma_k \sigma _\mathrm{C}^{2},\nonumber\\[-2mm]
	&\qquad\qquad\qquad\qquad\qquad\qquad\qquad\qquad\qquad\quad \forall k\in \mathcal{K}.
\end{align}
Define real variable $\Delta_k\in \mathbb{R}^{1\times 1},\ \forall k\in\mathcal{K}$ and Hermitian matrix $\bm{\Delta }_\mathrm{S}\in \mathbb{C} ^{N_\mathrm{S}\times N_\mathrm{S}}$ with $N_\mathrm{S}^2$ real-valued unknowns. We consider the following system of linear equations:
\begin{align}
	&\sum_{k=1}^K\Delta_k\tilde{\bm{v}}_k^H\bm{A}_m\tilde{\bm{v}}_k+\mathrm{tr}( \tilde{\bm{V}}_\mathrm{S}^H\bm{A}_m\tilde{\bm{V}}_\mathrm{S}\bm{\Delta }_\mathrm{S})=0,\forall m\in\mathcal{M}\label{rank_deduction_1}\\[-1mm]
	&\Delta _k|\bm{h}_{k}^{H}\tilde{\bm{v}}_k|^2-\gamma _k(\sum_{j\ne k}\Delta _j|\bm{h}_{k}^{H}\tilde{\bm{v}}_j|^2+\bm{h}_{k}^{H}\tilde{\bm{V}}_\mathrm{S}\mathbf{\Delta }_\mathrm{S}\tilde{\bm{V}}_{S}^{H} \bm{h}_k)\nonumber\\[-2mm]
	&=0,\ \forall k\in\mathcal{K}.\label{rank_deduction_2}
\end{align}
Since $\bm{A}_m,\ \forall m\in\mathcal{M}$ is a Hermitian matrix, there are $M+K$ real-valued equations in (\ref{rank_deduction_1}) and (\ref{rank_deduction_2}) in total. Note that there are $K+N_\mathrm{S}^2$ real-valued unknowns. Thus, if $N_{\mathrm{S}}>\sqrt{M}$, there exists a non-zero solution of $(\{\Delta _k\}_{k=1}^K,\bm{\Delta }_\mathrm{S})$ to the linear equations. Denote $\xi_i,\ i=1,...,N_\mathrm{S}$ as the eigenvalues of $\bm{\Delta }_\mathrm{S}$. Define $\xi _0$ as
\begin{align}
	\xi_0=\underset{\{\Delta_k\} _{k=1}^{K},\{\xi_i\}_{i=1}^{N_{\mathrm{S}}}}{\max}\{| \Delta _k|,\!k\!=\!1,...,K,|\xi_i |,i\!=\!1,...,N_{\mathrm{S}}\}.\label{rank_deduction_judge}
\end{align}
Then, we have
\begin{align}
	1-{{\Delta_k}\big/{\xi_0}}\geq0,\ \forall k\in\mathcal{K}
\end{align}
and
\begin{align}
	\bm{I}_{N_\mathrm{S}}-\frac{1}{\xi _0}\bm{\Delta }_\mathrm{S}\succeq \bm{0}.
\end{align}
Based on these, we can construct new matrices
\begin{align}\label{rank_reduction_construct_Rk}
	\hat{\bm{R}}_k=(1-\frac{\Delta_k}{\xi_0})\tilde{\bm{v}}_k\tilde{\bm{v}}_k^H,\ \forall k\in\mathcal{K}
\end{align}
and 
\begin{align}\label{rank_reduction_construct_RS}
	\hat{\bm{R}}_\mathrm{S}=\tilde{\bm{V}}_\mathrm{S}(\bm{I}_{N_\mathrm{S}}-\frac{1}{\xi _0}\bm{\Delta }_\mathrm{S})\tilde{\bm{V}}_\mathrm{S}^H.
\end{align}
In the following, we show that $(\{\hat{\bm{R}}_k\}_{k=1}^K,\hat{\bm{R}}_\mathrm{S})$ is feasible for (P2) and achieves the optimal value.

\vspace{-0mm}
Due to (\ref{rank_deduction_1}), we have $\beta _m\mathrm{tr}\big(\bm{A}_m\big(\sum_{k=1}^K\hat{\bm{R}}_k+\hat{\bm{R}}_\mathrm{S}\big)\big) +\delta _m=\beta _m\mathrm{tr}\big( \bm{A}_m\big(\sum_{k=1}^K\tilde{\bm{R}}^\star_k+\tilde{\bm{R}}^\star_\mathrm{S}\big)\big) +\delta _m,\ \forall m\in\mathcal{M}$. Suppose $(\tilde{t}^\star,\{\tilde{\bm{R}}_k^\star\}_{k=1}^K,\tilde{\bm{R}}^\star_\mathrm{S})$ is an optimal solution to (P2-R), by setting $\hat{t}=\tilde{t}^\star$, $(\hat{t},\{\hat{\bm{R}}_k\}_{k=1}^K,\hat{\bm{R}}_\mathrm{S})$ satisfies the constraints in (\ref{P2_PCRB}) and (\ref{rank_k}), and achieves the same objective value of (P2-R) as well as (P2). Moreover, based on (\ref{rank_deduction_2}), we have
\begin{align}\label{rank_reduction_rate_hold}
	&\bm{h}_k^H\hat{\bm{R}}_k\bm{h}_k-\gamma_k(\sum_{j\ne k}\bm{h}_k^H\hat{\bm{R}}_j\bm{h}_k+\bm{h}_k^H\hat{\bm{R}}_\mathrm{S}\bm{h}_k+\sigma _\mathrm{C}^{2})\\[-1mm]
	=&\bm{h}_k^H\tilde{\bm{R}}^\star_k\bm{h}_k-\gamma_k(\sum_{j\ne k}\bm{h}_k^H\tilde{\bm{R}}_j^\star\bm{h}_k+\bm{h}_k^H\tilde{\bm{R}}_\mathrm{S}^\star\bm{h}_k+\sigma _\mathrm{C}^{2})\nonumber\\[-2mm]
	\geq &0,\ \forall k\in\mathcal{K}.\nonumber
\end{align}
Thus, (\ref{P2_rate}) holds for $(\{\hat{\bm{R}}_k\}_{k=1}^K,\hat{\bm{R}}_\mathrm{S})$. Furthermore, it can be shown from (\ref{rank_reduction_construct_Rk}) and (\ref{rank_reduction_construct_RS}) that $\hat{\bm{R}}_k\succeq\bm{0},\ \forall k\in\mathcal{K}$ and $\hat{\bm{R}}_\mathrm{S}\succeq\bm{0}$, thus (\ref{P2_PSD_C}) and (\ref{P2_PSD_S}) hold for $(\{\hat{\bm{R}}_k\}_{k=1}^K,\hat{\bm{R}}_\mathrm{S})$. Since $\mathrm{tr}(\bm{\varPsi}_k^\star\tilde{\bm{R}}_k^\star)=\tilde{\bm{v}}_k^H\bm{\varPsi}_k^\star\tilde{\bm{v}}_k=0$, we have
\begin{align}
	\mathrm{tr}(\bm{\varPsi}_k^\star\hat{\bm{R}}_k)=(1-{{\Delta_k}\big/{\xi_0}})\tilde{\bm{v}}_k^H\bm{\varPsi}_k^\star\tilde{\bm{v}}_k=0,\, \forall k\in\mathcal{K}.\label{slack_cons_1}
\end{align}
Similarly,
\begin{align}
	\mathrm{tr}(\bm{\varPsi}_\mathrm{S}^\star\hat{\bm{R}}_\mathrm{S})=\mathrm{tr}(\bm{\varPsi}_\mathrm{S}^\star\tilde{\bm{V}}_\mathrm{S}(\bm{I}_{N_\mathrm{S}}-{{1}\big/{\xi_0}}\bm{\Delta }_\mathrm{S})\tilde{\bm{V}}_\mathrm{S}^H)=0.\label{slack_cons_2}
\end{align}
Through right multiplying (\ref{KKT_2}) by $(1-\frac{\Delta_k}{\xi_0})\tilde{\bm{v}}_{k}$ and left multiplying (\ref{KKT_2}) by $\tilde{\bm{v}}_{k}^{H}$, right multiplying (\ref{KKT_3}) by $\tilde{\bm{V}}_{\mathrm{S}}( \bm{I}_{N_{\mathrm{S}}}-\frac{1}{\xi _0}\bm{\Delta }_{\mathrm{S}}) $ and left multiplying (\ref{KKT_3}) by $\tilde{\bm{V}}_{\mathrm{S}}^{H}$, and summing all equations in (\ref{KKT_2}) as well as (\ref{KKT_3}) after multiplication, we have
\begin{align}
	\sum\nolimits_{k=1}^K{\Delta_k\|\tilde{\bm{v}}_{k}\|^2}+\mathrm{tr}( \tilde{\bm{V}}_{\mathrm{S}}^{H} \tilde{\bm{V}}_{\mathrm{S}}\bm{\Delta}_{\mathrm{S}}) =0,
\end{align}
where (\ref{rank_deduction_1}), (\ref{rank_deduction_2}), (\ref{slack_cons_1}), and (\ref{slack_cons_2}) are used. Thus, $\mathrm{tr}\big(\sum_{k=1}^K{\hat{\bm{R}}_k}+\hat{\bm{R}}_\mathrm{S}\big)\leq P$ and (\ref{P2_power}) holds. Therefore, the newly constructed solution is feasible for (P2) and achieves the optimal value.

Note that we must have $\hat{\bm{R}}_k\neq\bm{0},\ \forall k\in\mathcal{K}$. Otherwise, inequalities in (\ref{rank_reduction_rate_hold}) are violated with $\gamma_k>0,\ \forall k\in\mathcal{K}$. Since $\sum_{k=1}^K\mathrm{rank}(\hat{\bm{R}}_k)+\mathrm{rank}(\hat{\bm{R}}_\mathrm{S})\leq K+\mathrm{rank}(\tilde{\bm{R}}^\star_\mathrm{S})-1$ and $\mathrm{rank}(\hat{\bm{R}}_k)=1$, we have $\mathrm{rank}(\hat{\bm{R}}_S)\leq \mathrm{rank}(\tilde{\bm{R}}^\star_\mathrm{S})-1$, namely,
the number of sensing beams needed is guaranteed to be reduced in each step. If the constructed $\hat{\bm{R}}_\mathrm{S}$ yields $\mathrm{rank}(\hat{\bm{R}}_\mathrm{S})>\sqrt{M}$, we can always repeat the above rank reduction method until $\mathrm{rank}(\hat{\bm{R}}_\mathrm{S})\leq \sqrt{M}$. Thus, (P2) always has an optimal solution $(\{\bm{R}_k^\star\}_{k=1}^K,\bm{R}_\mathrm{S}^\star)$ with $\mathrm{rank}(\bm{R}_\mathrm{S}^\star)\leq \sqrt{M}$. The optimal $\bm{S}^\star$ to (P1) can be obtained via $\bm{R}_\mathrm{S}^\star=\bm{S}^\star\bm{S}^{\star H}$ with $\mathrm{rank}(\bm{R}_{\mathrm{S}}^\star)=\mathrm{rank}(\bm{S}^\star\bm{S}^{\star H})\leq \sqrt{M}$. This thus completes the proof of Theorem 2.
\vspace{-5mm}
\section{Proof of Proposition 2}\label{proof_lemma_case_I}
\vspace{-1mm}
According to (\ref{KKT_2}) and (\ref{KKT_3}), we have
\begin{align}
	\bm{\varPsi}_k^\star=\bm{\varPsi}_\mathrm{S}^\star=\mu^\star \bm{I}_{N_t}-\bm{U}_\mathrm{M}^\star,\quad \forall k\in \mathcal{K},
\end{align}
which means that $\mathrm{rank}(\bm{\varPsi }_k^\star)=\mathrm{rank}(\bm{\varPsi }_\mathrm{S}^\star)=N_t-1,\forall k\in \mathcal{K}$, and $\bm{q}_1$ is the orthogonal basis of the null space of both $\{\bm{\varPsi }_k^\star\}_{k=1}^K$ and $\bm{\varPsi }_\mathrm{S}^\star$. Hence, the optimal solution to (P2-R) can be expressed as $\bm{R}_k^\star=P_{k,1}^\mathrm{C}\bm{q}_1\bm{q}_{1}^{H},\ \forall k\in\mathcal{K}$ and $\bm{R}_\mathrm{S}^\star=P_{1}^\mathrm{S}\bm{q}_1\bm{q}_{1}^{H}$. For any optimal solution $(\{\bm{R}_k^\star\}_{k=1}^K,\bm{R}_\mathrm{S}^\star)$ to (P2-R), we can always construct another solution $\tilde{\bm{R}}_{k}^{\star}=( P_{k,1}^\mathrm{C}+P_{1,k}^\mathrm{S} ) \bm{q}_1\bm{q}_{1}^{H}$ and $\tilde{\bm{R}}_\mathrm{S}^\star=\bm{0}$. The new solution $(\{\tilde{\bm{R}}_k^\star\}_{k=1}^K,\tilde{\bm{R}}_\mathrm{S}^\star)$ can achieve the same optimal value of (P2-R) and satisfy all constraints in (P2-R). Thus, $(\{\tilde{\bm{R}}_k^\star\}_{k=1}^K,\tilde{\bm{R}}_\mathrm{S}^\star)$ is also an optimal solution to (P2-R). Since $\mathrm{rank}(\tilde{\bm{R}}_k^\star)=1,\forall k\in\mathcal{K}$, $(\{\tilde{\bm{R}}_k^\star\}_{k=1}^K,\tilde{\bm{R}}_\mathrm{S}^\star)$ satisfies all constraints in (P2) and achieves the optimal value of (P2). This indicates that \emph{no} dedicated sensing beam is needed. An optimal solution to (P1) is given by $\bm{w}_{k}^{\star}=\sqrt{P_{k,1}^\mathrm{C}+P_{1,k}^\mathrm{S}}\bm{q}_1,\forall k\in \mathcal{K}$, $\bm{S}^\star=\bm{0}$. Thus, Proposition 2 is proved.
\vspace{-5mm}
\section{Proof of Proposition 3}\label{proof_prop_case_II}
The KKT optimality condition (\ref{KKT_2}) can be written as
\begin{align}\label{KKT_2_case2_all}
	\bm{\varPsi }_k^\star=\begin{cases}
		\mu ^{\star}\boldsymbol{I}_{N_t}\!\!-\!\!\boldsymbol{U}_{\mathrm{M}}^{\star}\!+\!\sum_{i=1}^{\mathrm{card}(\mathcal{K}_{\mathrm{A}}\!)}{\!\!\gamma _i\nu _{i}^{\star}\boldsymbol{h}_i\boldsymbol{h}_{i}^{H}}\!-\!\nu _{k}^{\star}(\gamma _k\!\!+\!\!1)\boldsymbol{h}_k\boldsymbol{h}_{k}^{H},\\
		\qquad\qquad\qquad\qquad\qquad\qquad\qquad\quad  \forall k\in \mathcal{K}_{\mathrm{A}}&\\
		\mu ^{\star}\boldsymbol{I}_{N_t}\!\!-\!\!\boldsymbol{U}_{\mathrm{M}}^{\star}\!+\!\sum_{i=1}^{\mathrm{card}(\mathcal{K}_{\mathrm{A}}\!)}{\!\gamma _i\nu _{i}^{\star}\boldsymbol{h}_i\boldsymbol{h}_{i}^{H}},  \forall k\in \mathcal{K}_{\mathrm{I}}.
	\end{cases}
	\vspace{-2mm}
\end{align}
In this case, we have
\begin{align}
	\mu^\star\!+\!\nu_k^\star(\gamma_k\!+\!1)\tilde{\bm{q}}_i^H\bm{h}_k\bm{h}_k^H\tilde{\bm{q}}_i&\!=\!\tilde{\bm{q}}_i^H(\tilde{\bm{U}}_\mathrm{M}^\star+\nu_k^\star(\gamma_k\!+\!1)\bm{h}_k\bm{h}_k^H)\tilde{\bm{q}}_i\nonumber\\
	&\!\leq\! \lambda_{\max}(\tilde{\bm{U}}_\mathrm{M}^\star+\nu_k^\star(\gamma_k\!+\!1)\bm{h}_k\bm{h}_k^H)\nonumber\\
	&\!=\!\mu^\star,1\leq i\leq \tilde{N},\forall k\in \mathcal{K}_{\mathrm{A}}.
\end{align}
Given $\nu_k^\star>0,\,\forall k\in \mathcal{K}_{\mathrm{A}}$, we have
\begin{align}\label{h_q_bar}
	\bm{h}_k^H\tilde{\bm{q}}_i=0,\quad 1\leq i\leq \tilde{N},\,\forall k\in \mathcal{K}_{\mathrm{A}}.
\end{align}

Based on (\ref{KKT_2_case2_all}) and (\ref{h_q_bar}), we have $N_t-\tilde{N}-1\le \mathrm{rank}(\bm{\varPsi }_k^\star)\le N_t-\tilde{N},\,\forall k\in \mathcal{K}_{\mathrm{A}}$. For $1\le k\le \mathrm{card}(\mathcal{K}_{\mathrm{A}})$, if $\mathrm{rank}(\bm{\varPsi }_k^\star)=N_t-\tilde{N}$, then $\bm{R}_k^\star=\sum_{n=1}^{\tilde{N}}{P_{k,n}^\mathrm{C}\tilde{\bm{q}}_n\tilde{\bm{q}}_n^H}$. Due to (\ref{h_q_bar}), we have $\bm{h}_k^H\bm{R}_k^\star\bm{h}_k=0$, which violates the communication rate constraint. Hence, $\mathrm{rank}(\bm{\varPsi }_k^\star)=N_t-\tilde{N}-1,\ \forall k\in \mathcal{K}_{\mathrm{A}}$. Denote $[\tilde{\bm{J}},\bm{f}_k]\in \mathbb{C}^{N_t \times (\tilde{N}+1)}$ as the orthogonal basis of the null space of $\bm{\varPsi}_k^\star$ with $\bm{f}_k^H\tilde{\bm{J}}=\bm{0},\,\forall k\in \mathcal{K}_{\mathrm{A}}$. The optimal $\bm{R}_k^\star,\,\forall k\in \mathcal{K}_{\mathrm{A}}$ to (P2-R) can be expressed as
\begin{align}
	\bm{R}_k^\star\!=\!P_{f,k}\bm{f}_k\bm{f}_k^H\!+\!\sum\nolimits_{n=1}^{\tilde{N}}{P_{k,n}^\mathrm{C}\tilde{\bm{q}}_n\tilde{\bm{q}}_n^H},\  \forall k\in \mathcal{K}_{\mathrm{A}},
\end{align}
with $P_{k,n}^\mathrm{C}\geq 0,\,n=1,...,\tilde{N}$.

Due to (\ref{KKT_2_case2_all}), we can obtain that $\mathrm{rank}(\bm{\varPsi}_k^\star)= N_t-\tilde{N},\ \forall k\in \mathcal{K}_\mathrm{I}$, and $\tilde{\bm{J}}$ is the orthogonal basis of the null space of $\bm{\varPsi}_k^\star$, $\forall k\in \mathcal{K}_\mathrm{I}$. The optimal $\bm{R}_k^\star$ with $\mathrm{card}(\mathcal{K}_{\mathrm{A}})< k\le K$ can thus be given by
\begin{align}
	\bm{R}_k^\star&=\sum_{n=1}^{\tilde{N}}{P_{k,n}^\mathrm{C}\tilde{\bm{q}}_n\tilde{\bm{q}}_n^H},\quad \forall k\in \mathcal{K}_\mathrm{I},
\end{align}
with $P_{k,n}^\mathrm{C}\geq 0,\,n=1,...,\tilde{N}$. Since $\bm{\varPsi}_\mathrm{S}^\star=\bm{\varPsi}_k^\star,\,\forall k\in \mathcal{K}_\mathrm{I}$, the optimal $\bm{R}_\mathrm{S}^\star$ can be expressed as
\begin{align}
	\bm{R}_\mathrm{S}^\star&=\sum_{n=1}^{\tilde{N}}{P_{n}^\mathrm{S}\tilde{\bm{q}}_n\tilde{\bm{q}}_n^H},
\end{align}
where $P_n^\mathrm{S}\!\ge\! 0,\,n\!=\!1,...,\tilde{N}$ and $\sum_{n=1}^{\tilde{N}}{\!\big(\sum_{k=1}^K{P_{k,n}^\mathrm{C}}\!+\!P_{n}^\mathrm{S}\big)}\!=\!P$.

In the following, we show that we can construct another solution $(\{\tilde{\bm{R}}_k^\star\}_{k=1}^K,\tilde{\bm{R}}_\mathrm{S}^\star)$ that is also an optimal solution to (P2) with $\mathrm{rank}(\tilde{\bm{R}}_k^\star)=1$ in the following two cases.

Firstly, we consider the re-construction when $\tilde{N}=1$ and $1\le \mathrm{card}(\mathcal{K}_{\mathrm{A}})<K$. In this case, we set
\begin{align}
	\tilde{\bm{R}}_k^\star&=P_{k,f}\bm{f}_k\bm{f}^H_k,\quad 1\le k\le \mathrm{card}(\mathcal{K}_{\mathrm{A}}),\\
	\tilde{\bm{R}}_{\mathrm{card}(\mathcal{K}_{\mathrm{A}})+1}^\star&=(\sum\nolimits_{n=1}^{\mathrm{card}(\mathcal{K}_{\mathrm{A}}) +1}{P_{n,1}^\mathrm{C}}+P_1^\mathrm{S})\tilde{\bm{q}}_1\tilde{\bm{q}}_1^H,\\
	\tilde{\bm{R}}_k^\star&=P_{k,1}^\mathrm{C}\tilde{\bm{q}}_1\tilde{\bm{q}}_1^H,\quad\ \mathrm{card}(\mathcal{K}_{\mathrm{A}})\! +\!2\leq k\leq K,\\
	\tilde{\bm{R}}_\mathrm{S}^\star&=\bm{0}.
\end{align}
It can be verified that the solution $(\{\tilde{\bm{R}}_k^\star\}_{k=1}^K,\tilde{\bm{R}}_\mathrm{S}^\star)$ achieves the optimal value to (P2) and satisfies the constraints in (\ref{P2_PCRB}) and (\ref{P2_power})-(\ref{P2_PSD_S}). Due to the fact that $\bm{h}_k^H\tilde{\bm{J}}=\bm{0},\ k\in \mathcal{K}_\mathrm{A}$, the constraints in (\ref{P2_rate}) hold for $1\le k\le \mathrm{card}(\mathcal{K}_{\mathrm{A}})$. For $k\in \mathcal{K}_\mathrm{I}$, we have
\begin{align}
	&\bm{h}_{\mathrm{card}(\mathcal{K}_{\mathrm{A}})+1}^{H}\tilde{\bm{R}}_{\mathrm{card}(\mathcal{K}_{\mathrm{A}})+1}^{\star}\bm{h}_{\mathrm{card}(\mathcal{K}_{\mathrm{A}})+1}\\
	&\ge \bm{h}_{\mathrm{card}(\mathcal{K}_{\mathrm{A}})+1}^{H}\bm{R}_{\mathrm{card}(\mathcal{K}_{\mathrm{A}})+1}^{\star}\bm{h}_{\mathrm{card}(\mathcal{K}_{\mathrm{A}})+1}\nonumber\\
	&\ge\gamma_{\mathrm{card}(\mathcal{K}_{\mathrm{A}})+1}(\sum\nolimits_{j\ne \mathrm{card}(\mathcal{K}_{\mathrm{A}})+1}\!\!\!{\bm{h}_{\mathrm{card}(\mathcal{K}_{\mathrm{A}})+1}^{H}\bm{R}_{j}^{\star}\bm{h}_{\mathrm{card}(\mathcal{K}_{\mathrm{A}})+1}}\nonumber\\
	&+\!\bm{h}_{\mathrm{card}(\mathcal{K}_{\mathrm{A}})+1}^{H}\bm{R}_\mathrm{S}^{\star}\bm{h}_{\mathrm{card}(\mathcal{K}_{\mathrm{A}})+1}+\sigma _\mathrm{C}^{2})\nonumber\\
	&\ge\gamma _{\mathrm{card}(\mathcal{K}_{\mathrm{A}})+1}(\sum\nolimits_{j\ne \mathrm{card}(\mathcal{K}_{\mathrm{A}})+1}\!\!\!{\bm{h}_{\mathrm{card}(\mathcal{K}_{\mathrm{A}})+1}^{H}\tilde{\bm{R}}_{j}^{\star}\bm{h}_{\mathrm{card}(\mathcal{K}_{\mathrm{A}})+1}}\nonumber\\
	&+\sigma _\mathrm{C}^{2})\nonumber
\end{align}
and
\begin{align}
	&\bm{h}_{k}^{H}\tilde{\bm{R}}_{k}^{\star}\bm{h}_k=\bm{h}_{k}^{H}\bm{R}_{k}^{\star}\bm{h}_k\\
	&\ge \gamma _k(\sum_{j\ne k}{\bm{h}_{k}^{H}\bm{R}_{j}^{\star}\bm{h}_k}+\bm{h}_{k}^{H}\bm{R}_\mathrm{S}^{\star}\bm{h}_k+\sigma _\mathrm{C}^{2})\nonumber\\
	&=\gamma _k(\sum_{j\ne k}{\bm{h}_{k}^{H}\tilde{\bm{R}}_{j}^{\star}\bm{h}_k}+\sigma _\mathrm{C}^{2}),\,  \mathrm{card}(\mathcal{K}_{\mathrm{A}})+1< k\leq K.\nonumber
\end{align}
Thus, $(\{\tilde{\bm{R}}_k^\star\}_{k=1}^K,\tilde{\bm{R}}_\mathrm{S}^\star)$ is an optimal solution to (P2), which means \emph{no} dedicated sensing beam is needed. The optimal solution to (P1) can be constructed as $\bm{S}^\star=\bm{0}$,
\begin{equation}
	\!\!\!\!\!\!\!\!	{\bm{w}}_k^\star=\begin{cases}
		\sqrt{P_{k,f}}\bm{f}_k,\quad  & k\!\in\! \mathcal{K}_\mathrm{A}\\
		\sqrt{\sum_{i=1}^{\mathrm{card}(\!\mathcal{K}_{\mathrm{A}}\!) \!+\!1}{P_{i,1}^\mathrm{C}}\!\!+\!\!P_1^\mathrm{S}}\tilde{\bm{q}}_1,\!\!\!\!\!\! &k\!=\!\mathrm{card}(\mathcal{K}_{\mathrm{A}}\!)+1\\
		\sqrt{P_{k,1}^\mathrm{C}}\tilde{\bm{q}}_1,\quad & k\!\in\! \mathcal{K} _\mathrm{I}\backslash \{\mathrm{card}(\mathcal{K}_{\mathrm{A}}\!)\!\!+\!\!1\}.
	\end{cases} \!\!\!\!\!\!\!
\end{equation}

Secondly, if $\tilde{N}=1$ and $\mathrm{card}(\mathcal{K}_{\mathrm{A}}) =K$, then we have $\tilde{\bm{R}}_k^\star=P_{k,f}\bm{f}_k\bm{f}^H_k,\ k\in \mathcal{K}$ and $\tilde{\bm{R}}_\mathrm{S}^\star=\big(\sum_{i=1}^{K}{P_{i,1}^\mathrm{C}}+P_1^\mathrm{S}\big)\tilde{\bm{q}}_1\tilde{\bm{q}}_1^H$, which means \emph{at most} \emph{one} dedicated sensing beam is needed. The optimal solution to (P1) can be constructed as
${\bm{w}}_k^\star=\sqrt{P_{k,f}}\bm{f}_k,\forall k\in \mathcal{K}$ and ${\bm{S}}^\star=\sqrt{\sum_{i=1}^{K}{P_{i,1}^\mathrm{C}}+P_1^\mathrm{S}}\tilde{\bm{q}}_1$. This thus completes the proof of Proposition 3.

\vspace{-5mm}
\section{Proof of Proposition 4}\label{proof_prop_case_III}
In Case III, based on (\ref{KKT_2}), we have $\mathrm{rank}(\bm{\varPsi}_k^\star)=N_t$ and $\bm{R}_k^\star=\bm{0}$ for all $k$'s that satisfy $\lambda_{\max}\big( \bm{U}_\mathrm{M}^\star-\sum_{i=1}^{\mathrm{card}(\mathcal{K}_{\mathrm{A}})}{\gamma _i\nu _{i}^{\star}\bm{h}_i\bm{h}_{i}^{H}}+\nu_k^\star(\gamma_k+1)\bm{h}_k\bm{h}_k^H \big)=\lambda _{\max}\big(\bm{U}_\mathrm{M}^\star-\sum_{i=1}^{\mathrm{card}(\mathcal{K}_{\mathrm{A}})}{\gamma _i\nu _{i}^{\star}\bm{h}_i\bm{h}_{i}^{H}}\big)$, which contradicts with the rate constraints. Therefore, the condition $\lambda _{\max}\big( \bm{U}_\mathrm{M}^\star-\sum_{i=1}^{\mathrm{card}(\mathcal{K}_{\mathrm{A}})}{\gamma _i\nu _{i}^{\star}\bm{h}_i\bm{h}_{i}^{H}}+\nu_k^\star(\gamma_k+1)\bm{h}_k\bm{h}_k^H \big)> \lambda _{\max}\big(\bm{U}_\mathrm{M}^\star-\sum_{i=1}^{\mathrm{card}(\mathcal{K}_{\mathrm{A}})}{\gamma _i\nu _{i}^{\star}\bm{h}_i\bm{h}_{i}^{H}}\big)$ should hold for all users to ensure that $\bm{\varPsi}_k^\star$ is singular. Consequently, we have $\mathrm{card}(\mathcal{K}_{\mathrm{A}})=K$ and $\mathcal{K}_{\mathrm{I}}=\varnothing$, i.e., all communication rate constraints are active. According to (\ref{KKT_3}), we have $\mathrm{rank}(\bm{\varPsi}_\mathrm{S}^\star)=N_t$ and $\bm{R}_\mathrm{S}^\star=\bm{0}$. Moreover, since $\bm{\varPsi}_k^\star=\bm{\varPsi}_\mathrm{S}^\star-\nu_k^\star(\gamma_k+1)\bm{h}_k\bm{h}_k^H,\ \forall k\in\mathcal{K}$, we have $N_t-1\leq \mathrm{rank}(\bm{\varPsi}_k^\star)\leq N_t$. Since $\mathrm{det}(\bm{\varPsi}_k^\star)=0$, we can obtain that $\mathrm{rank}(\bm{\varPsi}_k^\star)=N_t-1$. Thus, $\mathrm{rank}(\bm{R}_k^\star)=1,\forall k\in \mathcal{K}$. The optimal solution to (P1) can be obtained via $\bm{R}_k^\star=\bm{w}_k^\star{\bm{w}_k^\star}^H,\forall k\in \mathcal{K}$ and $\bm{S}^\star=\bm{0}$. This thus completes the proof of Proposition 4.


\vspace{-5mm}
\section{Proof of Lemma 2}\label{Appendix_lemma_P1_2}
Consider this problem with only one dual-functional communication beam. The problem can be written as:
\begin{align}
	\mbox{(P2-HS)}\quad\max_{\scriptstyle \boldsymbol{R}_1^\mathrm{C}\succeq \mathbf{0}:\atop \scriptstyle \mathrm{tr}( \boldsymbol{R}_1^\mathrm{C}) \leq P,\mathrm{rank}(\bm{R}_1^\mathrm{C})=1}\quad&\mathrm{tr}( \boldsymbol{A}_1\boldsymbol{R}_1^\mathrm{C}) \\[-2mm]
	\mathrm{s.t.}\qquad\qquad& \boldsymbol{h}_1^H\boldsymbol{R}_1^\mathrm{C}\boldsymbol{h}_1\ge \gamma_1\sigma _{\mathrm{C}}^{2}.
\end{align}
Denote (P2-HS-R) as the relaxed version of (P2-HS) by dropping the rank-one constraint on $\bm{R}_1^{\mathrm{C}}$. According to \cite{huang2009rank}, the SDR from (P2-HS) to (P2-HS-R) is guaranteed to be tight, thus (P2-HS) is equivalent to (P2-HS-R). Moreover, for any feasible solution to (P2-R) denoted by $(\bm{R}_1,\bm{R}_\mathrm{S})$, we can always construct a feasible solution $\boldsymbol{R}_1^\mathrm{C}=\bm{R}_1+\bm{R}_\mathrm{S}$ to (P2-HS-R) that achieves the same objective value as that of (P2-R). Hence, the optimal solution to (P2-R) is also a feasible solution to (P2-HS-R), which indicates that the optimal value to (P2-R) is less than or equal to that of (P2-HS-R). On the other hand, since the optimal solution to (P2-HS-R) denoted by ${\bm{R}_1^{\mathrm{C}}}^\star$ satisfies $\mathrm{tr}(\bm{A}_1{\bm{R}_1^{\mathrm{C}}}^\star)=\mathrm{tr}(\bm{A}_1(\bm{R}_1^\star+\bm{R}_\mathrm{S}^\star))$ with $\bm{R}_1^\star={\bm{R}_1^{\mathrm{C}}}^\star$ and $\bm{R}_\mathrm{S}^\star=\bm{0}$, $(\bm{R}_1^\star,\bm{R}_\mathrm{S}^\star)$ is feasible for (P2-R) and achieves the same objective value as the optimal value to (P2-HS-R). Thus, the optimal value of (P2-R) is no smaller than that of (P2-HS-R). Hence, (P2-R) and (P2-HS-R) share the same optimal value, indicating (P2-R) is equivalent to (P2-HS-R) and consequently (P2-HS). Since (P2) is equivalent to (P2-R), (P2) is consequently equivalent to (P2-HS). For any optimal ${\bm{R}_1^{\mathrm{C}}}^\star$ to (P2-HS), $(\bm{R}_1^\star={\bm{R}_1^{\mathrm{C}}}^\star, \bm{R}_\mathrm{S}^\star=\bm{0})$ is an optimal solution to (P2). Thus, \emph{no} sensing beam is needed with homogeneous targets and $K=1$. Lemma 2 is thus proved.

\vspace{-4mm}
\section{Proof of Equivalence Between (P3) and (P4)}\label{proof_equivalence}
Given any feasible solution $(\{\bm{R}_k\}_{k=1}^K,\bm{R}_\mathrm{S})$ to (P3'), by constructing $y_m'=\big(1+\beta _m\mathrm{tr}\big(\bm{A}_m\big(\sum_{k=1}^K{\bm{R}_k}+\bm{R}_\mathrm{S}\big)\big)+\delta_m\big)^{-\frac{1}{2}},\ \forall m\in\mathcal{M}$, $(\{\bm{R}_k\}_{k=1}^K,\bm{R}_\mathrm{S},\bm{y}^\prime)$ is feasible for (P4) and achieves the same objective value as that of (P3'), thus the optimal value of (P4) is no smaller than that of (P3'). Moreover, since the optimal solution to (P4) denoted by $(\{\bm{R}_k^\star\}_{k=1}^K,\bm{R}_\mathrm{S}^\star,\bm{y}^\star)$ satisfies $\sum_{m=1}^M\sqrt{1-y_m^{\star2}}=\sum_{m=1}^M\big(1+\frac{1}{\beta_m\mathrm{tr}\left(\bm{A}_m\left(\sum_{k=1}^K{\bm{R}_k^\star}+\bm{R}_\mathrm{S}^\star\right)\right)+\delta_m}\big)^{-\frac{1}{2}}$, $(\{\bm{R}_k^\star\}_{k=1}^K,\bm{R}_\mathrm{S}^\star)$ is feasible for (P3') and achieves the same objective value as the optimal value of (P4). Thus, the optimal value of (P3') is no smaller than that of (P4). Hence, (P4) and (P3') have the same optimal value, which implies that (P3) is equivalent to (P4) by noting that \hbox{(P3) is equivalent to (P3').}

\begin{table}[t]
	\centering
		\caption{Number of Sensing Beams Needed and Further Reduced under Tighter Bounds.}
		\label{tab:beam_reduction}
		\vspace{-4mm}
		\text{(a) Min-max problem}
		\vspace{-0mm}
		\scalebox{1}{
			\begin{tabular}{|c|c|c|c|}
				\hline
				\multicolumn{2}{|c|}{\makecell{Solution obtained with\\ only the
						general rank\\ reduction method}} &
				\multicolumn{2}{c|}{\begin{tabular}[c]{@{}c@{}}\makecell{Solution reduced with\\ discussion of the three cases in\\ different rate regimes}\end{tabular}} \\
				\hline
				\begin{tabular}[c]{@{}c@{}}Number of sensing\\ beams needed\end{tabular} &
				\begin{tabular}[c]{@{}c@{}}Percentage\\ (\%)\end{tabular} &
				\begin{tabular}[c]{@{}c@{}}Number of sensing\\ beams reduced\end{tabular} &
				\begin{tabular}[c]{@{}c@{}}Percentage\\ (\%)\end{tabular} \\
				\hline
				0 & 57.16 & \multicolumn{2}{c|}{N.A.} \\
				\hline
				1 & 33.83 & 1 & 32.11 \\
				\hline
				\multirow{2}{*}{2} & \multirow{2}{*}{8.11} & 1 & 0 \\
				\cline{3-4}
				&  & 2 & 9.30 \\
				\hline
				3 & 0.90 & 1/2/3 & 0 \\
				\hline
	\end{tabular}}
	\par\vspace{1mm}
		\text{(b) Min-sum problem}
		\vspace{0mm}
		\scalebox{1}{
			\begin{tabular}{|c|c|c|c|}
				\hline
				\multicolumn{2}{|c|}{\makecell{Solution obtained with\\ only the
						general rank\\ reduction method}
				} &
				\multicolumn{2}{c|}{\makecell{Solution reduced with\\ discussion of the three cases in\\ different rate regimes}} \\
				\hline
				\begin{tabular}[c]{@{}c@{}}Number of sensing\\ beams needed\end{tabular} &
				\begin{tabular}[c]{@{}c@{}}Percentage\\ (\%)\end{tabular} &
				\begin{tabular}[c]{@{}c@{}}Number of sensing\\ beams reduced\end{tabular} &
				\begin{tabular}[c]{@{}c@{}}Percentage\\ (\%)\end{tabular} \\
				\hline
				0 & 33.68 & \multicolumn{2}{c|}{N.A.} \\
				\hline
				1 & 29.17 & 1 & 99.97 \\
				\hline
				\multirow{2}{*}{2} & \multirow{2}{*}{23.52} & 1 & 1.80 \\
				\cline{3-4}
				&  & 2 & 87.57 \\
				\hline
				\multirow{3}{*}{3} & \multirow{3}{*}{13.63} & 1 & 0 \\
				\cline{3-4}
				&  & 2 & 5.64 \\
				\cline{3-4}
				&  & 3 & 78.27 \\
				\hline
	\end{tabular}}
\end{table}

\vspace{-4.5mm}
\section{Numerical Results on Optimal Beamforming Designs under General and Tighter Bounds}\label{Appendix_numerical_results}
To demonstrate the necessity and the effectiveness of our proposed Propositions 2-4 and Proposition 6, Table \ref{tab:beam_reduction} shows the percentage distribution of the number of sensing beams needed and further reduced via discussion of the different cases through Monte Carlo simulations. Specifically, we generate $100$ independent system setup realizations with $M\le15$, where communication channels and the targets’ PDFs are also generated randomly. For each realization, the rate requirements range from $0.5$ bps/Hz to the maximum achievable rates, yielding more than $10,000$ realizations in total. As shown in the table, discussing the three cases reduces the number of sensing beams needed in a large fraction of realizations, which further validates the importance of the discussion of these cases (regimes).
	
\end{document}